%% file: P3Mpaper_repre.tex
\documentstyle[prb,aps,amsfonts,floats,twocolumn,epsfig]{revtex}

\begin{document}
\title{How to mesh up Ewald sums (I): A theoretical and numerical \\
       comparison of various particle mesh routines}
\author{Markus Deserno, Christian Holm}
\address{Max-Planck-Institut f{\"u}r Polymerforschung, 
Ackermannweg 10, 55128 Mainz, Germany}
\date{April 27, 1998}
\maketitle



\newcommand{\erfc}{{\mbox{erfc}}}
\newcommand{\infd}{{\mbox{d}}}
\newcommand{\mod}{{\mbox{{~}mod{~}}}}

\newcommand{\MM}{{\Bbb{M}}}
\newcommand{\NN}{{\Bbb{N}}}
\newcommand{\RR}{{\Bbb{R}}}
\newcommand{\ZZ}{{\Bbb{Z}}}

\newcommand{\VECd}{{\mathbf{d}}}
\newcommand{\VECe}{{\mathbf{e}}}
\newcommand{\VECk}{{\mathbf{k}}}
\newcommand{\VECl}{{\mathbf{l}}}
\newcommand{\VECm}{{\mathbf{m}}}
\newcommand{\VECn}{{\mathbf{n}}}
\newcommand{\VECr}{{\mathbf{r}}}
\newcommand{\VECD}{{\mathbf{D}}}
\newcommand{\VECE}{{\mathbf{E}}}
\newcommand{\VECF}{{\mathbf{F}}}
\newcommand{\VECR}{{\mathbf{R}}}

\newcommand{\CALC}{{\mathcal C}}
\newcommand{\CALI}{{\mathcal I}}
\newcommand{\CALL}{{\mathcal L}}
\newcommand{\CALM}{{\mathcal M}}
\newcommand{\CALR}{{\mathcal R}}

\newcommand{\exa}{{^{\mbox{\footnotesize exa}}}}
\newcommand{\maxi}{{_{\mbox{\scriptsize max}}}}
\newcommand{\Mesh}{{_{\mbox{\scriptsize M}}}}
\newcommand{\nd}{{^{\mbox{\footnotesize nd}}}}
\newcommand{\opt}{{_{\mbox{\footnotesize opt}}}}
\newcommand{\self}{{^{\mbox{\scriptsize self}}}}
\newcommand{\th}{{^{\mbox{\footnotesize th}}}}


\hspace{2cm}\begin{abstract}
\begin{centering}
\parbox{16cm}{
\vspace{-1cm}
\hfill
\parbox{14.2cm}{
Standard Ewald sums, which calculate e.g.\ the electrostatic 
energy or the force in periodically closed systems of charged 
particles, can be efficiently speeded up by the use of the Fast 
Fourier Transformation (FFT). 
In this article we investigate three algorithms for the 
FFT-accelerated Ewald sum, which attracted a widespread attention,
namely, the so-called par\-tic\-le-par\-tic\-le--par\-tic\-le-mesh 
(P$^{3}$M), particle mesh Ewald (PME) and smooth PME method.
We present a unified view of the underlying techniques and
the various ingredients which comprise those routines.
Additionally, we offer detailed accuracy measurements, which shed 
some light on the influence of several tuning parameters and also
show that the existing methods -- although similar in spirit -- 
exhibit remarkable differences in accuracy.
We propose combinations of the individual components, mostly relying
on the P$^{3}$M approach, which we regard as most flexible.
}
}
\end{centering}
\end{abstract}



\narrowtext


\section*{Introduction}

A challenging task in every computer simulation of particles which 
are subject to periodic boundary conditions and long range 
interactions is the efficient calculation of quantities like the 
interparticle forces or the interaction energies. 
The famous Ewald sum \cite{Ewald,LPS} does a remarkable job in 
splitting the very slowly (not even unconditionally) converging 
sum over the Coulomb potential into two sums which converge 
exponentially fast. Still, this method suffers from two deficits:
First, it is computationally demanding, since one part of the problem 
is solved in reciprocal space, thereby implying the need for several 
Fourier transformations.
Second, the algorithm scales like $N^2$ (with $N$ being the number of
charged particles in the simulation box) or at best like $N^{3/2}$, 
if one uses cutoffs which are optimized with respect to the splitting 
parameter\cite{PPL}.

Several methods have been proposed to tackle the first problem, e.g.\ 
tabulation of the complete Ewald potential \cite{SD} or the use of 
polynomial approximations, in particular expansion of the non-spherical
contributions to the Ewald potential in cubic harmonics \cite{SD,AD}. 
Apart from the difficulty of a computational overhead which might 
strongly increase with the desired accuracy, all these methods do not 
solve the second problem: the unfavorable scaling with particle number.

The essential idea is {\em not} to avoid the Fourier transforms but 
to modify the problem in a way that permits an employment of the {\em 
Fast Fourier Transformation}\cite{PTVF} (FFT), thereby reducing the 
complexity of the reciprocal part of the Ewald sum to essentially order 
$N\log N$. If the real space cutoff is chosen small enough, this 
scaling applies to the complete Ewald sum. 
Since the FFT is a grid transformation, there are discretization 
problems to be solved and corresponding discretization errors to be 
minimized.

At present there exist several mesh implementations of the Ewald 
sum  -- similar in spirit but different in detail. In this article we 
will focus on the original par\-tic\-le-par\-tic\-le--par\-tic\-le-mesh 
(P$^{3}$M) method of Hockney and Eastwood \cite{HE} and two variants, 
namely, the particle mesh Ewald (PME) method of Darden {\em et.\,al}.\ 
\cite{DYP} and an upgrade of the latter by Essmann {\em et.\,al}.\ 
\cite{EPBDLP}, which we will refer to as SPME (the ``S'' stands for 
``smooth'').

There have been some uncertainties in the literature concerning the
relative performance of these methods and it has been shown 
previously\cite{Gunny} that the P$^{3}$M approach -- the oldest of 
the three -- is actually the most accurate one and should be the 
preferred choice.
However, since in this reference the PME method was combined with
a disadvantageous charge assignment scheme and the more recent
SPME could not be considered, we found it worthwhile to test 
again these three methods under similar well posed and reproducible
conditions and a larger number of tuning parameters.

\begin{figure}[t]
\vspace{4.2cm}
\end{figure}

The original literature on particle mesh routines is mostly not easy to 
digest for the layman, obscured by the fact that the various authors approach 
the problem from different directions and use different notations. 
In this article we try to present a unified view of the common methods 
and analyze in detail the ingredients comprising them.
By this we want to uncover the large number of possibilities for
combining the different parts, thus allowing a judicious balance
of accuracy, speed and ease of implementation. Moreover, we show that due 
to some subtle interdependencies not all combinations are advantageous, 
although they might appear promising at first sight.

This paper is structured as follows: 
In the first section we briefly review the idea of the Ewald sum and provide 
the most important formulas. Then we describe in some detail the steps which 
must be carried out if FFT algorithms are to be employed for the Fourier 
transformation, namely: charge assignment onto a mesh, solving Poisson's 
equation on that mesh, differentiating the potential to obtain the forces and 
interpolating the mesh based forces back to the particles.
All these steps can be performed in different ways and in the following
section we investigate in detail their accuracy, in particular, we 
compare P$^{3}$M, PME and SPME with respect to their root
mean square error in the force.
Based on our theoretical and numerical investigations, we find that the most 
accurate and versatile routine is the P$^{3}$M method, supplemented by 
ingredients (dealing with the differentiation) of the other two approaches.

The important task of an optimal tuning of the parameters -- especially 
the Ewald parameter $\alpha$ -- is satisfactorily solved for the standard
Ewald and the PME method, since there exist accurate analytic estimates 
for the root mean square error in the force. We will tackle the
corresponding problem for P$^{3}$M in a forthcoming publication \cite{wir}.


\section*{The Ewald sum}

There are many examples of long range interactions which can be 
treated by Ewald techniques, but in this article we will solely be concerned 
with Coulomb point charges, i.e.\ with an interaction potential $1/r$. 
Consider therefore a system of $N$ particles with charges $q_{i}$ at 
positions $\VECr_{i}$ in an overall neutral and (for simplicity) cubic 
simulation box of length $L$ and volume $V_{b}=L^{3}$. 
If periodic boundary conditions are applied, the total electrostatic 
energy of the box is given by

\begin{equation}\label{Boxenergie}
E = \frac{1}{2}\sum_{i,j=1}^{N}\sum_{\VECn\in\ZZ^{3}}^{\prime}
\frac{q_{i}q_{j}}{|\VECr_{ij}+\VECn L|} 
\end{equation}

The sum over $\VECn$ takes into account the periodic images
of the charges and the prime indicates that in the case $i=j$ 
the term $\VECn=0$ must be omitted. 
Of course $\VECr_{ij}=\VECr_{i}-\VECr_{j}$, and our unit 
conventions are shortly described in Appendix 
\ref{unitconventions}.

Strictly speaking, since this sum is only {\em conditionally} 
convergent, its value is not well defined unless one spe\-ci\-fies 
the precise way in which the cluster of simulation boxes is supposed 
to fill the $\RR^{3}$, i.e.\ its shape (e.g.\ approximately spherical 
\cite{BBT}) and the conditions outside the cluster (e.g.\ vacuum or 
some dielectric).
A thorough discussion is given elsewhere\cite{LPS,Caillol}.

The slowly decaying long range part of the Coulomb potential renders
a straightforward summation of Eqn.\ (\ref{Boxenergie}) impracticable.
The trick is to split the problem into two parts by the following 
trivial identity:
\begin{equation}\label{EwaldAufspaltung}
\frac{1}{r} = \frac{f(r)}{r} + \frac{1-f(r)}{r}
\end{equation}
The underlying idea is to distribute the two main complications 
of the Coulomb potential -- its rapid variation at small $r$ and 
its slow decay at large $r$ -- between the two terms by a suitable
choice of $f$. In particular:
\begin{itemize}
\item The first part $\frac{f(r)}{r}$ should be negligible 
  (or even zero) beyond some cutoff $r\maxi$, so that summation 
  up to the cutoff is a good approximation to (or the exact result 
  of) this contribution to the total electrostatic potential.
\item The second part $\frac{1-f(r)}{r}$ should be a slowly varying 
  function for {\em all} $r$, so that its Fourier transform can be 
  represented by only a few $\VECk$-vectors with $|\VECk|\le 
  k\maxi$. This permits an efficient calculation of this contribution 
  to the total electrostatic potential in reciprocal space.
\end{itemize}
Since the field equations are linear, the sum of these two contributions 
gives the solution for the potential of the original problem.
However, the two requirements on the function $f$ mentioned 
above leave a large freedom of choice \cite{H}. 
The traditional selection is the complementary error function 
$\erfc(r) := 2\pi^{-1/2}\int_{r}^{\infty}\infd t \exp(-t^{2})$, 
which results in the well known Ewald formula for the
electrostatic {\em energy} of the box:
\begin{equation}\label{EwaldAnteile}
E = E^{(r)} + E^{(k)} + E^{(s)} + E^{(d)}
\end{equation}
where the contribution from real space $E^{(r)}$, the
contribution from reciprocal space $E^{(k)}$, the self 
energy $E^{(s)}$ and the dipole correction $E^{(d)}$ are 
given by
\begin{eqnarray}
E^{(r)} & = & \frac{1}{2} \sum_{i,j} \sum_{\VECm\in\ZZ^{3}}^{\prime}
q_{i}q_{j} \frac{\erfc(\alpha|\VECr_{ij}+\VECm L|)}
{|\VECr_{ij}+\VECm L|} \label{Realraumanteil} \\
E^{(k)} & = & \frac{1}{2}\frac{1}{L^{3}} \sum_{\VECk\ne 0}
\frac{4\pi}{k^{2}} e^{-k^{2}/4\alpha^{2}}
|\tilde{\rho}(\VECk)|^{2} \label{Impulsraumanteil} \\
E^{(s)} & = & -\frac{\alpha}{\sqrt{\pi}}\sum_{i}q_{i}^{2}
\label{Selbstenergie} \\
E^{(d)} & = &
\frac{2\pi}{(1+2\epsilon')L^{3}}\left(\sum_{i}q_{i}\VECr_{i}\right)^{2}
\label{Dipolkorrektur}
\end{eqnarray}
and the Fourier transformed charge density $\tilde{\rho}(\VECk)$
is defined as
\begin{equation}\label{FLadungsdichte}
\tilde{\rho}(\VECk) = 
\int_{V_{b}}\infd^{3}r \; \rho(\VECr)e^{-i\;\VECk\cdot\VECr} =
\sum_{j=1}^{N}q_{j}\;e^{-i\,\VECk\cdot\VECr_{j}}
\end{equation}
The inverse length $\alpha$, which we will refer to as the
{\em Ewald parameter}, tunes the relative weight of the real 
space and the reciprocal space contribution, but the final result 
is of course independent of $\alpha$. The $\VECk$-vectors form 
the discrete set $\{2\pi\VECn/L:\VECn\in\ZZ^{3}\}$.

The form (\ref{Dipolkorrektur}) given for the dipole correction 
assumes that the set of periodic replications of the simulation box 
tends in a spherical way towards an infinite cluster and that the 
medium outside this sphere is a homogeneous dielectric\cite{LPS,Caillol} 
with dielectric constant $\epsilon'$.
Note that the case of a surrounding vacuum corresponds to 
$\epsilon'=1$ and that the dipole correction vanishes for metallic 
boundary conditions, since then $\epsilon'=\infty$.
Note also that this term is independent of $\alpha$, which again 
shows that it is not specific to the Ewald sum but more generally
reflects the problems inherent to the conditional convergence
of the sum in Eqn.\ (\ref{Boxenergie}). 
Some complications regarding the correct implementation of this term
are discussed by Caillol\cite{Caillol}.

The advantage of rewriting Eqn.\ (\ref{Boxenergie}) this way is that
the exponentially converging sums over $\VECm$ and $\VECk$ in
(\ref{Realraumanteil},\ref{Impulsraumanteil}) allow the introduction 
of relatively small cutoffs without much loss in accuracy. Typically 
one chooses $\alpha$ large enough as to employ the minimum image 
convention in Eqn.\ (\ref{Realraumanteil}). 
It is important to realize that at given real- and reciprocal space 
cutoffs there exists an {\em optimal} $\alpha$ such that the accuracy 
of the approximated Ewald sum is as high as possible. This optimal 
value can be determined easily with the help of the excellent estimates 
for the cutoff errors derived by Kolafa and Perram\cite{KP} -- 
essentially by demanding that the real- and reciprocal space contribution 
to the error should be equal. 

Some more insight into the Ewald sum can be gained by the following
considerations. Let $\tilde{g}(\VECk):=4\pi/k^{2}$ be the Fourier 
transformed Green function of the Coulomb potential $1/r$  and 
$\tilde{\gamma}(\VECk):=\exp(-k^{2}/4\alpha^{2})$. Then
Eqn.\ (\ref{Impulsraumanteil}) can be rewritten as follows:
\begin{eqnarray}
E^{(k)} & = 
& \frac{1}{2}\sum_{j}q_{j}
\left(
\frac{1}{L^{3}}\sum_{\VECk\ne 0} 
\tilde{g}(\VECk) \tilde{\gamma}(\VECk) \tilde{\rho}(\VECk) 
e^{i\;\VECk\cdot\VECr_{j}}
\right) \nonumber\\
{} & =: & \frac{1}{2}\sum_{j} q_{j} \, \phi^{(k)}(\VECr_{j}) 
\label{Impulsraumanteil_}
\end{eqnarray}
Here $\phi^{(k)}(\VECr_{j})$ is the electrostatic potential at 
the point $\VECr_{j}$ due to the {\em second} term in Eqn.\ 
(\ref{EwaldAufspaltung}) and by definition it is clear that 
its Fourier transform is given by
\begin{equation}\label{elektrostatischesPotential}
\tilde{\phi}^{(k)}(\VECk) = \tilde{g}(\VECk) \tilde{\gamma}(\VECk) 
\tilde{\rho}(\VECk)
\end{equation}
As is known, products in reciprocal space correspond to convolutions 
in real space.
Hence Eqn.\ (\ref{elektrostatischesPotential}) shows that the reciprocal 
space contribution to the electrostatic potential is created by a 
charge distribution which is obtained from the original point charge 
distribution by a convolution with a ``smearing function'' $\gamma(\VECr)$. 

For the standard Ewald sum $\gamma(\VECr)$ is a Gaussian, i.e.\ 
$\gamma(\VECr)=\alpha^{3}\pi^{-3/2}\exp(-\alpha^{2}r^{2})$, 
but this is merely a consequence of choosing the splitting function 
$f$ in Eqn.\ (\ref{EwaldAufspaltung}) to be the complementary error 
function. 
In fact, an alternative method\cite{AT} to motivate the splitting which 
was done in Eqn.\ (\ref{EwaldAufspaltung}) is to replace the point 
charge distribution $\rho$ by a screened charge distribution 
$\rho-\rho\star\gamma$ and compensate this screening by adding the 
smeared charge distribution $\rho\star\gamma$.
(The star denotes the convolution operation.) From a mathematical 
point of view these two interpretations are perfectly equivalent: 
Instead of splitting the potential one splits the charge density.

At this point a word of caution seems appropriate: Whereas the 
electrostatic {\em potential} depends linearly on the charge density,
the electrostatic {\em energy} does not. 
Thus, calculating the energies resulting from the charge densities
$\rho-\rho\star\gamma$ and $\rho\star\gamma$ and adding these 
contributions together would {\em not} give the energy of the charge density 
$\rho$. Consequently, $E^{(k)}$ is not the electrostatic
energy of a charge density $\rho\star\gamma$ but the {\em Fourier 
space contribution} to the electrostatic energy of the charge density
$\rho$. We want to make this subtle point more clear by writing down 
the energy explicitly. If we denote with $\phi_{\rho}$ the potential 
originating from $\rho$, we have due to the linear dependence of 
$\phi_{\rho}$ on $\rho$ an equation like
$\phi_{\rho} = \phi_{\rho-\rho\star\gamma} + \phi_{\rho\star\gamma}$.
Hence we can obtain for the electrostatic energy the following expression:
\begin{eqnarray}
E'  & = & \frac{1}{2} \int \infd^{3}r \; \rho(\VECr) \, 
\phi_{\rho}(\VECr) \nonumber \\
{} & = & \frac{1}{2} \int \infd^{3}r \; \rho(\VECr) \,
\Big[\phi_{\rho-\rho\star\gamma}(\VECr) + 
\phi_{\rho\star\gamma}(\VECr)\Big] \nonumber \\
{} & = & \frac{1}{2} \int \infd^{3}r \; \rho(\VECr) \,
\phi_{\rho-\rho\star\gamma}(\VECr)
+  \frac{1}{2} \int \infd^{3}r \; \rho(\VECr) \,
\phi_{\rho\star\gamma}(\VECr) \label{Energiesubtilitaet}
\end{eqnarray}
The two terms in the last line are the real space and the Fourier space
contribution to the energy, but neither of them can be interpreted as the 
energy of a charge distribution $\rho-\rho\star\gamma$ or $\rho\star\gamma$! 
Moreover, the quantity $E'$ contains unphysical self energy contributions, 
i.e.\ energy due to the interaction of a charge (or $\gamma$-smeared charge) 
with itself. In the actual Ewald sum the self energy contribution of the real 
space part is canceled by omitting the term $\VECm=0$ for $i=j$ in Eqn.\ 
(\ref{Realraumanteil}), whereas the self energy contribution of
$E^{(k)}$ must be subtracted separately (this is the origin of
the term $E^{(s)}$).  

Finally, the force $\VECF_{i}$ on particle $i$ is obtained by 
differentiating the electrostatic potential energy $E$ with respect 
to $\VECr_{i}$, i.e.\ 
\begin{equation}\label{F=-dEdr}
\VECF_{i} = -\frac{\partial}{\partial\VECr_{i}} E
\end{equation}
Using Eqns.\ (\ref{EwaldAnteile} -- \ref{FLadungsdichte}) one obtains
the following Ewald formula for the forces:
\begin{equation}\label{EwaldKraftAnteile}
\VECF_{i} = \VECF_{i}^{(r)} + \VECF_{i}^{(k)} + \VECF_{i}^{(d)}
\end{equation}
with the real space, Fourier space and dipole contributions respectively
given by:
\begin{eqnarray}
\VECF_{i}^{(r)} & = & q_{i} \sum_{j} q_{j} \sum_{\VECm\in\ZZ^{3}}^{\prime}
\bigg( \frac{2\alpha}{\sqrt{\pi}}\exp(-\alpha^{2}|\VECr_{ij}+\VECm L|^{2}) 
+ \nonumber \\
{~} & {~} & + \frac{\erfc(\alpha|\VECr_{ij}+\VECm L|)}{|\VECr_{ij}+\VECm L|}
\bigg) \frac{\VECr_{ij}+\VECm L}{|\VECr_{ij}+\VECm L|^{2}}
\label{Realraumanteil_Kraft} \\
\VECF_{i}^{(k)} & = & \frac{q_{i}}{L^{3}}\sum_{j}q_{j}\sum_{\VECk\ne 0}
\frac{4\pi\VECk}{k^{2}}\exp\left(-\frac{k^{2}}{4\alpha^{2}}\right)
\sin(\VECk\cdot\VECr_{ij}) \label{Impulsraumanteil_Kraft} \\
\VECF_{i}^{(d)} & = & -\frac{4\pi q_{i}}{(1+2\epsilon')L^{3}}
\sum_{j}q_{j}\VECr_{j} \label{Dipolanteil_Kraft}
\end{eqnarray}
Since the self energy from Eqn.\ (\ref{Selbstenergie}) is independent 
of particle positions, it does not contribute to the force.


\section*{Ewald summation on a grid}

Performing the Fourier transformations inherent to the reciprocal
space part of the Ewald sum by FFT routines is by no means a 
straightforward business. 
First, the point charges with their continuous coordinates have 
to be replaced by a grid based charge density, because the FFT 
is a {\em discrete} and {\em finite} Fourier transformation.
Second, it is neither obvious nor true that the best grid 
approximation to the continuum solution of Poisson's equation 
is achieved by using the {\em continuum} Green function. 
Third, there are at least three ways for implementing the 
differentiation needed in Eqn.\ (\ref{F=-dEdr}), which differ 
in accuracy and speed.
And fourth, the procedure of assigning the forces calculated on
the mesh back to the actual particles can -- under certain 
circumstances -- lead to unwanted violations of Newton's third 
law, which can be anything between harmless and disastrous.

The four steps involved in a particle mesh calculation are
sources for various kinds of errors, originating e.g.\ from
discretization, interpolation or aliasing problems (with the 
latter we want to denote inaccuracies resulting from the fact 
that a {\em finite} grid cannot represent arbitrarily large 
$\VECk$-vectors). 
Since these contributions are not independent of each other 
(reducing one might enhance another), the only reasonable 
demand is the minimization of the {\em total} error at given 
computational effort.

One of our aims is to compile some of the possibilities for 
each step, in order to draw a comparison between the three 
mesh implementations mentioned in the introduction 
-- PME, SPME and P$^{3}$M\@.
Like the Ewald sum, all these algorithms can be extended to a 
{\em triclinic} simulation cell by reverting to general dual 
basis vectors and one can also use a different number of grid 
points along each direction. However, in order to keep the 
notation simple, we restrict to the case of a cubic box and
employ the same number of mesh points in each direction. 
How the generalizations can be done is described e.g.\ in
the references on PME \cite{DYP} or SPME \cite{EPBDLP}.


\subsection*{Charge Assignment}

The actual {\em procedure} of assigning the charges to the
grid can be written down very easily. We will first discuss 
the one-dimensional case, i.e.\ particles with coordinates
$x\in[0;L]\subset\RR$ have to be assigned to the mesh points 
$x_{p}\in\MM=\{p\,h : p=0,\ldots,N\Mesh-1\}$, where $N\Mesh$ 
is the {\em number} of mesh points and $h:=L/N\Mesh$ is their 
{\em spacing}. 
To keep the notation simple, we will abstain from 
{\em ex\-pli\-cit\-ly} taking into account that any $x$-value, which 
is outside $[0;L]$, has to be folded back into this interval in order 
to conform to periodic boundary conditions. Rather, we assume that 
this is done as necessary, i.e.\ all calculations are to be 
understood ``modulo $L$''. 

Define the even function $W(x)$ such that the fraction of 
charge which is assigned to the mesh point $x_{p}$ due to a 
unit charge at position $x$ is given by $W(x-x_{p})$. 
If the charge density of the system is $\rho(x)$, then the 
{\em mesh based} charge density $\rho\Mesh$, defined at the 
mesh points $x_{p}$, can be written as the following convolution:
\begin{equation}\label{chargeassignment}
\rho\Mesh(x_{p}) = \frac{1}{h} \int_{0}^{L} \infd x \; 
W(x_{p}-x)\,\rho(x)
\end{equation}
The prefactor $1/h$ merely ensures that $\rho\Mesh$ is in fact
a {\em density}. Henceforth we will refer to any such $W$ as a 
{\em charge assignment function}.

The important question is: What {\em properties} should
$W(x)$ have in order to be a suitable choice?
The following wish list summarizes some desirable features:
\begin{itemize}
\item Charge conservation, i.e.\ the fractional charges of one 
  particle, which have been distributed to the surrounding grid
  points, 
  sum up to the total charge of that particle. 
\item Finite and if possible small support, because the computational
  cost increases with the number of mesh points among which the 
  charge of each particle is distributed.
  (The support of a real-valued function $f$ defined 
  on $X$ is (the closure of) the set $\{x\in X:f(x)\ne 0\}$, 
  i.e.\ basically the range of values for which the function 
  is nonzero.) 
\item Localization of discretization errors, i.e.\ inaccuracies 
  in the force between two particles due to the discretization 
  should become small with increasing particle separation.
\item Large degree of smoothness, i.e.\ the fractional charge
  of particle $i$ which is assigned to some mesh point $x_{p}$
  should be a smoothly varying function of the position of
  particle $i$. 
\item Minimization of aliasing errors, i.e.\ since on a finite
  grid there is only space for a limited number of $\VECk$-vectors,
  the charge assignment function should decay sufficiently 
  rapidly in Fourier space. 
\item Easy and transparent implementation.
\end{itemize}

It is important to realize that these characteristics cannot be
achieved all at the same time. Although some properties are
positively correlated (e.g., a large degree of smoothness
implies a fast decay in reciprocal space and thus minimizes 
aliasing errors), some other properties exclude each other 
(e.g., minimization of aliasing errors implies a sufficient 
localization in reciprocal space which is incompatible with a 
small support in real space).
Thus, a good charge assignment function is always a compromise
between these different demands. 

\begin{figure}[t]
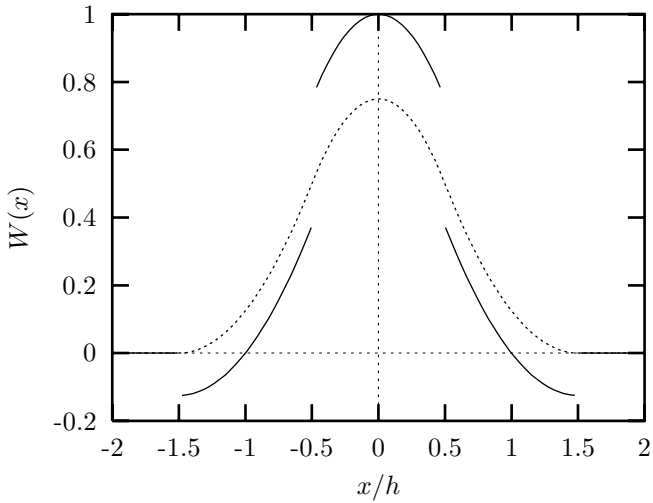

\hspace{-0.7cm}\parbox{8.6cm}{\include{f1}}
\caption{Third order charge assignment function for the Lagrange
  scheme\cite{DYP,P} (solid line) and the spline scheme\cite{HE} (dotted 
  line, see also Eqn.\ (\ref{HEspline})). 
  Both assignment functions have support $[-\frac{3}{2}h;\frac{3}{2}h]$ 
  and are piecewise quadratic.
  While for all odd assignment orders the Lagrange assignment function 
  is discontinuous (for even assignment order it is continuous but not 
  differentiable), the spline assignment function is in general $P-2$ 
  times continuously differentiable by construction. Note, however, that 
  the Lagrange assignment function is optimized with respect to a different 
  property, see Appendix \ref{GreenandLagrange}.}\label{lagrange_spline}
\end{figure}

We also want to stress that the choice of $W(x)$ is {\em not} 
independent of the other decisions made for the mesh implementation. 
In Appendix \ref{GreenandLagrange} we show that if one sticks to 
the {\em continuum} Coulomb Green function in the mesh calculation, 
the requirement of localization of discretization errors is enough 
to restrict the charge assignment to a {\em Lagrange interpolation 
scheme}. This is in fact the combination used for PME\cite{DYP,P} 
(see also Appendix \ref{GreenandLagrange}). 
Hence, other influence functions can only be competitive if the 
Coulomb Green function is somehow adjusted at the same time.

The choice for the charge assignment function of Hockney and Eastwood
is as follows: In a $P\th$ order assignment scheme (i.e.\ the charge 
of one particle is distributed between its $P$ nearest mesh points) 
define the Fourier transformed charge assignment function as
\begin{equation}\label{HEsplinek}
\tilde{W}^{(P)}(k) = h \left(\frac{\sin(kh/2)}{kh/2}\right)^{P}
\end{equation}
Transforming this back to real space gives
\begin{equation}\label{HEspline}
W^{(P)}(x) =
(\underbrace{\chi_{[-\frac{1}{2},\frac{1}{2}]}
  \star\cdots\star\chi_{[-\frac{1}{2},\frac{1}{2}]}}_{P 
  \mbox{\footnotesize -fold convolution}})(x/h)
\end{equation}
with $\chi_{[-\frac{1}{2},\frac{1}{2}]}$ being the characteristic 
function of the interval $[-\frac{1}{2},\frac{1}{2}]$, i.e.\ the function 
that is 1 within this interval and 0 outside. Thus, e.g.\ by the central 
limit theorem, the $P\th$ order charge assignment function resembles 
(in this case) for increasing $P$ more and more closely a centered 
Gaussian (with a variance $P$ times as large as the variance of 
$W^{(1)}$), but it has {\em finite} support $[-\frac{Ph}{2},\frac{Ph}{2}]$. 
This assignment function is very smooth for large $P$, since it is a 
spline of order $P$ and thus $P-2$ times differentiable \cite{Schoenberg}.
As a matter of convenience we decided to tabulate the corresponding
charge fractions $W^{(P)}_{p}(x)=W^{(P)}(x-x_{p})$ for 
$P\in\{1,\ldots,7\}$ in Appendix \ref{splinepolies}. 

In Fig.\ \ref{lagrange_spline} the third order assignment functions for
the Lagrange and the spline interpolation scheme are plotted. Note that
while the spline function is in general $P-2$ times continuously 
differentiable, the Lagrange assignment function is not as smooth:
Generally, for even assignment orders it is continuous, but the derivative 
is not, while for odd assignment orders it is discontinuous right away.
Incidentally, for $P=1$ and $P=2$ both schemes coincide.

The SPME method uses in essence the same charge assignment
functions as the P$^{3}$M-method, but this is discussed more
appropriately in the next section.

Charge assignment in more than one dimension can be achieved 
by a simple factorization approach. E.g., the three-dimensional 
charge assignment function $W(\VECr)$ can be written as
\begin{equation}\label{productassignment}
W(\VECr) = W(x)W(y)W(z)
\end{equation}
This is certainly not the only possibility\cite{HE}, but it is 
computationally advantageous. 

The generalization of Eqn.\ (\ref{chargeassignment}) to three dimensions 
can be written as
\begin{eqnarray}
\rho\Mesh(\VECr_{p}) & = & \frac{1}{h^{3}} \int_{L^{3}} 
\infd^{3} r \; W(\VECr_{p}-\VECr)\,\rho(\VECr) \label{chargeassignment3d} \\
{~} & = & \frac{1}{h^{3}} \sum_{i=1}^{N} 
q_{i} W(\VECr_{p}-\VECr_{i}) \label{chargeassignment3d_}
\end{eqnarray} 
In the last equation the reader should not confuse the coordinate of 
particle $i$, $\VECr_{i}$, with the coordinate of mesh 
point $p$, $\VECr_{p}$. 


\subsection*{Solving Poisson's equation}

For the standard Ewald sum the Fourier space contribution to the 
electrostatic energy is given by Eqn.\ (\ref{Impulsraumanteil}). 
How is this equation to be modified, now that we are working on 
a discrete mesh? 

The simplest approach is used in the PME method \cite{DYP}, where 
it is assumed that this equation is appropriate in the discrete 
case as well. The only difference is that the Fourier transformed 
charge density $\tilde{\rho}$ from Eqn.\ (\ref{FLadungsdichte}) 
is replaced by the {\em finite Fourier transform} of the mesh based 
charge density, $\hat{\rho}\Mesh$, which we define as
\begin{equation}\label{FFT}
\hat{\rho}\Mesh(\VECk) := h^{3} \sum_{\VECr_{p}\in\MM}
\rho\Mesh(\VECr_{p})\;e^{-i\,\VECk\cdot\VECr_{p}}
\end{equation}
where $\sum_{\VECr_{p}\in\MM}$ is the sum over the 
(three-di\-men\-sio\-nal) mesh in real space and the
$\VECk$-vectors are from the corresponding Fourier space 
mesh. 
Of course, the way back to real space is also done by a (inverse) 
finite Fourier transform. 
(In order to distinguish between the usual and the finite Fourier
transform, we indicate the latter by a hat and not by a tilde.)
As discussed in the previous subsection and Appendix 
\ref{GreenandLagrange}, the usage of the continuum Coulomb Green 
function is best accompanied by a Lagrange interpolation scheme for 
the charge assignment. See e.g.\ Petersen\cite{P} for a tabulation
of the corresponding polynomials and their implementation.

\vspace{0.5cm}

A second algorithm, the SPME method, was presented by Essmann 
{\em et.\,al}.\ \cite{EPBDLP}\@. It uses a smooth charge assignment 
scheme and hence an adjusted Green function. The reasoning is as follows:
Starting with Eqns.\ (\ref{Impulsraumanteil},\ref{FLadungsdichte})
it is argued that charge assignment onto the mesh can equivalently
be viewed as interpolating exponentials of the form $\exp(ikx)$ at 
discrete grid points. This problem has a particularly elegant solution,
the so-called {\em exponential Euler splines} \cite{Schoenberg}.
If $x$ is the continuous particle coordinate, we have for even $P$
(a recipe for the treatment of odd $P$ can be found in the original
SPME reference\cite{EPBDLP}):
\begin{equation}\label{Eulerexponentialspline}
e^{ikx} 
\approx b(k)\sum_{l\in\ZZ}M^{(P)}(x-lh)
e^{iklh} 
\end{equation}
\begin{equation}\label{Eulerexponentialspline2}
\mbox{with} \qquad b(k) = \frac{e^{ikPh}}
{\sum_{l=1}^{P-1}M^{(P)}(lh)e^{iklh}}
\end{equation}
The function $M^{(P)}$ is a cardinal-B-spline of order $P$, and Essmann 
{\em et.\,al}.\cite{EPBDLP} give a recursive definition. They also point 
out that (in our notation for $h=1$) the function $M^{(P)}$ is identical 
to the probability distribution of the sum of $P$ independent random 
variables, each distributed uniformly on the unit interval. Since this 
distribution is given by the $P$-fold convolution of the characteristic 
function $\chi_{[0;1]}$ with itself, we can see by comparison with Eqn.\ 
(\ref{HEspline}) that the charge assignment functions from the SPME and 
the P$^{3}$M method are in fact identical up to a translation: 
$M^{(P)}(x)=W^{(P)}(x-\frac{Ph}{2})$. Of course, $M^{(P)}$ has finite 
support $[0;Ph]$, so the sum in Eqn.\ (\ref{Eulerexponentialspline}) 
is actually finite.

Note that $M^{(P)}$ is not an even function, and in the original
reference on SPME \cite{EPBDLP} the charge assignment differs
slightly from our Eqn.\ (\ref{chargeassignment3d_}). However, the 
only effect of the translation is that the original system is 
represented by a {\em shifted} mesh system, which is from a 
practical point of view irrelevant, because this shift is undone 
in the back-interpolation (if accomplished with the same assignment 
function).

Now the following approximation for $E^{(k)}$ can be derived by 
inserting Eqns.\ (\ref{Eulerexponentialspline},
\ref{Eulerexponentialspline2}) into Eqn.\ (\ref{Impulsraumanteil}):
\begin{equation}\label{EPBDLPEnergie}
E^{(k)} \approx \frac{1}{2} \sum_{\VECr_{p}\in\MM} 
h^{3}\rho\Mesh(\VECr_{p})\big[\rho\Mesh\star G\big](\VECr_{p})
\end{equation}
Here the star denotes the finite convolution
\begin{equation}
\big[\rho\Mesh\star G\big](\VECr_{p}) = h^{3} 
\sum_{\VECr_{q}\in\MM}\rho\Mesh(\VECr_{q})G(\VECr_{p}-\VECr_{q})
\end{equation}
(again, the periodic closure is not written down ex\-pli\-cit\-ly)
and the function $G$ is given by its finite Fourier transform
\begin{equation}\label{EPBDLPinfluencefunction1}
\hat{G}(\VECk) =
B(\VECk)\sum_{\VECm\in\ZZ^{3}}\frac{4\pi}
{(\VECk+\frac{2\pi}{h}\VECm)^{2}} \tilde{\gamma}(\VECk+\frac{2\pi}{h}\VECm)
\end{equation}
with $B(\VECk):=|b(k_{x})b(k_{y})b(k_{z})|^{2}$.
Following Hockney and Eastwood we will refer to $G$ as the 
{\em influence function}. The nice thing about $G$ is that it is by 
construction independent of particle coordinates and can therefore
be precomputed.

Eqn.\ (\ref{EPBDLPEnergie}) can be made plausible in the following way: 
$G$ plays the role of a Coulomb Green function which has incorporated 
the ``smearing'' with the Gaussian $\gamma$. Hence, its convolution with 
the mesh based point charge density gives the mesh based electrostatic 
potential of $\gamma$-smeared charges. Multiplying this with the mesh based 
charge $h^{3}\rho\Mesh$ and summing over all mesh points gives the Fourier 
space contribution to the electrostatic energy up to a factor $1/2$, which 
merely cancels some double counting. This should be compared to Eqn.\ 
(\ref{Impulsraumanteil_}) or the second term in Eqn.\ 
(\ref{Energiesubtilitaet}).

As pointed out before, a charge assignment different from the Lagrange
interpolation scheme can only be competitive, if the Coulomb Green function 
is changed at the same time. The replacement of the usual (and smeared) 
Green function $g\star\gamma$ with the influence function $G$, which
essentially differs by the additional prefactor $B$ in Fourier space, 
achieves exactly that.
Conversely, PME uses a Lagrange interpolation scheme together with the 
unchanged Coulomb Green function, i.e.\ 
Eqns.\ (\ref{EPBDLPEnergie},\ref{EPBDLPinfluencefunction1})
with $B\equiv 1$. 

Finally, the alias sum occurring in Eqn.\ (\ref{EPBDLPinfluencefunction1}) 
is substituted according to the following rule\cite{EPBDLP}:
If $N\Mesh$ is the number of mesh points in each direction and the 
vector $\VECk$ on the left hand side is given by $\VECk=2\pi\VECn/L$,
$\VECn\in\{0,\ldots,N\Mesh-1\}^{3}$, then define $\hat{G}(\VECk) =
B(\VECk)\tilde{g}(\VECk')\tilde{\gamma}(\VECk')$, where $\VECk'= 
2\pi\VECn'/L$ and $n_{i}'=n_{i}$ for $0\le n_{i}\le N\Mesh/2$ and 
$n_{i}'=n_{i}-N\Mesh$ otherwise ($i=x,y,z$). 

\vspace{0.5cm}

A third possibility -- the so-called P$^{3}$M method -- was presented 
by Hockney and Eastwood \cite{HE}: 
Their objective was an optimization of the influence function $G$ in
Eqn.\ (\ref{EPBDLPEnergie}), which causes the final result of 
the {\em mesh} calculation to be as close as possible to the 
ori\-gi\-nal {\em continuum} problem. So in order to proceed one 
first has to make the statement ``as close as possible'' more
quantitative, and this can be done as follows: 

Take two particles with coordinates $\VECr_{1}$ and $\VECr_{2}$ 
and define $\VECr:=\VECr_{1}-\VECr_{2}$. The true force between 
these particles should be a function of $\VECr$ only, but in any 
mesh implementation the actual force also depends on the positions 
of the particles relative to the mesh, say, on the position of the 
first particle within its mesh cell (i.e.\ the original translational 
symmetry is broken by the mesh). 
This suggests the following measure for the error: integrate the 
square of the difference between the calculated force $\VECF$ and 
the true reference force $\VECR$ over all values of $\VECr$ and 
average this quantity over all positions of e.g.\ the first particle 
within one particular mesh cell:
\begin{equation}\label{P3MQ}
Q := \frac{1}{V_{c}}\int_{V_{c}}\infd^{3}r_{1}\int_{V_{b}}\infd^{3}r
\left[\VECF(\VECr;\VECr_{1})- 
  \VECR(\VECr)\right]^{2}
\end{equation}
Here $V_{c}=h^{3}$ is the volume of one mesh cell. 
The solution of Poisson's equation is accomplished in essence by Eqn.\ 
(\ref{EPBDLPEnergie}) (it is only written down somewhat differently), 
and the derivative in Eqn.\ (\ref{F=-dEdr}) is performed by applying 
finite difference operators to the mesh based electrostatic potential
(see below).
Since the discretization error $Q$ can be regarded as a functional of 
$\hat{G}$, the {\em optimal} influence function $\hat{G}\opt$ can be 
obtained by setting the {\em functional derivative} of $Q$ with respect 
to $\hat{G}$ to zero, i.e.\ 
\begin{equation}\label{Funktionalableitung}
\left.\frac{\delta Q}{\delta\hat{G}}\right|_{\hat{G}=
\hat{G}_{\mbox{\tiny opt}}} = 0
\end{equation}
Starting from this idea, Hockney and Eastwood were able to derive 
the following expression for $\hat{G}\opt$ \cite{HE}:
\begin{equation}\label{optimalinfluencefunction}
\hat{G}\opt(\VECk) = 
\frac{\tilde{\VECD}(\VECk)\cdot\sum_{\VECm\in\ZZ^{3}}
\tilde{U}^{2}(\VECk+\frac{2\pi}{h}\VECm)\tilde{\VECR}
(\VECk+\frac{2\pi}{h}\VECm)}
{|\tilde{\VECD}(\VECk)|^{2}\left[\sum_{\VECm\in\ZZ^{3}}
\tilde{U}^{2}(\VECk+\frac{2\pi}{h}\VECm)\right]^{2}}
\end{equation}
Here $\tilde{\VECD}(\VECk)$ is the Fourier transform of the employed 
differentiation operator (see next section and Appendix \ref{optidiff}), 
$\tilde{U}(\VECk)=\tilde{W}(\VECk)/V_{c}$ is the Fourier transform 
of the charge assignment function divided by the volume of one 
mesh cell and $\tilde{\VECR}(\VECk)$ is the Fourier transform of 
the true reference force, given by
\begin{equation}\label{referenceforce}
\tilde{\VECR}(\VECk) = -i\VECk\tilde{g}(\VECk)\tilde{\gamma}(\VECk)
\end{equation}
Note that this differs from the expression from the book of
Hockney and Eastwood \cite{HE}, who use $\tilde{\gamma}^{2}$ 
instead of $\tilde{\gamma}$. The reason is that Eqn.\ 
(\ref{referenceforce}) describes the true reference force between 
a $\gamma$-smeared charge and a point charge, while Hockney and 
Eastwood choose a slightly different approach in which they need 
the force between two $\gamma$-smeared charges. Also, we keep the 
factor $4\pi$ in the Fourier transformed Green function $\tilde{g}$ 
and do not to hide it somewhere else.

The alias sums over $\VECm$ in Eqn.\ (\ref{optimalinfluencefunction}) 
are typically well converged for $|\VECm|\le 2$ and the sum in the 
denominator could even be done analytically. Again we want to
emphasize that the calculation of the influence function has to be
done only once prior to the actual simulation and thus does not produce
any runtime overhead. 
Note also that the expression (\ref{EPBDLPinfluencefunction1}) differs 
from the optimal form (\ref{optimalinfluencefunction}) and hence cannot 
be optimal.

A final word concerning the implementation: Although the convolution 
$\rho\Mesh\star G$ in Eqn.\ (\ref{EPBDLPEnergie}) is a nice and compact 
notation, the whole purpose of these particle mesh routines is to employ 
the {\em convolution theorem} and use efficient FFT routines to calculate 
$\rho\Mesh\star G$. The central steps are thus:
\begin{itemize}
\item Calculate the finite Fourier transform $\hat{\rho}\Mesh$ of the 
  mesh based charge density $\rho\Mesh$.
\item Multiply $\hat{\rho}\Mesh$ with the precomputed Fourier space
  representation of the influence function, $\hat{G}$. 
\item Apply an inverse finite Fourier transform to this product to end
  up with the finite convolution of $\rho\Mesh$ with $G$. 
  Formally this can be symbolized as
  \begin{equation}\label{Meshpotential}
    \rho\Mesh\star G = \; \stackrel{\longleftarrow}{\mbox{FFT}}\left[
      \stackrel{\longrightarrow}{\mbox{FFT}}[\rho\Mesh] \;\times\;
      \stackrel{\longrightarrow}{\mbox{FFT}}[G]\right]
  \end{equation}
  Note that in this way one only needs $\hat{G}$ to calculate
  $\rho\Mesh\star G$ but actually never $G$ itself. 
\end{itemize}

This is the important part which all particle mesh algorithms have in 
common. The various methods differ e.g.\ in their choice of $G$,
the assignment function $W$ or the implementation of the derivative
in Eqn.\ (\ref{F=-dEdr}). 


\subsection*{Differentiation}

After the calculation of the electrostatic energy, the forces on 
the particles are obtained by differentiation according to Eqn.\ 
(\ref{F=-dEdr}). However, for the Fourier space part of particle 
mesh methods there are several possibilities to implement this 
procedure. In other words: there exist several possible substitutes 
for Eqn.\ (\ref{Impulsraumanteil_Kraft}), in particular
\begin{enumerate}
\item Differentiation in Fourier space.
\item Analytic differentiation of the assignment function in real space.
\item Discrete differentiation on the mesh in real space.
\end{enumerate}
Differentiation in Fourier space is easy, since it merely involves 
a multiplication with the Fourier transformed differentiation 
operator $\tilde{\VECD}(\VECk)$, which is a fast, local and accurate 
operation. Although one might want to use Fourier transforms of
discrete difference operators -- allowing for the fact that one is
actually working on a mesh -- the best results are obtained when
the Fourier transform of the usual differential operator, namely
$i\VECk$, is employed. Therefore we will refer to this method as
$i\VECk$-differentiation.
The basic idea is not to calculate the mesh based electrostatic 
potential $\phi^{(k)}(\VECr_{p})$ via Eqn.\ (\ref{Meshpotential}) but 
the mesh based electric field $\VECE(\VECr_{p})$ by the following 
simple change to this equation:
\begin{eqnarray}
\VECE(\VECr_{p}) & = & -\frac{\partial}{\partial\VECr_{p}}
\phi^{(k)}(\VECr_{p}) =
-\frac{\partial}{\partial\VECr_{p}} 
\big[\rho\Mesh\star G\big](\VECr_{p}) \nonumber \\
{~} & = & - \stackrel{\longleftarrow}{\mbox{FFT}}
\left[ i\VECk \;\times\; \hat{\rho}\Mesh \;\times\;\hat{G} 
\right](\VECr_{p}) \label{MeshEfeld}
\end{eqnarray}
This method is employed in the PME algorithm and, as shown later, 
leads to the most accurate force calculations, if it is used in
conjunction with the optimal influence function from Eqn.\ 
(\ref{optimalinfluencefunction}).
Note, however, that since $\VECk$ is a vector, there are in fact
{\em three} inverse three-dimensional Fourier transforms to be
calculated in Eqn.\ (\ref{MeshEfeld}), which is obviously 
computationally demanding. 

The electrostatic energy calculated on the mesh depends on the particle 
coordinates through the arguments of the charge assignment function $W$. 
As the creators of the SPME method point out\cite{EPBDLP}, a smooth 
charge assignment scheme permits an {\em analytic} differentiation of 
the energy, since the quantity $\rho\Mesh$, which contains the particle 
coordinates $\VECr_{i}$, depends in a differentiable way on the $\VECr_{i}$.
Using Eqns.\ (\ref{F=-dEdr},\ref{EPBDLPEnergie}) and the fact that $G$ is 
independent of particle coordinates and an even function (since $\hat{G}$ 
is even), one can derive
\begin{eqnarray}
\VECF_{i} & \approx & -\frac{\partial}{\partial\VECr_{i}}
\frac{1}{2} \sum_{\VECr_{p}\in\MM} 
h^{3}\rho\Mesh(\VECr_{p})\big[\rho\Mesh\star G\big](\VECr_{p}) \nonumber \\
{} & = & -\frac{1}{2}h^{6} \sum_{\VECr_{p},\VECr_{q}\in\MM}
\bigg(\frac{\partial\rho\Mesh}{\partial\VECr_{i}}(\VECr_{p})\rho\Mesh(\VECr_{q})
+\rho\Mesh(\VECr_{p})\frac{\partial\rho\Mesh}{\partial\VECr_{i}}(\VECr_{q})\bigg)
\times\nonumber\\
{} & {} & \qquad\qquad\times G(\VECr_{p}-\VECr_{q}) \nonumber \\
{} & = & -h^{6}\sum_{\VECr_{p},\VECr_{q}\in\MM}
\frac{\partial\rho\Mesh}{\partial\VECr_{i}}(\VECr_{p})\rho\Mesh(\VECr_{q}) 
\times\nonumber \\
{} & {} & \qquad\qquad\times \frac{1}{2}\Big(G(\VECr_{p}-\VECr_{q}) + 
G(\VECr_{q}-\VECr_{p})\Big) \nonumber \\
{} & = & - h^{3} \sum_{\VECr_{p}\in\MM}
\frac{\partial \rho\Mesh}{\partial \VECr_{i}}(\VECr_{p})
\big[\rho\Mesh\star G\big](\VECr_{p}) \label{EPBDLPKraft}
\end{eqnarray}
From Eqn.\ (\ref{chargeassignment3d}) it is obvious that the array
$[\partial \rho\Mesh/\partial \VECr_{i}](\VECr_{p})$ is essentially
obtained by a charge assignment scheme which uses the {\em gradient}
of the assignment function $W$ and can thus be calculated conveniently 
at the same time as $\rho\Mesh$.
Since only one Fourier transform back to real space is necessary, this 
procedure is indeed very fast. Unfortunately, this differentiation scheme 
leads to a small random particle drift, since momentum is not conserved 
any more (see next section). Although the {\em total} momentum of the 
simulation box can be kept constant by subtracting the mean force 
$\frac{1}{N}\sum_{i}\VECF_{i}$ from each particle, the small reduction 
in the accuracy of the particle forces due to these {\em local} random 
fluctuations can only be compensated marginally by this {\em global} 
correction.

A third possibility for implementing the derivative in Eqn.\ 
(\ref{F=-dEdr}) is the use of finite difference operators, which 
calculate the force on one mesh point from the potential at the 
neighboring mesh points. This is basically the method which is 
favored by Hockney and Eastwood for P$^{3}$M\@.
Higher accuracy is achieved by considering not only the nearest 
neighbors but also mesh points farther away, i.e.\ using linear 
combinations of nearest neighbor, next nearest neighbor etc.\ 
difference operators. In Appendix \ref{optidiff} we show, how 
these approximations are constructed systematically.
In the P$^{3}$M-method the Fourier transforms of these operators 
are needed for the calculation of the optimal influence function
(\ref{optimalinfluencefunction}).
This approach as well needs only one Fourier transformation back 
to real space, like in the method of analytic differentiation.
But unlike the latter, it conserves momentum (if the difference
operators are chosen correctly\cite{HE}) and thus has no 
problems with spurious particle drifts and resulting errors in 
the force. However, using the neighboring points is a nonlocal 
approach and increasing its accuracy can only be done by taking
into account more neighbors -- which makes it even more nonlocal
and more costly. 

Obviously, there is no unique optimal way for doing the 
differentiation. Each approach joins together advantages 
and drawbacks which have to be balanced against each other 
under the constraint of required accuracy and available 
computational resources. 
Let us make just one example: If the required accuracy is not very 
high, using only nearest neighbors for the discrete differentiation 
on the mesh might be accurate enough. Certainly, multiplication 
in Fourier space by $i\VECk$ gives better results, but let us 
assume that this approach is actually slower due to the two 
additional Fourier transformations. However, if the required 
accuracy increases, the finite difference approximation calls 
for more neighbors and thus becomes more and more costly,
whereas the $i\VECk$-approach right away gives the best result 
possible by discrete differentiation. This is because increasing 
the order of the differentiation scheme means that in Fourier 
space the transformed operators approximate $i\VECk$ to higher 
and higher truncation order (actually, that is how these 
approximations are constructed, see Appendix \ref{optidiff}). 
In other words, accepting the two additional Fourier transformations 
can be competitive. 
Moreover: The method of analytic differentiation could be 
faster than the discrete difference method even for $J=1$. 
Thus, in cases where the latter is less accurate than analytic 
differentiation, there is no reason for using it.

Whether there exists a break-even point between these methods and 
-- if yes -- where it is located can depend on the tuning
parameters like mesh size and interpolation order as well as on the 
details of the implementation or the computational facilities one
is working with. A general statement seems to be difficult.


\subsection*{Back-interpolation}

At some stage of any particle mesh method a back-interpolation
of the mesh based results to the actual particles is necessary.
As we have seen in the last subsection, this can be done before
or after the Fourier transformation back to real space, i.e.\ 
the mesh points can contain either the potential or the 
components of the electric field.

Basically, this back-interpolation is done in a similar way as the 
distribution of the charges to the mesh at the beginning of the
 calculation: via some assignment function $W$.
E.g., the force on particle $i$ is given by
\begin{equation}\label{ikKraft}
\VECF_{i} = q_{i} \sum_{\VECr_{p}\in\MM} \VECE(\VECr_{p}) W(\VECr_{i}-\VECr_{p})
\end{equation}
with $\VECE(\VECr_{p})$ being the electric field on mesh point $\VECr_{p}$
from Eqn.\ (\ref{MeshEfeld}). The interpretation of Eqn.\ (\ref{ikKraft}) 
is the following: Due to the discretization each particle is replaced by 
several ``sub-particles'', which are located at the surrounding mesh points 
and carry a certain fraction of the charge of the original particle. The 
force on each sub-particle is given by its charge times the electric field 
at its mesh point, and the force on the original particle is the sum of the 
forces of its sub-particles. 

From a technical point of view it is convenient to use the 
{\em same} function $W$ for the assignment onto and from the 
mesh, because if in the first step one does not only calculate 
the {\em total charge} accumulated at some mesh point but 
additionally memorizes, to what extend the {\em individual 
particles} contributed to this charge, the interpolation 
back can be done without a single function call to $W$.

However, there is also a more subtle reason which suggests a 
symmetric interpolation, and this is related to the conservation 
of momentum. As demonstrated by Hockney and Eastwood\cite{HE}, 
the force which a particle acts onto itself is zero and Newton's 
third law is obeyed (up to machine precision), if
\begin{itemize}
\item charge assignment and force interpolation are done by
  the same function $W$ and
\item the approximations to the derivatives are correctly
  space centered.
\end{itemize}
The second requirement states that if the electric field
at some mesh point $\VECr_{p}$ can formally be written as
$\sum_{\VECr_{q}} \VECd(\VECr_{p},\VECr_{q})
\rho\Mesh(\VECr_{q})$, then $\VECd(\VECr_{p},\VECr_{q})=
-\VECd(\VECr_{q},\VECr_{p})$.

The method of analytical differentiation mentioned in the last 
subsection does not use mesh based derivatives and this is the 
approach chosen in the implementation of SPME \cite{EPBDLP}. 
In their paper the authors state that the sum of the electrostatic
forces on the atoms is not zero but a random quantity
of the order of the rms error in the force. We believe 
that these fluctuations have their origin in a violation
of the above conditions, although strictly speaking 
these are only {\em sufficient} conditions for momentum
conservation.

For a more detailed discussion of related effects and the
connection between momentum conserving and energy 
conserving methods see Hockney and Eastwood\cite{HE}.


\section*{Investigating the accuracy}

An investigation of the errors connected with particle 
mesh Ewald methods is important for several reasons.
First, the complete procedure of discretization introduces
new sources of errors in addition to the ones originating 
from real and reciprocal space cutoffs.
Second, comparing the efficiency of different mesh methods
is only fair if it is done at the same level of accuracy.
And third, the tuning parameters should be chosen in such
a way as to run the algorithm at its optimal operation
point. 

However, there is no unique or optimal {\em measure} of accuracy.
If molecular dynamics simulation are performed, the main
interest lies in errors connected with the {\em force}, while
in Monte Carlo simulations one is concerned with
accurate {\em energies}. In the simulation of ensemble
averages it is the {\em global} accuracy -- measured e.g.\ by
root mean square quantities -- which is important, but in
the simulation of rare events {\em local} accuracy and maximal
errors are also relevant. Errors in the force can be
due to their {\em magnitude} or due to their {\em direction}. 
And finally, one might be interested in {\em absolute} or 
{\em relative} errors.

Whatever quantity one decides to look at, it can be investigated
as a function of {\em system} parameters like particle separation
or distribution, {\em tuning} parameters like $\alpha$, mesh size 
or interpolation order and {\em components} of the algorithm, e.g.\ 
interpolation or differentiation scheme or splitting function 
$f(r)$. 
Obviously this gives rise to a very large number of combinations.
In other words: The corresponding parameter space is large and 
nontrivial, i.e.\ general statements concerning the performance
of one method can usually not be extracted from low-dimensional 
cuts through this space, because different methods scale differently 
with respect to their parameters.

Nevertheless, we want to present some numerical accuracy measurements
at important points of this parameter space for the following reasons:
As we pointed out, there are several options for the implementation 
of each step of a mesh calculation -- e.g.\ three ways for doing the 
derivative in Eqn.\ (\ref{F=-dEdr}). 
This freedom of choice and its impact on the overall accuracy 
has not been systematically investigated so far, although a 
qualitative understanding of at least {\em typical} influences of 
the different parts on the performance permits a judicious assessment 
and comparison of the resulting algorithms, in particular P$^{3}$M, 
PME and SPME\@.  
We want to show which combinations are attractive and which should
definitely be avoided. 
And finally we want to present easily reproducible measurements
which should allow the reader a comparison with his own 
implementations of particle mesh Ewald routines.
However, we will {\em not} present large accurate tables, which 
provide an easy way for tuning these algorithms under all 
circumstances.
On the contrary, we want to encourage any potential user to
perform some of these simple measurements on his own and thereby 
not only gaining insight but also the possibility to optimize
his tuning parameters. 
We want to stress that parameters which are only roughly 
estimated or even historically handed down should be taken 
with great care.


\subsection*{One possible measure of accuracy}

In this article we will solely be concerned with one measure
of accuracy, namely the root mean square (rms) error in the
force, given by
\begin{equation}\label{rmsforce}
\Delta F := \sqrt{\frac{1}{N}\sum_{i=1}^{N}
\left(\VECF_{i}-\VECF^{\mbox{\footnotesize exa}}_{i}\right)^{2}}
\end{equation}
Where $\VECF_{i}$ is the force on particle number $i$ calculated via
some mesh method and $\VECF^{\mbox{\footnotesize exa}}_{i}$ is the 
{\em exact} force on that particle, calculable e.g.\ by a well 
converged standard Ewald sum. 
There exist error estimates for the real space and Fourier space
contribution to this error for the standard Ewald sum \cite{KP}
and for the PME method \cite{P} which greatly simplify
the determination of the optimal value of $\alpha$.


\subsection*{Error as a function of $\alpha$}

We investigated the rms error (\ref{rmsforce}) for a system
of 100 particles (50 carry a positive and 50 a negative unit 
charge), which were randomly placed within a simulation box 
of length $L=10$, as a function of the Ewald parameter $\alpha$. 
In order to make our results fully reproducible, we describe in 
Appendix \ref{modelsystem}, how our actual random configuration
was generated.

For small $\alpha$ the result of the Ewald sum (or
any of the described particle mesh methods) is dominated by the 
real space contribution (\ref{Realraumanteil}) while for large 
$\alpha$ it is the Fourier part (\ref{Impulsraumanteil}) which is 
important. 
This is a simple consequence of the fact that in the real space
sum $\alpha$ occurs in the {\em numerator} of the exponential
function (or -- to be precise -- of the complementary error
function) while in the Fourier space sum it occurs in the
{\em denominator} and thus influences the decay of both
contributions in a converse way. 
Hence, {\em at given cutoffs}, the same applies to the errors. 
Since with increasing $\alpha$ the real space contribution becomes 
more accurate while the Fourier space contribution degrades in 
accuracy, one can expect an optimal $\alpha$ to exist at which
the {\em total} error is minimal. This is approximately at the
point where real space and Fourier space errors are equal.
Since the different mesh methods we investigate all coincide
in the treatment of the real space part, their errors should
all be the same for sufficiently small $\alpha$.

\begin{figure}[!t]
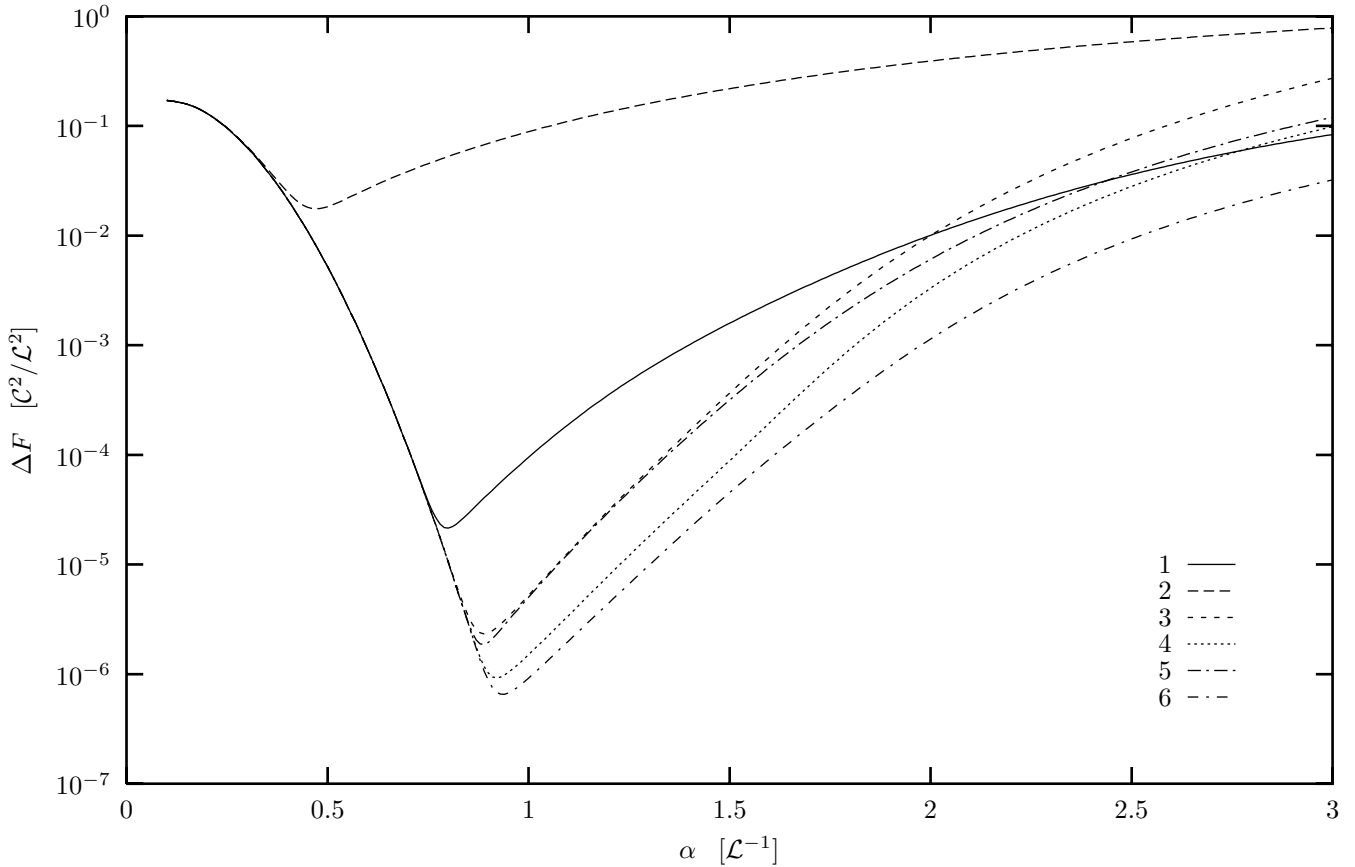

\hspace{-0.3cm}\parbox{2cm}{\include{f2}}
\begin{minipage}{18cm}
\caption{
  Comparison of different mesh methods: The rms error 
  $\Delta F$ from Eqn.\ (\ref{rmsforce}) for a system of 100 
  charged particles randomly distributed within a cubic box of 
  length $L=10$ (see Appendix \ref{modelsystem}) is shown as a
  function of the Ewald parameter $\alpha$ for 6 mesh algorithms, 
  which all share $N\Mesh=32$, $P=7$ and $r\maxi=4$. 
  Line 1 is PME\@. Line 2 corresponds to an algorithm which is 
  obtained from PME by retaining the continuum Green function 
  but changing to the spline charge assignment. Lines 3 and 4 
  are analytically and $i\VECk$-differentiated SPME respectively 
  and line 5 and 6 are analytically and $i\VECk$-differentiated 
  P$^{3}$M respectively. Note the logarithmic vertical scale in 
  this and the following figures.
}\label{alphafehler}
\end{minipage}
\end{figure}

In Fig.\ \ref{alphafehler} we plot the rms error of the force as a 
function of $\alpha$, which was obtained by investigating our system 
with various mesh methods. They all share a mesh size of $N\Mesh=32$ 
(and thus have $32^{3}$ mesh points in total), an interpolation order 
$P=7$ and a real space cutoff $r\maxi=4$.
We find indeed the general features described above, like a low 
accuracy for very small or very large values of $\alpha$ and an
optimal value in-between. However, the various methods differ
considerably in their accuracy. (Note that in this and the following
figures the vertical scale is logarithmic!)

The solid line corresponds to PME. This method comprises some elements
which make one think about possible improvements: a not very smooth 
charge assignment scheme (namely, the Lagrange interpolation) and the use 
of the plain continuum Green function. There is clearly no {\em obvious}
advantageous replacement for the latter, but it is easy to replace
the Lagrange scheme by the smooth spline interpolation (by just changing 
the assignment function). Yet, the result of this supposed improvement, 
shown in line 2, is in fact disappointing. As we mentioned several times, 
the continuum Green function is best accompanied by a Lagrange interpolation 
scheme, because this leads to a cancellation of certain discretization 
errors. Changing the assignment scheme destroys this effect and the 
resulting error shatters the desired improvement in accuracy completely. 

Upgrading PME requires a proper treatment of both elements -- charge 
assignment {\em and} Green function. This is in fact what the remaining 
two algorithms (SPME and P$^{3}$M) accomplish. Since they both use a
smooth spline interpolation, they are both potential candidates for
analytic differentiation. In fact, the SPME method, as described in the 
original publication\cite{EPBDLP}, chooses this implementation of the 
derivative, because it is very fast (line 3). Nevertheless, the 
$i\VECk$-method is still possible and leads to an even better result 
(line 4), which admittedly has to be payed with two additional FFT calls. 
Analytically differentiated P$^{3}$M gives an error almost identical
to analytically differentiated SPME, but if one implements the 
$i\VECk$-derivative, P$^{3}$M improves a little bit on SPME. 
From a theoretical point of view the latter is not too much surprising:
After all, if P$^{3}$M uses an optimal differentiation (in view of 
accuracy) and an optimal influence function, it can be expected to 
constitute a kind of lower bound for the error. However, if the optimal 
differentiation is replaced by the analytic differentiation, a new
source of error appears (namely, the random force fluctuations described
in the section on back-interpolation). If this contribution dominates,
the fact that P$^{3}$M uses a better influence function than SPME cannot
make a large difference.
In our case the analytically differentiated SPME is a factor 9.2 more 
accurate than PME, while the $i\VECk$-differentiated P$^{3}$M method
is more accurate than PME by a factor of about 33. However, one must 
realize that SPME and P$^{3}$M have different execution times, since
P$^{3}$M needs two additional FFT calls compared to SPME. But apart from the 
analytically differentiated curves all methods summarized in Fig.\ 
\ref{alphafehler} need {\em exactly} the same time for a mesh calculation. 
This comes from the fact that the methods differ only in parts which 
normally are tabulated anyway, like the influence function.

There is another surprising thing to note about SPME: For the chosen values 
of $N\Mesh$ and $P$ the curves for PME and (analytically differentiated) 
SPME {\em intersect}, i.e.\ the latter is not necessarily more accurate. 
It could be argued that at least for the {\em optimal} value of $\alpha$ 
SPME is better, but this optimal $\alpha$ of course depends on the real 
space cutoff $r\maxi$ as well. If this cutoff is decreased, the real space 
contribution to the error is increased. In fact, using the estimate of Kolafa
and Perram\cite{KP} one finds that at the intersection point of PME and 
SPME this contribution will have the same size as the Fourier space 
contribution for (in this case) $r\maxi\approx 1$. Thus, for even smaller 
values of $r\maxi$ PME would actually be more accurate than SPME\@.

Now that we have compared various particle mesh {\em methods}, we want
to examine in a little more detail some {\em parts} of the algorithm.
We will always use the P$^{3}$M method for illustration.
Corresponding plots for PME or SPME would look qualitatively very 
similar and hence are not presented in the sequel.

\begin{figure}[t]
\vspace{14cm}
\end{figure}

In Figure \ref{alphafehler_P3M1} we took all parameters of the 
$i\VECk$-differentiated P$^{3}$M method from Fig.\ 
\ref{alphafehler} but varied the charge assignment order from 
$P=1$ to $P=7$. Increasing $P$ improves the accuracy by more 
than three orders of magnitude (from $P=1$ to $P=7$). However, 
the reward for going from $P$ to $P+1$ is larger for small $P$. 
Note also that the optimal value of $\alpha$ depends on $P$. 

In Figure \ref{alphafehler_P3M2} we fix the order of the charge assignment 
scheme to $P=3$ and vary the number $N\Mesh$ of Fourier mesh points. 
Qualitatively the behavior is similar to Fig.\ \ref{alphafehler_P3M1}: 
Improving the method reduces the error and shifts the optimal $\alpha$ to 
the right. 
Note that from a computational point of view Figs. \ref{alphafehler_P3M1} 
and \ref{alphafehler_P3M2} are sort of conjugate: The accuracy
depends on both $N\Mesh$ and $P$, but increasing one parameter
does not influence the performance of the other.
In other words: The charge assignment scales as $P^{3}$ independent
of $N\Mesh$ and the FFT scales as $(N\Mesh\log N\Mesh)^{3}$ independent 
of $P$. Optimal performance requires a suitable combination of 
$N\Mesh$ and $P$.

\begin{figure}[!t]
\hspace{-0.7cm}\parbox{8.6cm}{\include{f3}}
\caption{Influence of the charge assignment order: The rms error
  $\Delta F$ for our model system from Appendix \ref{modelsystem} is 
  calculated for the $i\VECk$-differentiated P$^{3}$M method with 
  $N\Mesh=32$ and $r\maxi=4$. From top to bottom the order $P$ of the 
  (spline) charge assignment scheme is increased from
  1 to 7.}\label{alphafehler_P3M1}

\vspace{1cm}

\hspace{-0.7cm}\parbox{8.6cm}{\include{f4}}
\caption{Influence of the mesh size: The rms error $\Delta F$ for our 
  model system from Appendix \ref{modelsystem} is calculated for the 
  $i\VECk$-differentiated P$^{3}$M method with $P=3$ and $r\maxi=4$. 
  From top to bottom the mesh size $N\Mesh$ is given by 4,8,16,32,64 
  and 128. (Note that the total number of mesh points in this 
  three-dimensional system is given by
  $(N\Mesh)^{3}$.)}\label{alphafehler_P3M2}
\end{figure}

\begin{figure}[!t]
\hspace{-0.7cm}\parbox{8.6cm}{\include{f5}}
\caption{Influence of the differentiation scheme: The rms error 
  $\Delta F$ for our model system from Appendix \ref{modelsystem} is 
  calculated for the P$^{3}$M method with $N\Mesh=32$, $P=7$ and $r\maxi
  =4$. Shown are 6 mesh based approximations to the differentiation 
  operator $i\VECk$ (from top to bottom: $\Delta^{(1)},\ldots,\Delta^{(6)}$, 
  see Appendix \ref{optidiff}) as well as the result for $i\VECk$ itself 
  (lowest curve, solid line).}\label{alphafehler_P3M3}

\vspace{0.2cm}

\hspace{-0.7cm}\parbox{8.6cm}{\include{f6}}
\caption{Comparison of a mesh method with the standard Ewald sum:
  The rms error $\Delta F$ for the Ewald (line 1) and PME 
  (line 2) method are calculated for our model system from Appendix 
  \ref{modelsystem}. The parameters for PME are the same as in Fig.\ 
  \ref{alphafehler} and the Fourier space cutoff for the Ewald sum 
  was set to $k\maxi=20\times 2\pi/L$. This value is interesting to 
  compare with the PME method, because it corresponds to the same 
  number of $\VECk$-vectors (since $\frac{4}{3}\pi\,20^{3}\approx 
  32^{3}$).
  Also shown is the estimate for the real space error\cite{KP} 
  (line 3), the Fourier space error for Ewald (line 4, we used 
  the slightly better estimate from Petersen\cite{P}) and the 
  Fourier space error for PME\cite{P} (line 5). 
  Note that the estimates for the Ewald sum can hardly be 
  distinguished from line 1.}\label{fehlerformel}

\vspace{0.1cm}
\end{figure}

Next we investigated the differentiation scheme. To this end we employed 
the P$^{3}$M method with $N\Mesh=32$ and $P=7$ and used various orders 
$J$ of the mesh based approximation to the difference operator 
(see Appendix \ref{optidiff}). (Actually, the calculations were done by 
a multiplication in Fourier space with the transformed approximations 
$\tilde{\VECD}^{(J)}$.)
The result is shown in Fig.\ \ref{alphafehler_P3M3}, which looks pretty 
much like Fig.\ \ref{alphafehler_P3M1} but was generated quite differently. 
With increasing order of the difference approximation the errors decrease. 
However, the result of the $i\VECk$-differentiation scheme forms a lower 
bound to the error of this method. (After all, $i\VECk$ is the Fourier
representation of the exact differential operator, and in the standard
Ewald sum the differentiation is also done this way, compare Eqn.\ 
(\ref{Impulsraumanteil_Kraft}).)
Here the bound is reached in the minimum at $J=7$, so further improving 
the differentiation order is of no use at all. Of course, if the accuracy 
of the lower bound is smaller (e.g., because the charge assignment order 
is lower) the $i\VECk$-bound will be reached already by smaller values of 
$J$.
Note that in this example the method of analytic differentiation gives
approximately the same accuracy as a fifth order difference scheme
(compare to Fig.\ \ref{alphafehler}).
Since analytic differentiation is much faster, it should be preferred 
to the finite difference approach in cases where the latter is less 
accurate anyway.

The last part of this section deals with the determination of the optimal 
$\alpha$-value. There exist rather good estimates for the real- and 
reciprocal space error of the standard Ewald sum\cite{KP} and 
the reciprocal space error of the PME method\cite{P}. 
The optimal $\alpha$-value of these two methods and the corresponding 
accuracy can be obtained very precisely by just calculating the 
intersection point of the real- and corresponding reciprocal space 
estimates. Their high quality is clearly demonstrated in Fig.\ 
\ref{fehlerformel}.
The existence of these formulas is certainly a big advantage of the PME 
method, since it permits an a priori determination of the optimal operation 
point as a function of system specifications (like box length, particle 
number or valence) or method parameters (like mesh size or assignment 
order).
The elaboration of a similar error estimate for the P$^{3}$M method is
currently pursued and will be presented in a forthcoming publication 
\cite{wir}.
This is basically the last step which is missing to advocate P$^{3}$M
as the most accurate and versatile Ewald mesh method.

A final word concerning the accuracy of mesh methods compared to the 
Ewald sum:
The optimal $\alpha$ value for a standard Ewald summation of our system 
with $k\maxi=20\times 2\pi/L$ (which thus has the same number of
$\VECk$-vectors, because $\frac{4}{3}\pi\,20^{3}\approx32^{3}$) is 
approximately 1.25 and the corresponding total error is of the order 
$5\times10^{-12}$ (see Fig.\ \ref{fehlerformel}). 
Although much optimization effort has been put into mesh methods in order 
to reduce errors, we must face the fact that one generally loses many 
orders of magnitude in accuracy due to discretization. So if high accuracy 
is essential but speed is not an issue, the conventional Ewald method is 
unsurpassed: it is much easier to program and the desired accuracy can
be increased up to machine precision without any additional programming 
effort. However, it would be misleading to infer that particle mesh methods 
sacrifice accuracy in favor of speed, because due to the more advantageous 
scaling with particle number (essentially $N\log N$ compared to $N^{3/2}$) 
there will always be a critical number $N^{\ast}$, such that the mesh method 
will be faster than the Ewald sum for particle numbers $N>N^{\ast}$. See 
e.g.\ Petersen\cite{P} for a discussion of the break-even value $N^{\ast}$ 
for PME\@. 


\subsection*{Error as a function of minimum image distance}

Instead of calculating the rms error for a complete configuration, it is 
also worthwhile to investigate it as a function of the minimum image distance 
$r$ between just {\em two} particles. This is a possibility to monitor
the distance dependence of the accuracy for the various methods.
Thus, we randomly created a pair of particles inside the simulation box 
(again, $L=10$) with {\em given} minimum image separation $r$ and calculated 
the rms error from Eqn.\ (\ref{rmsforce}). This was repeated for $5\times 10^{4}$ 
separations equally spaced between 0 and $\frac{1}{2}\sqrt{3}L$, which is 
the largest possible minimum image separation. As this is done at constant
$\alpha$ for each method, the real space contribution to the force always 
cancels when performing the difference in Eqn.\ (\ref{rmsforce}), so this 
plot is only sensitive to the {\em Fourier} contribution and it is not necessary 
to specify a real space cutoff. A grid with $N\Mesh=32$ was chosen and the 
charge assignment order was set to $P=7$. However, as can be seen from
Fig.\ \ref{alphafehler}, different methods have their optimal operation
point at different values of $\alpha$. Therefore we found it more sensible
to compare the different methods at their {\em individual} optimal value of
$\alpha$, which can be obtained from Fig.\ \ref{alphafehler}. Although the 
curves in Fig.\ \ref{alphafehler} correspond to a system which contains 100 
particles (and not just 2), we believe that this has no influence on the 
optimal value of $\alpha$, since e.g.\ the error formulas for the Ewald sum 
derived by Kolafa and Perram\cite{KP} show, that the real space and the 
Fourier space contribution to $\Delta F$ display the same dependence on 
particle number.

Note that the Coulomb problem in the given periodic geometry lacks
spherical symmetry and due to the existence of a grid also the translational 
symmetry is broken. So -- strictly speaking -- $\Delta F$ is not just a 
function of $r$ but also depends on the orientation of the particles and 
their location within the box. This ma\-ni\-fests itself in the fact that the 
measured points $\Delta F(r)$ do not collapse onto a single smooth curve but 
show some scatter. Since we are not interested in this effect, we averaged 
the scatter by binning 50 points together at one time and additionally 
performing a Gaussian smoothing (with width 0.1). This just makes the data 
easier to plot and digest. 

\begin{figure}[!t]
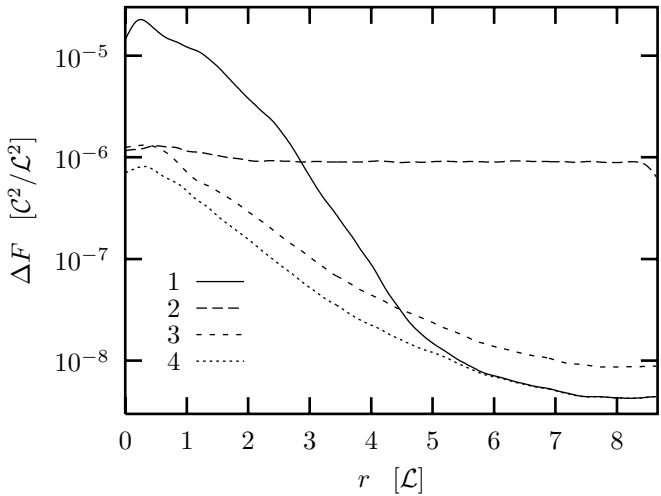

\hspace{-0.7cm}\parbox{8.6cm}{\include{f7}}
\caption{Distance dependency: The rms error $\Delta F$ as a 
  function of the minimum image separation $r$ between two 
  particles is shown for several mesh methods, which all share 
  $N\Mesh=32$ and $P=7$.
  Line 1 corresponds to PME, line 2 and 3 are analytically and 
  $i\VECk$-differentiated SPME respectively and line 4 is 
  $i\VECk$-differentiated P$^{3}$M. 
  For each method $\alpha$ was individually set to its optimal 
  value from Fig. \ref{alphafehler}: $\alpha_{1}=0.8$, $\alpha_{2}
  =0.89$, $\alpha_{3}=0.92$ and $\alpha_{4}=0.94$.}\label{rfehler}
\end{figure}

The result of this measurement is shown in Fig.\ \ref{rfehler}. 
Several interesting things can be observed: All algorithms
produce their largest errors at small distances and get considerably
more accurate at larger values of $r$ -- with one exception: The
analytically differentiated SPME method almost immediately 
settles to a (comparatively large) constant error. Since the only
difference between line 2 and 3 is the differentiation scheme, it 
must be the random force fluctuations discussed in the section
on back-interpolation which are responsible for this effect.
Note that PME at some distance gives better results than
$i\VECk$-differentiated SPME. Also it is most surprising that
at large distances PME and P$^{3}$M give identical errors,
although they differ considerably in the charge assignment scheme as
well as in the employed Coulomb Green function. Finally, the 
$i\VECk$-differentiated P$^{3}$M method is most accurate for all 
distances. In this case this is not so much surprising, because the 
quantity $Q$ from Eqn.\ (\ref{P3MQ}), with respect to which P$^{3}$M 
is optimized, is essentially the integral over any of these curves 
in Fig.\ \ref{rfehler}, weighted with the probability density of the 
minimum image distance $r$.

\section*{Conclusions}

Based on our theoretical considerations and the results of our
numerical experiments we draw the following conclusions:

\begin{itemize}
\item The error of all Ewald calculations -- be it the standard Ewald
  sum or any particle mesh method -- depends very sensitively on the
  Ewald parameter $\alpha$. Hence, finding the optimal value of $\alpha$
  is not merely an option but {\em absolutely essential}. Abstaining from 
  a proper $\alpha$-tuning results (at best) in wasting accuracy and 
  (at worst) in the calculation of wrong forces or energies.
  For the standard Ewald sum and for PME there exist estimates for the
  real- and reciprocal space contribution to the error, which allow an
  a priori determination of the optimal value for $\alpha$, depending on
  the relevant system parameters. 
  For the P$^{3}$M method we will tackle this problem in a forthcoming 
  publication \cite{wir}.
\item At given computational effort the errors produced by different methods
  do not just vary marginally but by orders of magnitude, so putting some
  effort into this topic is certainly worth the trouble.
\item Generally, the total error is a combination of several contributions. 
  If one of them dominates, there is no point in improving the other parts 
  of the method. Assume for instance that we are using a finite difference 
  scheme for the derivative and that the overall accuracy is actually limited 
  by the discretization errors resulting from a low order charge assignment 
  scheme. In this case, increasing the order $J$ of the differentiation scheme
  would be useless. E.g., in Fig.\ \ref{alphafehler_P3M3} going beyond $J=7$ 
  would not yield any improvement. If in this figure the assignment order
  $P$ was 3 and not 7, it would even suffice to use $J=2$ (compare with Fig.\ 
  \ref{alphafehler_P3M1}).
\item Different methods scale differently with respect to their parameters.
  E.g., since the $\alpha$-dependency of the error for PME is not the same
  as for SPME (there is not just a constant factor between them), the error 
  curves can intersect (see Fig.\ \ref{alphafehler}). The (almost trivial)
  consequence is that if one method is more accurate than another method in 
  a specific region of the space of tuning parameters, this need not be the
  case in another region or even for all choices of parameters.
\item There exist many possibilities for combining the various parts of a mesh
  calculation -- like charge assignment or differentiation scheme. This
  freedom of choice can be exploited to suit ones particle mesh algorithm to 
  already existing constraints in the complete simulation program.
  However, these combinations should always be tested thoroughly, since
  naive ``improvements'' can turn out to be disastrous (see line 2 in
  Fig.\ \ref{alphafehler}). There are, so to speak, several incompatible 
  roads towards  optimization, and one step away from a local optimum is 
  in general a disimprovement.
\item If method A is at the same computational effort 10 times more accurate 
  than method B, this can be advantageous even if one is happy with the 
  accuracy of B: Almost surely method A -- tuned down to the accuracy of B -- 
  will be faster than B, because, e.g., the number of mesh points could be 
  reduced.
\item If one wants to use the continuum Green function, a Lagrange 
  interpolation scheme should be used. However, our tests show that
  a combination of the smooth spline interpolation with an appropriately 
  adjusted Green function -- like in the P$^{3}$M and SPME approach -- 
  should be preferred, since this can be made more accurate.
  We recommend the P$^{3}$M approach, because it uses the analytically 
  derived optimal influence function from Eqn.\ 
  (\ref{optimalinfluencefunction}), which minimizes the force errors, 
  and -- due to its smooth charge assignment -- permits all investigated 
  differentiation schemes.
  In particular, the $i\VECk$ method is the most accurate implementation 
  of the derivative (and comes most closely to the Ewald method, see Eqn.\ 
  (\ref{Impulsraumanteil_Kraft})), whereas the analytic differentiation
  (introduced by Essmann {\em et.\,al}.\ \cite{EPBDLP} originally
  for the SPME method) -- although somewhat less accurate -- is a fast
  and attractive alternative.
\item Compared to a standard Ewald sum, which uses the same number of 
  $\VECk$-vectors, all mesh algorithms are much less accurate (see Fig.\ 
  \ref{fehlerformel}). However, the accuracy which is actually {\em needed}
  in a simulation is typically not too large, since most simulations
  employ at the same time some kind of thermostat, and it is a waste of 
  time to calculate the electrostatic forces much more accurate than the
  random fluctuations of the thermostat.
\end{itemize}

{\em Note added in proof.} After the submission of our paper T.\ Darden 
kindly brought to our attention a publication\cite{DTP} where he performed a 
numerical comparison of the P$^{3}$M to the SPME method, leading to results 
which are in agreement with our findings.


\section*{Acknowledgments}

Both authors are grateful to W.\ F.\ van Gunsteren for initiating
this research, U. Micka and Q. Spreiter for fruitful discussions, 
and to K.\ Kremer for encouragement and helpful comments. 
C.\ H.\ further thanks the DFG for financial support.


\begin{appendix}


\section{Some comments on units}\label{unitconventions}

Different people and communities prefer different conventions 
for units, especially if it comes to electrostatics. In this
small appendix we present our choice.

We write the Coulomb potential generated by a point particle
with charge $q$ located at the position $\VECr_{0}$ as:
\begin{equation}
\phi(\VECr) = \frac{q}{|\VECr-\VECr_{0}|}
\end{equation}
Thus, its dimension is charge divided by length. In other words,
if we measure all lengths in multiples of some unit length
$\CALL$ and all charges in multiples of some unit charge $\CALC$,
the dimension of the electrostatic potential is $\CALC/\CALL$.
As a consequence, the dimension of electrostatic energy is
$\CALC^{2}/\CALL$ and of electrostatic force is $\CALC^{2}/\CALL^{2}$.

In this article there is no need in specifying $\CALC$ or $\CALL$, 
and if it comes to the final result (be it formulas or numbers), 
it can always be embellished with prefactors like 
$1/4\pi\varepsilon_{0}$. 

We give just one example: If one chooses $\CALL=$ {\AA} = $10^{-10}\,$m,
$\CALC=e_{0}\approx1.6022\times10^{-19}\,$C and includes the standard-SI-prefactor
$1/4\pi\varepsilon_{0}$, the numerical value of the original expression
$q_{1}q_{2}/r^{2}$ gives the force in units of $2.3071\times10^{-8}\,$N. 

Another common unit of force -- especially among chemists -- is
kcal mol$^{-1}$ {\AA}$^{-1}$. Obviously we have
\begin{displaymath}
\frac{\mbox{kcal}}{\mbox{mol}\,\mbox{{\AA}}} \approx
\frac{4.186\times10^{3}\,\mbox{J}}{6.02217\times10^{23}\;10^{-10}\,\mbox{m}}
\approx
6.9510\times10^{-10}\,\mbox{N}
\end{displaymath}
Thus, if one prefers to measure forces in units of  kcal mol$^{-1}$ 
{\AA}$^{-1}$, one only has to multiply the numerical value of 
the original $q_{1}q_{2}/r^{2}$ by a factor of approximately 331.9.


\section{Continuum Green function and Lagrange 
  interpolation scheme}\label{GreenandLagrange}

In this appendix we show, how the implemented Green function and
the charge assignment scheme are related to each other. 
More specifically, we demonstrate that the use of the {\em continuum} 
version of the Coulomb Green function, as it appears in the conventional 
Ewald sum, suggests a so called {\em Lagrange interpolation scheme}, 
because this leads to a nice cancellation of certain discretization errors.
We closely follow the notation of Hockney and Eastwood\cite{HE}.

We consider only the one-dimensional case.
The electrostatic potential at position $x'$ due to a unit charge 
residing at position $x$ is {\em not} just a function of $|x'-x|$ 
but also depends on the distances of this charge from its neighboring 
mesh points. This artifact of the mesh can be quantified as follows:
Let $g(x)$ be the continuum Coulomb Green function and 
$W_{p}(x)=W(x-x_{p})$ the charge assigned to mesh point $p$ at 
position $x_{p}$ due to a unit charge at position $x$. The electrostatic
potential at position $x'$ can then be written as
\begin{equation}\label{phiGitter}
\phi(x') = \sum_{p=1}^{P} W_{p}(x) \, g(x'-x_{p})
\end{equation}
where the sum is taken over all $P$ mesh points to which
the particle at position $x$ contributed some fraction of
its charge, i.e.\ the $P$ mesh points which are closest 
to $x$. Taylor expanding $g(x'-x_{p})$ about $(x'-x)$ gives:
\begin{equation}\label{phiTaylor}
\phi(x') = \sum_{p=1}^{P} W_{p}(x) 
  \sum_{n=0}^{\infty}\frac{(x-x_{p})^{n}}{n!}g^{(n)}(x-x')
\end{equation}
It is possible to cancel the artificial terms in the $n$-sum 
(i.e.\ the ones which depend on $x-x_{p}$) up to order $P$ by 
choosing the charge fractions $W_{p}(x)$ such that
\begin{equation}\label{BedingungHE}
\sum_{p=1}^{P} W_{p}(x) \, (x-x_{p})^{n-1} = \delta_{1,n}
\quad,\quad n=1,\ldots,P
\end{equation}
By induction with respect to $n$ one can show that this may 
equivalently be expressed as
\begin{equation}\label{BedingungICH1}
\sum_{p=1}^{P} W_{p}(x) \, x_{p}^{n-1} = x^{n-1}
\quad,\quad n=1,\ldots,P
\end{equation}
This system of $P$ linear equations has a unique solution for
the $W_{p}(x)$ since the coefficient matrix $x_{p}^{n-1}$ is
a Vandermonde matrix for the distinct points $x_{1},\ldots,x_{P}$
and hence has full rank.
The $W_{p}(x)$ are thus polynomials of degree $P-1$.
Since in particular Eqn.\ (\ref{BedingungICH1}) must be true
at the mesh points, it follows
\begin{equation}\label{BedingungICH2}
\sum_{p=1}^{P} W_{p}(x_{q}) \, x_{p}^{n-1} = x_{q}^{n-1}
\quad,\quad n,q=1,\ldots,P
\end{equation}
which -- again due to the invertibility of $x_{p}^{n-1}$ -- 
can only be true if
\begin{equation}\label{Lagrange}
W_{p}(x_{q}) = \delta_{pq}\quad,\quad p,q=1,\ldots,P
\end{equation}
Equation (\ref{Lagrange}) suffices to determine the polynomials 
$W_{p}(x)$. They are referred to as the {\em fundamental 
polynomials for the Lagrange interpolation problem} \cite{LT}.
Petersen\cite{P} tabulates them for $P=3,\ldots,7$ and
their implementation is explained in detail.
If one needs these assignment functions for higher values of $P$,
one has to solve the system of linear equations (\ref{BedingungICH1})
or the interpolation problem (\ref{Lagrange}).


\section{Systematic difference approximations to the 
  differential operator}\label{optidiff}

In this appendix we show, how mesh approximations for the
differential operator $\infd/\infd x$ can systematically be 
written as convex combinations of difference operators.
In this way one can implement optimal combinations of these
operators into the program right-away, so an empirical tuning 
of the coefficients\cite{HE,Gunny} is no longer necessary. 
We describe the idea only for one dimension, the generalization 
to higher dimensions can be done easily via the Cartesian 
components. 

First we define the $j\th$-neighbor centered difference 
operator $\Delta_{j}$ by
\begin{equation}
(\Delta_{j}\,f)(x) := \frac{f(x+jh)-f(x-jh)}{2jh}
\end{equation}
where $h$ is the mesh spacing and $x$ some mesh point. 
Applying this operator on a function $f$ can be written
as the convolution $D_{j}\star f$, where $D_{j}(x)$ 
is defined as
\begin{equation}
D_{j}(x) := \frac{\delta(x+jh)-\delta(x-jh)}{2jh}
\end{equation}
From the convolution theorem it follows that in Fourier
space the derivative is given by $\tilde{D}_{j}(k)\tilde{f}(k)$,
where it is easily verified that the Fourier transform of
$D_{j}$ is
\begin{equation}\label{Fourierdiff}
\tilde{D}_{j}(k) = i\frac{\sin(jkh)}{jh}
\end{equation}
(Note that in the limit $h\downarrow 0$ this reduces to the
Fourier representation of $\infd/\infd x$, namely $ik$.)

Since one can expect to achieve better approximations 
for the differential operator by using linear combinations 
of the difference operators $\Delta_{j}$, we define a 
$J\th$-order difference operator by
\begin{equation}\label{NteOrdnungDifferenz}
\Delta^{(J)} := \sum_{j=1}^{J}c_{j}\,\Delta_{j}
\end{equation}
Using the Fourier representation of the differential 
and the $j\th$-neighbor centered difference operator 
from Eqn.\ (\ref{Fourierdiff}), we demand
\begin{eqnarray}
\sum_{j=1}^{J} c_{j} \, i\frac{\sin(jkh)}{jh} & = & 
ik + {\mathcal O}((kh)^{2J+1}) \\
\mbox{or}\qquad
\sum_{j=1}^{J} c_{j}\,\cos(jkh) & = & 1 + 
{\mathcal O}((kh)^{2J}) \label{optdiff}
\end{eqnarray}
where the second equation follows from differentiating the 
first. Taylor expanding the cosine in Eqn.\ (\ref{optdiff}) 
and equating coefficients gives $J$ linear equations for 
the $J$ unknowns $c_{j}$. The first few are given in 
Table (\ref{optdiff_table}). 

Note that in the case of the $2\nd$-order approximation 
the weighting $(\frac{4}{3},-\frac{1}{3})$, which empirically 
was found to be optimal \cite{HE}, is reproduced.

\begin{table}
\begin{tabular}{ccccccc}
order $J$ & $c_{1}$ & $c_{2}$ & $c_{3}$ & $c_{4}$ & $c_{5}$ & $c_{6}$ \\ \hline
  1   &  1   &        &       &       &       &        \\
  2   &  4/3 &  -1/3  &       &       &       &        \\
  3   &  3/2 &  -3/5  &  1/10 &       &       &        \\
  4   &  8/5 &  -4/5  &  8/35 & -1/35 &       &        \\
  5   &  5/3 & -20/21 &  5/14 & -5/63 & 1/126 &        \\ 
  6   & 12/7 & -15/14 & 10/21 & -1/7  & 2/77  & -1/462 \\
\end{tabular}
\caption{Optimal form for the weighting coefficients of the $J\th$-order 
difference operator $\Delta^{(J)}$ from Eqn.\ (\ref{NteOrdnungDifferenz})
for several values of $J$.}\label{optdiff_table}
\end{table}


\section{The model system}\label{modelsystem}

The rms error in the force for a system of 100 particles randomly 
distributed in the simulation box is somewhat sensitive to details
of the generated configuration, e.g.\ the actual minimum distance. 
In order to make our measurements fully reproducible we decided to 
present our configuration as well.

We found it easier not to list the particle positions but to describe
the procedure which was used to generate them.
The coordinates of the 100 particles were constructed by first drawing 
300 random numbers $\CALR_{n}$ between 0 and 1.
If $L$ is the box length then particle 1 gets the coordinates 
$(L\CALR_{1},L\CALR_{2},L\CALR_{3})$, particle 2
gets $(L\CALR_{4},L\CALR_{5},L\CALR_{6})$ and so on. 
Moreover, particles with an even/odd number will get a positive/negative 
unit charge.

The choice of the random number generator is the following: If $a_{n}$
is a positive integer, define its successor $a_{n+1}$ via:
\begin{equation}\label{Generator1}
a_{n+1} := (1103515245 \; a_{n} + 12345) \mod 2^{32}
\end{equation}
Now define the pseudo random number $\CALR_{n}\in[0;1[$ by
\begin{equation}\label{Generator2}
\CALR_{n} := \frac{(a_{n}\div65536) \mod 32768}{32769}
\end{equation}
Where ``$\div$'' should denote an integer division which discards
any division rest. 
Choosing $a_{0}=1$ we obtained the sequence of random numbers 
$16838,5758,10113,\ldots$, of which the first 300 were used for 
positioning the particles, e.g.\ -- with $L=10$ -- the first 
particle has coordinates $(5.138\ldots,1.757\ldots,3.086\ldots)$ 
and a negative unit charge. The smallest minimum image distance 
is approximately 0.370264 and occurs between particle 46 and 
particle 98.

Incidentally, we did not choose this random number generator 
because it is particulary good (it is not), but it is very easy 
to implement. Many {\tt C} libraries provide a function {\tt rand}, 
which relies on Eqns.\ (\ref{Generator1},\ref{Generator2}).


\section{Charge assignment with splines}\label{splinepolies}

In this appendix we describe in a little more detail the procedure of
charge assignment and present the charge fractions which are needed 
for a $P\th$ order assignment scheme \`{a} la Hockney and Eastwood 
\cite{HE} (see Eqns.\ (\ref{HEsplinek},\ref{HEspline})).

Let the units be chosen such that the grid spacing is $1$. 
For any $P$ consecutive mesh points there exists an interval $\CALI$ 
of length $1$ such that the charge of a particle with coordinate
$x\in\CALI$ is distributed between these mesh points. By
simple shifting we can assume this interval to be 
$[-\frac{1}{2},+\frac{1}{2}]$. Then the $P$ mesh points
will lie at $-\frac{P-1}{2},-\frac{P-1}{2}+1,\ldots,\frac{P-1}{2}$.
For $P=3$ this is schematically shown in Fig.\ \ref{chargeassignment1}.
The charge fraction $W_{p}(x)$, which will be assigned to the mesh point
$x_{p}$, is related to the charge assignment function $W(x)$ via
$W_{p}(x)=W(x-x_{p})$.

The basic steps which have to be done for a particle with coordinate 
$x$ (generally not in $[-\frac{1}{2},+\frac{1}{2}]$) during a $P\th$ 
order charge assignment are thus:
\begin{enumerate}
\item Define $\bar{x}$ to be the coordinate of the particle's 
  nearest mesh point (if $P$ is odd) or the midpoint between 
  the two nearest mesh points (if $P$ is even).
\item Find the $P$ mesh points $x_{p}$ which are closest to $x$.
  They will be indexed by their relative position to $\bar{x}$,
  so $p\in\{-\frac{P-1}{2},-\frac{P-1}{2}+1,\ldots,\frac{P-1}{2}\}$
\item The fraction of charge which is assigned to each of these
  mesh points is given by $W_{p}(x-\bar{x})$.
\end{enumerate}
In this way the charge fractions are written as a function
of the separation $x-\bar{x}\in[-\frac{1}{2},+\frac{1}{2}]$.
Hockney and Eastwood refer to the cases $P$ = 1, 2 and 3 as 
NGP (nearest grid point), CIC (cloud in cell) and TSC (triangular 
shaped cloud) respectively. Generally, for $P\in\{1,\ldots,7\}$
the charge fractions $W_{p}^{(P)}(x)$ are given by the following 
polynomials:

\begin{figure}[!t]
\includegraphics{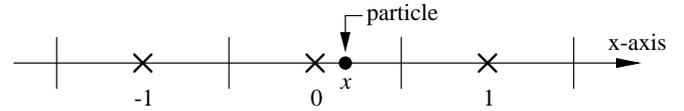}
\caption{Schematic picture for a three-point charge assignment. 
The crosses are the mesh points and the lines indicate the
(Wigner-Seitz) cell boundaries of each point (the mesh spacing
is $h=1$).
All particles with $x\in[-\frac{1}{2},+\frac{1}{2}]$ distribute 
their charge between the mesh points at -1, 0 and +1 and the 
corresponding charge fractions are $W_{p}(x)$, $p\in\{-1,0,+1\}$.
(After Hockney and Eastwood \cite{HE}.)}\label{chargeassignment1}
\end{figure}

$P=1$:
\begin{eqnarray*}
W^{(1)}_{ 0}(x) & = & 1
\end{eqnarray*}

$P=2$:
\begin{eqnarray*}
W^{(2)}_{-1/2}(x) & = & \frac{1}{2}(1-2x) \\
W^{(2)}_{+1/2}(x) & = & \frac{1}{2}(1+2x)
\end{eqnarray*}

$P=3$:
\begin{eqnarray*}
W^{(3)}_{-1}(x) & = & \frac{1}{8}(1-4x+4x^2) \\
W^{(3)}_{ 0}(x) & = & \frac{1}{4}(3-4x^2) \\
W^{(3)}_{+1}(x) & = & \frac{1}{8}(1+4x+4x^2)
\end{eqnarray*}

$P=4$:
\begin{eqnarray*}
W^{(4)}_{-3/2}(x) & = & \frac{1}{48}(1-6x+12x^2-8x^3) \\
W^{(4)}_{-1/2}(x) & = & \frac{1}{48}(23-30x-12x^2+24x^3) \\
W^{(4)}_{+1/2}(x) & = & \frac{1}{48}(23+30x-12x^2-24x^3) \\
W^{(4)}_{+3/2}(x) & = & \frac{1}{48}(1+6x+12x^2+8x^3)
\end{eqnarray*}

$P=5$:
\begin{eqnarray*}
W^{(5)}_{-2}(x) & = & \frac{1}{384}(1-8x+24x^2-32x^3+16x^4) \\
W^{(5)}_{-1}(x) & = & \frac{1}{ 96}(19-44x+24x^2+16x^3-16x^4) \\
W^{(5)}_{ 0}(x) & = & \frac{1}{192}(115-120x^2+48x^4) \\
W^{(5)}_{+1}(x) & = & \frac{1}{ 96}(19+44x+24x^2-16x^3-16x^4) \\
W^{(5)}_{+2}(x) & = & \frac{1}{384}(1+8x+24x^2+32x^3+16x^4) \\
\end{eqnarray*}

\newpage

\begin{minipage}{18cm}

$P=6$:
\begin{eqnarray*}
W^{(6)}_{-5/2}(x) & = & \frac{1}{3840}(1-10x+40x^2-80x^3+80x^4-32x^5) \\
W^{(6)}_{-3/2}(x) & = & \frac{1}{3840}(237-750x+840x^2-240x^3-240x^4+160x^5) \\
W^{(6)}_{-1/2}(x) & = & \frac{1}{1920}(841-770x-440x^2+560x^3+80x^4-160x^5) \\
W^{(6)}_{+1/2}(x) & = & \frac{1}{1920}(841+770x-440x^2-560x^3+80x^4+160x^5) \\
W^{(6)}_{+3/2}(x) & = & \frac{1}{3840}(237+750x+840x^2+240x^3-240x^4-160x^5) \\
W^{(6)}_{+5/2}(x) & = & \frac{1}{3840}(1+10x+40x^2+80x^3+80x^4+32x^5)
\end{eqnarray*}

$P=7$:
\begin{eqnarray*}
W^{(7)}_{-3}(x) & = & \frac{1}{46080}(1-12x+60x^2-160x^3+240x^4-192x^5+64x^6) \\
W^{(7)}_{-2}(x) & = & \frac{1}{23040}(361-1416x+2220x^2-1600x^3+240x^4+384x^5-192x^6) \\
W^{(7)}_{-1}(x) & = & \frac{1}{46080}(10543-17340x+4740x^2+6880x^3-4080x^4-960x^5+960x^6) \\
W^{(7)}_{ 0}(x) & = & \frac{1}{11520}(5887-4620x^2+1680x^4-320x^6) \\
W^{(7)}_{+1}(x) & = & \frac{1}{46080}(10543+17340x+4740x^2-6880x^3-4080x^4+960x^5+960x^6) \\
W^{(7)}_{+2}(x) & = & \frac{1}{23040}(361+1416x+2220x^2+1600x^3+240x^4-384x^5-192x^6) \\
W^{(7)}_{+3}(x) & = & \frac{1}{46080}(1+12x+60x^2+160x^3+240x^4+192x^5+64x^6) \\
\end{eqnarray*}

\end{minipage}


\end{appendix}


\end{document}

%% file: f1.tex
\setlength{\unitlength}{0.1bp}
\special{!
/gnudict 40 dict def
gnudict begin
/Color false def
/Solid false def
/gnulinewidth 5.000 def
/vshift -33 def
/dl {10 mul} def
/hpt 31.5 def
/vpt 31.5 def
/M {moveto} bind def
/L {lineto} bind def
/R {rmoveto} bind def
/V {rlineto} bind def
/vpt2 vpt 2 mul def
/hpt2 hpt 2 mul def
/Lshow { currentpoint stroke M
  0 vshift R show } def
/Rshow { currentpoint stroke M
  dup stringwidth pop neg vshift R show } def
/Cshow { currentpoint stroke M
  dup stringwidth pop -2 div vshift R show } def
/DL { Color {setrgbcolor Solid {pop []} if 0 setdash }
 {pop pop pop Solid {pop []} if 0 setdash} ifelse } def
/BL { stroke gnulinewidth 2 mul setlinewidth } def
/AL { stroke gnulinewidth 2 div setlinewidth } def
/PL { stroke gnulinewidth setlinewidth } def
/LTb { BL [] 0 0 0 DL } def
/LTa { AL [1 dl 2 dl] 0 setdash 0 0 0 setrgbcolor } def
/LT0 { PL [] 0 1 0 DL } def
/LT1 { PL [4 dl 2 dl] 0 0 1 DL } def
/LT2 { PL [2 dl 3 dl] 1 0 0 DL } def
/LT3 { PL [1 dl 1.5 dl] 1 0 1 DL } def
/LT4 { PL [5 dl 2 dl 1 dl 2 dl] 0 1 1 DL } def
/LT5 { PL [4 dl 3 dl 1 dl 3 dl] 1 1 0 DL } def
/LT6 { PL [2 dl 2 dl 2 dl 4 dl] 0 0 0 DL } def
/LT7 { PL [2 dl 2 dl 2 dl 2 dl 2 dl 4 dl] 1 0.3 0 DL } def
/LT8 { PL [2 dl 2 dl 2 dl 2 dl 2 dl 2 dl 2 dl 4 dl] 0.5 0.5 0.5 DL } def
/P { stroke [] 0 setdash
  currentlinewidth 2 div sub M
  0 currentlinewidth V stroke } def
/D { stroke [] 0 setdash 2 copy vpt add M
  hpt neg vpt neg V hpt vpt neg V
  hpt vpt V hpt neg vpt V closepath stroke
  P } def
/A { stroke [] 0 setdash vpt sub M 0 vpt2 V
  currentpoint stroke M
  hpt neg vpt neg R hpt2 0 V stroke
  } def
/B { stroke [] 0 setdash 2 copy exch hpt sub exch vpt add M
  0 vpt2 neg V hpt2 0 V 0 vpt2 V
  hpt2 neg 0 V closepath stroke
  P } def
/C { stroke [] 0 setdash exch hpt sub exch vpt add M
  hpt2 vpt2 neg V currentpoint stroke M
  hpt2 neg 0 R hpt2 vpt2 V stroke } def
/T { stroke [] 0 setdash 2 copy vpt 1.12 mul add M
  hpt neg vpt -1.62 mul V
  hpt 2 mul 0 V
  hpt neg vpt 1.62 mul V closepath stroke
  P  } def
/S { 2 copy A C} def
end
}
\begin{picture}(2789,1836)(0,0)
\special{"
gnudict begin
gsave
50 50 translate
0.100 0.100 scale
0 setgray
/Helvetica findfont 100 scalefont setfont
newpath
-500.000000 -500.000000 translate
LTa
600 507 M
2006 0 V
1603 251 M
0 1534 V
LTb
600 251 M
63 0 V
1943 0 R
-63 0 V
600 507 M
63 0 V
1943 0 R
-63 0 V
600 762 M
63 0 V
1943 0 R
-63 0 V
600 1018 M
63 0 V
1943 0 R
-63 0 V
600 1274 M
63 0 V
1943 0 R
-63 0 V
600 1529 M
63 0 V
1943 0 R
-63 0 V
600 1785 M
63 0 V
1943 0 R
-63 0 V
600 251 M
0 63 V
0 1471 R
0 -63 V
851 251 M
0 63 V
0 1471 R
0 -63 V
1102 251 M
0 63 V
0 1471 R
0 -63 V
1352 251 M
0 63 V
0 1471 R
0 -63 V
1603 251 M
0 63 V
0 1471 R
0 -63 V
1854 251 M
0 63 V
0 1471 R
0 -63 V
2105 251 M
0 63 V
0 1471 R
0 -63 V
2355 251 M
0 63 V
0 1471 R
0 -63 V
2606 251 M
0 63 V
0 1471 R
0 -63 V
600 251 M
2006 0 V
0 1534 V
-2006 0 V
600 251 L
LT0
600 507 M
20 0 V
21 0 V
20 0 V
20 0 V
20 0 V
21 0 V
20 0 V
20 0 V
20 0 V
21 0 V
20 0 V
20 0 V
863 347 M
21 3 V
20 4 V
20 7 V
20 8 V
21 11 V
20 13 V
20 15 V
21 16 V
20 20 V
20 21 V
20 23 V
21 25 V
20 28 V
20 29 V
20 32 V
21 33 V
20 36 V
20 38 V
20 40 V
21 42 V
20 44 V
20 46 V
20 48 V
21 51 V
1370 1509 M
20 46 V
21 42 V
20 37 V
20 34 V
20 29 V
21 25 V
20 21 V
20 16 V
20 13 V
21 8 V
20 4 V
20 0 V
20 -4 V
21 -8 V
20 -13 V
20 -16 V
20 -21 V
21 -25 V
20 -29 V
20 -34 V
20 -37 V
21 -42 V
20 -46 V
1856 980 M
21 -51 V
20 -48 V
20 -46 V
20 -44 V
21 -42 V
20 -40 V
20 -38 V
20 -36 V
21 -33 V
20 -32 V
20 -29 V
20 -28 V
21 -25 V
20 -23 V
20 -21 V
20 -20 V
21 -16 V
20 -15 V
20 -13 V
21 -11 V
20 -8 V
20 -7 V
20 -4 V
21 -3 V
2363 507 M
20 0 V
20 0 V
21 0 V
20 0 V
20 0 V
20 0 V
21 0 V
20 0 V
20 0 V
20 0 V
21 0 V
20 0 V
LT3
600 507 M
20 0 V
21 0 V
20 0 V
20 0 V
20 0 V
21 0 V
20 0 V
20 0 V
20 0 V
21 0 V
20 0 V
20 0 V
20 0 V
21 2 V
20 5 V
20 6 V
20 9 V
21 11 V
20 12 V
20 15 V
21 17 V
20 19 V
20 21 V
20 24 V
21 25 V
20 27 V
20 30 V
20 31 V
21 34 V
20 36 V
20 38 V
20 40 V
21 42 V
20 44 V
20 46 V
20 48 V
21 50 V
20 50 V
20 46 V
21 42 V
20 38 V
20 33 V
20 29 V
21 25 V
20 21 V
20 17 V
20 12 V
21 9 V
20 4 V
20 0 V
20 -4 V
21 -9 V
20 -12 V
20 -17 V
20 -21 V
21 -25 V
20 -29 V
20 -33 V
20 -38 V
21 -42 V
20 -46 V
20 -50 V
21 -50 V
20 -48 V
20 -46 V
20 -44 V
21 -42 V
20 -40 V
20 -38 V
20 -36 V
21 -34 V
20 -31 V
20 -30 V
20 -27 V
21 -25 V
20 -24 V
20 -21 V
20 -19 V
21 -17 V
20 -15 V
20 -12 V
21 -11 V
20 -9 V
20 -6 V
20 -5 V
21 -2 V
20 0 V
20 0 V
20 0 V
21 0 V
20 0 V
20 0 V
20 0 V
21 0 V
20 0 V
20 0 V
20 0 V
21 0 V
20 0 V
stroke
grestore
end
showpage
}
\put(1603,0){\makebox(0,0){$x/h$}}
\put(280,1018){%
\special{ps: gsave currentpoint currentpoint translate
270 rotate neg exch neg exch translate}%
\makebox(0,0)[b]{\shortstack{$W(x)$}}%
\special{ps: currentpoint grestore moveto}%
}
\put(2606,151){\makebox(0,0){2}}
\put(2355,151){\makebox(0,0){1.5}}
\put(2105,151){\makebox(0,0){1}}
\put(1854,151){\makebox(0,0){0.5}}
\put(1603,151){\makebox(0,0){0}}
\put(1352,151){\makebox(0,0){-0.5}}
\put(1102,151){\makebox(0,0){-1}}
\put(851,151){\makebox(0,0){-1.5}}
\put(600,151){\makebox(0,0){-2}}
\put(540,1785){\makebox(0,0)[r]{1}}
\put(540,1529){\makebox(0,0)[r]{0.8}}
\put(540,1274){\makebox(0,0)[r]{0.6}}
\put(540,1018){\makebox(0,0)[r]{0.4}}
\put(540,762){\makebox(0,0)[r]{0.2}}
\put(540,507){\makebox(0,0)[r]{0}}
\put(540,251){\makebox(0,0)[r]{-0.2}}
\end{picture}

%% file: f2.tex
\setlength{\unitlength}{0.1bp}
\special{!
/gnudict 40 dict def
gnudict begin
/Color false def
/Solid false def
/gnulinewidth 5.000 def
/vshift -33 def
/dl {10 mul} def
/hpt 31.5 def
/vpt 31.5 def
/M {moveto} bind def
/L {lineto} bind def
/R {rmoveto} bind def
/V {rlineto} bind def
/vpt2 vpt 2 mul def
/hpt2 hpt 2 mul def
/Lshow { currentpoint stroke M
  0 vshift R show } def
/Rshow { currentpoint stroke M
  dup stringwidth pop neg vshift R show } def
/Cshow { currentpoint stroke M
  dup stringwidth pop -2 div vshift R show } def
/DL { Color {setrgbcolor Solid {pop []} if 0 setdash }
 {pop pop pop Solid {pop []} if 0 setdash} ifelse } def
/BL { stroke gnulinewidth 2 mul setlinewidth } def
/AL { stroke gnulinewidth 2 div setlinewidth } def
/PL { stroke gnulinewidth setlinewidth } def
/LTb { BL [] 0 0 0 DL } def
/LTa { AL [1 dl 2 dl] 0 setdash 0 0 0 setrgbcolor } def
/LT0 { PL [] 0 1 0 DL } def
/LT1 { PL [4 dl 2 dl] 0 0 1 DL } def
/LT2 { PL [2 dl 3 dl] 1 0 0 DL } def
/LT3 { PL [1 dl 1.5 dl] 1 0 1 DL } def
/LT4 { PL [5 dl 2 dl 1 dl 2 dl] 0 1 1 DL } def
/LT5 { PL [4 dl 3 dl 1 dl 3 dl] 1 1 0 DL } def
/LT6 { PL [2 dl 2 dl 2 dl 4 dl] 0 0 0 DL } def
/LT7 { PL [2 dl 2 dl 2 dl 2 dl 2 dl 4 dl] 1 0.3 0 DL } def
/LT8 { PL [2 dl 2 dl 2 dl 2 dl 2 dl 2 dl 2 dl 4 dl] 0.5 0.5 0.5 DL } def
/P { stroke [] 0 setdash
  currentlinewidth 2 div sub M
  0 currentlinewidth V stroke } def
/D { stroke [] 0 setdash 2 copy vpt add M
  hpt neg vpt neg V hpt vpt neg V
  hpt vpt V hpt neg vpt V closepath stroke
  P } def
/A { stroke [] 0 setdash vpt sub M 0 vpt2 V
  currentpoint stroke M
  hpt neg vpt neg R hpt2 0 V stroke
  } def
/B { stroke [] 0 setdash 2 copy exch hpt sub exch vpt add M
  0 vpt2 neg V hpt2 0 V 0 vpt2 V
  hpt2 neg 0 V closepath stroke
  P } def
/C { stroke [] 0 setdash exch hpt sub exch vpt add M
  hpt2 vpt2 neg V currentpoint stroke M
  hpt2 neg 0 R hpt2 vpt2 V stroke } def
/T { stroke [] 0 setdash 2 copy vpt 1.12 mul add M
  hpt neg vpt -1.62 mul V
  hpt 2 mul 0 V
  hpt neg vpt 1.62 mul V closepath stroke
  P  } def
/S { 2 copy A C} def
end
}
\begin{picture}(5328,3196)(0,0)
\special{"
gnudict begin
gsave
50 50 translate
0.100 0.100 scale
0 setgray
/Helvetica findfont 100 scalefont setfont
newpath
-500.000000 -500.000000 translate
LTa
600 251 M
0 2894 V
LTb
600 251 M
63 0 V
4482 0 R
-63 0 V
600 664 M
63 0 V
4482 0 R
-63 0 V
600 1078 M
63 0 V
4482 0 R
-63 0 V
600 1491 M
63 0 V
4482 0 R
-63 0 V
600 1905 M
63 0 V
4482 0 R
-63 0 V
600 2318 M
63 0 V
4482 0 R
-63 0 V
600 2732 M
63 0 V
4482 0 R
-63 0 V
600 3145 M
63 0 V
4482 0 R
-63 0 V
600 251 M
0 63 V
0 2831 R
0 -63 V
1358 251 M
0 63 V
0 2831 R
0 -63 V
2115 251 M
0 63 V
0 2831 R
0 -63 V
2873 251 M
0 63 V
0 2831 R
0 -63 V
3630 251 M
0 63 V
0 2831 R
0 -63 V
4388 251 M
0 63 V
0 2831 R
0 -63 V
5145 251 M
0 63 V
0 2831 R
0 -63 V
600 251 M
4545 0 V
0 2894 V
-4545 0 V
600 251 L
LT0
4599 1078 M
180 0 V
752 2828 M
15 -2 V
15 -2 V
15 -3 V
15 -3 V
15 -4 V
15 -5 V
16 -6 V
15 -7 V
15 -8 V
15 -9 V
15 -9 V
15 -11 V
15 -11 V
16 -12 V
15 -12 V
15 -14 V
15 -14 V
15 -14 V
15 -16 V
16 -16 V
15 -16 V
15 -17 V
15 -18 V
15 -19 V
15 -19 V
15 -19 V
16 -21 V
15 -21 V
15 -21 V
15 -23 V
15 -22 V
15 -24 V
15 -24 V
16 -25 V
15 -25 V
15 -26 V
15 -26 V
15 -27 V
15 -28 V
16 -28 V
15 -29 V
15 -29 V
15 -30 V
15 -30 V
15 -32 V
15 -31 V
16 -32 V
15 -33 V
15 -34 V
15 -34 V
15 -34 V
15 -35 V
15 -36 V
16 -36 V
15 -37 V
15 -37 V
15 -38 V
15 -39 V
15 -39 V
16 -39 V
15 -41 V
15 -40 V
15 -40 V
15 -40 V
15 -39 V
15 -35 V
16 -30 V
15 -22 V
15 -13 V
15 -2 V
15 5 V
15 9 V
15 13 V
16 14 V
15 14 V
15 15 V
15 15 V
15 15 V
15 15 V
16 14 V
15 14 V
15 15 V
15 14 V
15 14 V
15 14 V
15 13 V
16 14 V
15 13 V
15 14 V
15 13 V
15 13 V
15 13 V
15 13 V
16 13 V
15 12 V
15 13 V
15 12 V
15 12 V
15 12 V
16 12 V
15 12 V
15 11 V
15 12 V
15 11 V
15 12 V
15 11 V
16 11 V
15 10 V
15 11 V
15 11 V
15 10 V
15 10 V
15 11 V
16 10 V
15 10 V
15 9 V
15 10 V
15 10 V
15 9 V
16 10 V
15 9 V
15 9 V
15 9 V
15 9 V
15 9 V
15 9 V
16 9 V
15 8 V
15 9 V
15 8 V
15 9 V
15 8 V
15 8 V
16 9 V
15 8 V
15 8 V
15 8 V
15 8 V
15 8 V
16 8 V
15 7 V
15 8 V
15 8 V
15 7 V
15 8 V
15 7 V
16 8 V
15 7 V
15 8 V
15 7 V
15 7 V
15 7 V
15 7 V
16 8 V
15 7 V
15 7 V
15 7 V
15 7 V
15 7 V
16 6 V
15 7 V
15 7 V
15 7 V
15 6 V
15 7 V
15 7 V
16 6 V
15 7 V
15 6 V
15 7 V
15 6 V
15 6 V
15 7 V
16 6 V
15 6 V
15 6 V
15 7 V
15 6 V
15 6 V
16 6 V
15 6 V
15 6 V
15 5 V
15 6 V
15 6 V
15 6 V
16 6 V
15 5 V
15 6 V
15 5 V
15 6 V
15 5 V
15 6 V
16 5 V
15 6 V
15 5 V
15 5 V
15 5 V
15 6 V
16 5 V
15 5 V
15 5 V
15 5 V
15 5 V
15 5 V
15 5 V
16 5 V
15 5 V
15 5 V
15 4 V
15 5 V
15 5 V
15 4 V
16 5 V
15 5 V
15 4 V
15 5 V
15 4 V
15 5 V
15 4 V
16 4 V
15 5 V
15 4 V
15 4 V
15 4 V
15 5 V
16 4 V
15 4 V
15 4 V
15 4 V
15 4 V
15 4 V
15 4 V
16 4 V
15 4 V
15 4 V
15 4 V
15 3 V
15 4 V
16 4 V
15 3 V
15 4 V
15 4 V
15 3 V
15 4 V
15 3 V
16 4 V
15 3 V
15 4 V
15 3 V
15 4 V
15 3 V
15 3 V
16 4 V
15 3 V
15 3 V
15 3 V
15 4 V
15 3 V
16 3 V
15 3 V
15 3 V
15 3 V
15 3 V
15 3 V
15 3 V
16 3 V
15 3 V
15 3 V
15 3 V
15 3 V
15 3 V
15 2 V
16 3 V
15 3 V
15 3 V
15 2 V
15 3 V
15 3 V
16 2 V
15 3 V
15 2 V
15 3 V
15 3 V
15 2 V
15 3 V
16 2 V
15 3 V
15 2 V
15 2 V
LT1
4599 978 M
180 0 V
752 2828 M
15 -2 V
15 -2 V
15 -3 V
15 -3 V
15 -4 V
15 -5 V
16 -6 V
15 -7 V
15 -8 V
15 -8 V
15 -10 V
15 -10 V
15 -11 V
16 -12 V
15 -12 V
15 -13 V
15 -14 V
15 -14 V
15 -15 V
16 -15 V
15 -16 V
15 -16 V
15 -17 V
15 -17 V
15 -17 V
15 -18 V
16 -18 V
15 -18 V
15 -17 V
15 -16 V
15 -16 V
15 -14 V
15 -12 V
16 -10 V
15 -7 V
15 -4 V
15 -1 V
15 0 V
15 3 V
16 4 V
15 6 V
15 6 V
15 6 V
15 7 V
15 8 V
15 7 V
16 7 V
15 7 V
15 7 V
15 7 V
15 7 V
15 7 V
15 7 V
16 7 V
15 6 V
15 7 V
15 6 V
15 6 V
15 6 V
16 6 V
15 6 V
15 6 V
15 6 V
15 6 V
15 5 V
15 6 V
16 5 V
15 6 V
15 5 V
15 5 V
15 6 V
15 5 V
15 5 V
16 5 V
15 5 V
15 5 V
15 5 V
15 4 V
15 5 V
16 5 V
15 5 V
15 4 V
15 5 V
15 4 V
15 5 V
15 4 V
16 4 V
15 5 V
15 4 V
15 4 V
15 4 V
15 4 V
15 5 V
16 4 V
15 3 V
15 4 V
15 4 V
15 4 V
15 4 V
16 4 V
15 3 V
15 4 V
15 4 V
15 3 V
15 4 V
15 3 V
16 4 V
15 3 V
15 4 V
15 3 V
15 3 V
15 4 V
15 3 V
16 3 V
15 3 V
15 4 V
15 3 V
15 3 V
15 3 V
16 3 V
15 3 V
15 3 V
15 3 V
15 3 V
15 3 V
15 3 V
16 2 V
15 3 V
15 3 V
15 3 V
15 3 V
15 2 V
15 3 V
16 3 V
15 2 V
15 3 V
15 3 V
15 2 V
15 3 V
16 2 V
15 3 V
15 2 V
15 3 V
15 2 V
15 3 V
15 2 V
16 3 V
15 2 V
15 2 V
15 3 V
15 2 V
15 2 V
15 3 V
16 2 V
15 2 V
15 2 V
15 3 V
15 2 V
15 2 V
16 2 V
15 2 V
15 2 V
15 2 V
15 3 V
15 2 V
15 2 V
16 2 V
15 2 V
15 2 V
15 2 V
15 2 V
15 2 V
15 2 V
16 2 V
15 2 V
15 1 V
15 2 V
15 2 V
15 2 V
16 2 V
15 2 V
15 2 V
15 1 V
15 2 V
15 2 V
15 2 V
16 2 V
15 1 V
15 2 V
15 2 V
15 1 V
15 2 V
15 2 V
16 2 V
15 1 V
15 2 V
15 1 V
15 2 V
15 2 V
16 1 V
15 2 V
15 2 V
15 1 V
15 2 V
15 1 V
15 2 V
16 1 V
15 2 V
15 1 V
15 2 V
15 1 V
15 2 V
15 1 V
16 2 V
15 1 V
15 1 V
15 2 V
15 1 V
15 2 V
15 1 V
16 2 V
15 1 V
15 1 V
15 2 V
15 1 V
15 1 V
16 2 V
15 1 V
15 1 V
15 2 V
15 1 V
15 1 V
15 1 V
16 2 V
15 1 V
15 1 V
15 1 V
15 2 V
15 1 V
16 1 V
15 1 V
15 2 V
15 1 V
15 1 V
15 1 V
15 1 V
16 1 V
15 2 V
15 1 V
15 1 V
15 1 V
15 1 V
15 1 V
16 1 V
15 2 V
15 1 V
15 1 V
15 1 V
15 1 V
16 1 V
15 1 V
15 1 V
15 1 V
15 1 V
15 1 V
15 1 V
16 1 V
15 1 V
15 1 V
15 1 V
15 1 V
15 1 V
15 1 V
16 1 V
15 1 V
15 1 V
15 1 V
15 1 V
15 1 V
16 1 V
15 1 V
15 1 V
15 1 V
15 1 V
15 1 V
15 1 V
16 1 V
15 0 V
15 1 V
15 1 V
LT2
4599 878 M
180 0 V
752 2828 M
15 -2 V
15 -2 V
15 -3 V
15 -3 V
15 -4 V
15 -5 V
16 -6 V
15 -7 V
15 -8 V
15 -9 V
15 -9 V
15 -11 V
15 -11 V
16 -12 V
15 -12 V
15 -14 V
15 -14 V
15 -14 V
15 -16 V
16 -16 V
15 -16 V
15 -17 V
15 -18 V
15 -19 V
15 -19 V
15 -19 V
16 -21 V
15 -21 V
15 -21 V
15 -23 V
15 -22 V
15 -24 V
15 -24 V
16 -25 V
15 -25 V
15 -26 V
15 -26 V
15 -27 V
15 -28 V
16 -28 V
15 -29 V
15 -29 V
15 -30 V
15 -30 V
15 -32 V
15 -31 V
16 -32 V
15 -33 V
15 -34 V
15 -34 V
15 -34 V
15 -35 V
15 -36 V
16 -36 V
15 -37 V
15 -37 V
15 -38 V
15 -39 V
15 -39 V
16 -40 V
15 -40 V
15 -41 V
15 -41 V
15 -42 V
15 -42 V
15 -43 V
16 -44 V
15 -44 V
15 -44 V
15 -44 V
15 -45 V
15 -43 V
15 -43 V
16 -40 V
15 -37 V
15 -30 V
15 -21 V
15 -12 V
15 -4 V
16 4 V
15 8 V
15 11 V
15 13 V
15 14 V
15 15 V
15 16 V
16 15 V
15 17 V
15 16 V
15 16 V
15 17 V
15 16 V
15 17 V
16 16 V
15 17 V
15 16 V
15 16 V
15 17 V
15 16 V
16 16 V
15 16 V
15 16 V
15 16 V
15 16 V
15 16 V
15 16 V
16 16 V
15 15 V
15 16 V
15 15 V
15 16 V
15 15 V
15 16 V
16 15 V
15 15 V
15 15 V
15 15 V
15 15 V
15 15 V
16 15 V
15 15 V
15 15 V
15 14 V
15 15 V
15 15 V
15 14 V
16 15 V
15 15 V
15 14 V
15 15 V
15 14 V
15 14 V
15 15 V
16 14 V
15 14 V
15 15 V
15 14 V
15 14 V
15 14 V
16 14 V
15 14 V
15 14 V
15 14 V
15 14 V
15 13 V
15 14 V
16 14 V
15 13 V
15 14 V
15 13 V
15 13 V
15 14 V
15 13 V
16 13 V
15 13 V
15 13 V
15 13 V
15 12 V
15 13 V
16 12 V
15 13 V
15 12 V
15 12 V
15 13 V
15 12 V
15 12 V
16 11 V
15 12 V
15 12 V
15 11 V
15 12 V
15 11 V
15 11 V
16 11 V
15 11 V
15 11 V
15 11 V
15 11 V
15 10 V
16 11 V
15 10 V
15 10 V
15 10 V
15 10 V
15 10 V
15 10 V
16 10 V
15 10 V
15 9 V
15 10 V
15 9 V
15 9 V
15 9 V
16 9 V
15 9 V
15 9 V
15 9 V
15 9 V
15 8 V
16 9 V
15 8 V
15 9 V
15 8 V
15 8 V
15 8 V
15 8 V
16 8 V
15 8 V
15 8 V
15 7 V
15 8 V
15 7 V
15 8 V
16 7 V
15 7 V
15 8 V
15 7 V
15 7 V
15 7 V
15 7 V
16 7 V
15 6 V
15 7 V
15 7 V
15 6 V
15 7 V
16 6 V
15 7 V
15 6 V
15 6 V
15 6 V
15 6 V
15 7 V
16 6 V
15 5 V
15 6 V
15 6 V
15 6 V
15 6 V
16 5 V
15 6 V
15 6 V
15 5 V
15 6 V
15 5 V
15 5 V
16 6 V
15 5 V
15 5 V
15 5 V
15 5 V
15 5 V
15 5 V
16 5 V
15 5 V
15 5 V
15 5 V
15 5 V
15 5 V
16 4 V
15 5 V
15 5 V
15 4 V
15 5 V
15 4 V
15 5 V
16 4 V
15 4 V
15 5 V
15 4 V
15 4 V
15 5 V
15 4 V
16 4 V
15 4 V
15 4 V
15 4 V
15 4 V
15 4 V
16 4 V
15 4 V
15 4 V
15 4 V
15 4 V
15 3 V
15 4 V
16 4 V
15 4 V
15 3 V
15 4 V
LT3
4599 778 M
180 0 V
752 2828 M
15 -2 V
15 -2 V
15 -3 V
15 -3 V
15 -4 V
15 -5 V
16 -6 V
15 -7 V
15 -8 V
15 -9 V
15 -9 V
15 -11 V
15 -11 V
16 -12 V
15 -12 V
15 -14 V
15 -14 V
15 -14 V
15 -16 V
16 -16 V
15 -16 V
15 -17 V
15 -18 V
15 -19 V
15 -19 V
15 -19 V
16 -21 V
15 -21 V
15 -21 V
15 -23 V
15 -22 V
15 -24 V
15 -24 V
16 -25 V
15 -25 V
15 -26 V
15 -26 V
15 -27 V
15 -28 V
16 -28 V
15 -29 V
15 -29 V
15 -30 V
15 -30 V
15 -32 V
15 -31 V
16 -32 V
15 -33 V
15 -34 V
15 -34 V
15 -34 V
15 -35 V
15 -36 V
16 -36 V
15 -37 V
15 -37 V
15 -38 V
15 -39 V
15 -39 V
16 -40 V
15 -40 V
15 -41 V
15 -41 V
15 -42 V
15 -43 V
15 -43 V
16 -44 V
15 -44 V
15 -45 V
15 -45 V
15 -46 V
15 -47 V
15 -46 V
16 -47 V
15 -47 V
15 -46 V
15 -43 V
15 -40 V
15 -34 V
16 -24 V
15 -14 V
15 -4 V
15 3 V
15 8 V
15 10 V
15 12 V
16 13 V
15 13 V
15 14 V
15 15 V
15 14 V
15 15 V
15 14 V
16 15 V
15 15 V
15 15 V
15 15 V
15 15 V
15 15 V
16 15 V
15 15 V
15 15 V
15 16 V
15 15 V
15 15 V
15 15 V
16 15 V
15 15 V
15 15 V
15 15 V
15 15 V
15 15 V
15 15 V
16 14 V
15 15 V
15 15 V
15 15 V
15 14 V
15 15 V
16 14 V
15 15 V
15 14 V
15 15 V
15 14 V
15 14 V
15 15 V
16 14 V
15 14 V
15 14 V
15 14 V
15 15 V
15 14 V
15 14 V
16 14 V
15 14 V
15 14 V
15 14 V
15 15 V
15 14 V
16 14 V
15 14 V
15 15 V
15 14 V
15 14 V
15 15 V
15 14 V
16 15 V
15 14 V
15 14 V
15 15 V
15 14 V
15 15 V
15 14 V
16 14 V
15 15 V
15 14 V
15 14 V
15 14 V
15 14 V
16 14 V
15 14 V
15 14 V
15 14 V
15 13 V
15 14 V
15 13 V
16 13 V
15 14 V
15 13 V
15 13 V
15 12 V
15 13 V
15 12 V
16 13 V
15 12 V
15 12 V
15 12 V
15 12 V
15 12 V
16 11 V
15 12 V
15 11 V
15 11 V
15 11 V
15 11 V
15 11 V
16 11 V
15 10 V
15 10 V
15 11 V
15 10 V
15 10 V
15 10 V
16 9 V
15 10 V
15 10 V
15 9 V
15 9 V
15 10 V
16 9 V
15 9 V
15 9 V
15 8 V
15 9 V
15 8 V
15 9 V
16 8 V
15 8 V
15 9 V
15 8 V
15 8 V
15 7 V
15 8 V
16 8 V
15 7 V
15 8 V
15 7 V
15 8 V
15 7 V
15 7 V
16 7 V
15 7 V
15 7 V
15 7 V
15 7 V
15 6 V
16 7 V
15 7 V
15 6 V
15 6 V
15 7 V
15 6 V
15 6 V
16 6 V
15 6 V
15 6 V
15 6 V
15 6 V
15 6 V
16 6 V
15 5 V
15 6 V
15 5 V
15 6 V
15 5 V
15 6 V
16 5 V
15 5 V
15 6 V
15 5 V
15 5 V
15 5 V
15 5 V
16 5 V
15 5 V
15 5 V
15 5 V
15 4 V
15 5 V
16 5 V
15 4 V
15 5 V
15 5 V
15 4 V
15 4 V
15 5 V
16 4 V
15 5 V
15 4 V
15 4 V
15 4 V
15 5 V
15 4 V
16 4 V
15 4 V
15 4 V
15 4 V
15 4 V
15 4 V
16 4 V
15 4 V
15 3 V
15 4 V
15 4 V
15 4 V
15 3 V
16 4 V
15 4 V
15 3 V
15 4 V
LT4
4599 678 M
180 0 V
752 2828 M
15 -2 V
15 -2 V
15 -3 V
15 -3 V
15 -4 V
15 -5 V
16 -6 V
15 -7 V
15 -8 V
15 -9 V
15 -9 V
15 -11 V
15 -11 V
16 -12 V
15 -12 V
15 -14 V
15 -14 V
15 -14 V
15 -16 V
16 -16 V
15 -16 V
15 -17 V
15 -18 V
15 -19 V
15 -19 V
15 -19 V
16 -21 V
15 -21 V
15 -21 V
15 -23 V
15 -22 V
15 -24 V
15 -24 V
16 -25 V
15 -25 V
15 -26 V
15 -26 V
15 -27 V
15 -28 V
16 -28 V
15 -29 V
15 -29 V
15 -30 V
15 -30 V
15 -32 V
15 -31 V
16 -32 V
15 -33 V
15 -34 V
15 -34 V
15 -34 V
15 -35 V
15 -36 V
16 -36 V
15 -37 V
15 -37 V
15 -38 V
15 -39 V
15 -39 V
16 -40 V
15 -40 V
15 -41 V
15 -42 V
15 -42 V
15 -42 V
15 -43 V
16 -44 V
15 -45 V
15 -45 V
15 -45 V
15 -46 V
15 -47 V
15 -46 V
16 -46 V
15 -43 V
15 -38 V
15 -28 V
15 -15 V
15 -1 V
16 7 V
15 14 V
15 16 V
15 17 V
15 18 V
15 18 V
15 18 V
16 17 V
15 18 V
15 17 V
15 17 V
15 17 V
15 17 V
15 17 V
16 17 V
15 17 V
15 16 V
15 17 V
15 16 V
15 16 V
16 16 V
15 16 V
15 16 V
15 16 V
15 16 V
15 16 V
15 15 V
16 16 V
15 15 V
15 15 V
15 16 V
15 15 V
15 15 V
15 15 V
16 15 V
15 15 V
15 14 V
15 15 V
15 15 V
15 14 V
16 15 V
15 14 V
15 14 V
15 15 V
15 14 V
15 14 V
15 14 V
16 14 V
15 14 V
15 13 V
15 14 V
15 14 V
15 13 V
15 14 V
16 13 V
15 14 V
15 13 V
15 13 V
15 13 V
15 13 V
16 13 V
15 13 V
15 13 V
15 12 V
15 13 V
15 12 V
15 13 V
16 12 V
15 12 V
15 13 V
15 12 V
15 12 V
15 11 V
15 12 V
16 12 V
15 12 V
15 11 V
15 12 V
15 11 V
15 11 V
16 11 V
15 11 V
15 11 V
15 11 V
15 11 V
15 10 V
15 11 V
16 10 V
15 11 V
15 10 V
15 10 V
15 10 V
15 10 V
15 10 V
16 10 V
15 10 V
15 9 V
15 10 V
15 9 V
15 10 V
16 9 V
15 9 V
15 9 V
15 9 V
15 9 V
15 9 V
15 9 V
16 8 V
15 9 V
15 8 V
15 9 V
15 8 V
15 8 V
15 9 V
16 8 V
15 8 V
15 8 V
15 7 V
15 8 V
15 8 V
16 7 V
15 8 V
15 7 V
15 8 V
15 7 V
15 7 V
15 7 V
16 8 V
15 7 V
15 7 V
15 6 V
15 7 V
15 7 V
15 7 V
16 6 V
15 7 V
15 6 V
15 7 V
15 6 V
15 6 V
15 7 V
16 6 V
15 6 V
15 6 V
15 6 V
15 6 V
15 6 V
16 6 V
15 5 V
15 6 V
15 6 V
15 5 V
15 6 V
15 5 V
16 6 V
15 5 V
15 5 V
15 6 V
15 5 V
15 5 V
16 5 V
15 5 V
15 5 V
15 5 V
15 5 V
15 5 V
15 5 V
16 5 V
15 5 V
15 5 V
15 4 V
15 5 V
15 4 V
15 5 V
16 5 V
15 4 V
15 4 V
15 5 V
15 4 V
15 4 V
16 5 V
15 4 V
15 4 V
15 4 V
15 5 V
15 4 V
15 4 V
16 4 V
15 4 V
15 4 V
15 4 V
15 3 V
15 4 V
15 4 V
16 4 V
15 4 V
15 3 V
15 4 V
15 4 V
15 3 V
16 4 V
15 3 V
15 4 V
15 3 V
15 4 V
15 3 V
15 4 V
16 3 V
15 4 V
15 3 V
15 3 V
LT5
4599 578 M
180 0 V
752 2828 M
15 -2 V
15 -2 V
15 -3 V
15 -3 V
15 -4 V
15 -5 V
16 -6 V
15 -7 V
15 -8 V
15 -9 V
15 -9 V
15 -11 V
15 -11 V
16 -12 V
15 -12 V
15 -14 V
15 -14 V
15 -14 V
15 -16 V
16 -16 V
15 -16 V
15 -17 V
15 -18 V
15 -19 V
15 -19 V
15 -19 V
16 -21 V
15 -21 V
15 -21 V
15 -23 V
15 -22 V
15 -24 V
15 -24 V
16 -25 V
15 -25 V
15 -26 V
15 -26 V
15 -27 V
15 -28 V
16 -28 V
15 -29 V
15 -29 V
15 -30 V
15 -30 V
15 -32 V
15 -31 V
16 -32 V
15 -33 V
15 -34 V
15 -34 V
15 -34 V
15 -35 V
15 -36 V
16 -36 V
15 -37 V
15 -37 V
15 -38 V
15 -39 V
15 -39 V
16 -40 V
15 -40 V
15 -41 V
15 -41 V
15 -42 V
15 -43 V
15 -43 V
16 -44 V
15 -44 V
15 -45 V
15 -45 V
15 -46 V
15 -47 V
15 -46 V
16 -48 V
15 -47 V
15 -47 V
15 -46 V
15 -45 V
15 -41 V
16 -35 V
15 -27 V
15 -17 V
15 -7 V
15 -1 V
15 5 V
15 9 V
16 10 V
15 12 V
15 12 V
15 13 V
15 14 V
15 13 V
15 14 V
16 14 V
15 15 V
15 14 V
15 14 V
15 15 V
15 14 V
16 14 V
15 15 V
15 14 V
15 14 V
15 15 V
15 14 V
15 14 V
16 15 V
15 14 V
15 14 V
15 14 V
15 15 V
15 14 V
15 14 V
16 14 V
15 14 V
15 15 V
15 14 V
15 14 V
15 14 V
16 14 V
15 14 V
15 14 V
15 14 V
15 14 V
15 14 V
15 14 V
16 14 V
15 14 V
15 14 V
15 14 V
15 13 V
15 14 V
15 13 V
16 14 V
15 13 V
15 14 V
15 13 V
15 13 V
15 13 V
16 14 V
15 13 V
15 13 V
15 12 V
15 13 V
15 13 V
15 13 V
16 13 V
15 12 V
15 13 V
15 12 V
15 13 V
15 12 V
15 13 V
16 12 V
15 12 V
15 12 V
15 13 V
15 12 V
15 12 V
16 12 V
15 12 V
15 12 V
15 12 V
15 11 V
15 12 V
15 12 V
16 12 V
15 11 V
15 12 V
15 11 V
15 11 V
15 12 V
15 11 V
16 11 V
15 11 V
15 11 V
15 11 V
15 11 V
15 11 V
16 10 V
15 11 V
15 11 V
15 10 V
15 10 V
15 11 V
15 10 V
16 10 V
15 10 V
15 9 V
15 10 V
15 10 V
15 9 V
15 10 V
16 9 V
15 10 V
15 9 V
15 9 V
15 9 V
15 9 V
16 9 V
15 8 V
15 9 V
15 9 V
15 8 V
15 8 V
15 9 V
16 8 V
15 8 V
15 8 V
15 8 V
15 8 V
15 8 V
15 7 V
16 8 V
15 7 V
15 8 V
15 7 V
15 7 V
15 8 V
15 7 V
16 7 V
15 7 V
15 6 V
15 7 V
15 7 V
15 7 V
16 6 V
15 7 V
15 6 V
15 7 V
15 6 V
15 6 V
15 6 V
16 6 V
15 6 V
15 6 V
15 6 V
15 6 V
15 6 V
16 6 V
15 5 V
15 6 V
15 5 V
15 6 V
15 5 V
15 6 V
16 5 V
15 5 V
15 5 V
15 6 V
15 5 V
15 5 V
15 5 V
16 5 V
15 5 V
15 4 V
15 5 V
15 5 V
15 5 V
16 4 V
15 5 V
15 4 V
15 5 V
15 4 V
15 5 V
15 4 V
16 5 V
15 4 V
15 4 V
15 4 V
15 4 V
15 5 V
15 4 V
16 4 V
15 4 V
15 4 V
15 4 V
15 3 V
15 4 V
16 4 V
15 4 V
15 4 V
15 3 V
15 4 V
15 4 V
15 3 V
16 4 V
15 3 V
15 4 V
15 3 V
stroke
grestore
end
showpage
}
\put(4539,578){\makebox(0,0)[r]{6}}
\put(4539,678){\makebox(0,0)[r]{5}}
\put(4539,778){\makebox(0,0)[r]{4}}
\put(4539,878){\makebox(0,0)[r]{3}}
\put(4539,978){\makebox(0,0)[r]{2}}
\put(4539,1078){\makebox(0,0)[r]{1}}
\put(2872,0){\makebox(0,0){$\alpha \;\;\; [{\mathcal L}^{-1}]$}}
\put(240,1698){%
\special{ps: gsave currentpoint currentpoint translate
270 rotate neg exch neg exch translate}%
\makebox(0,0)[b]{\shortstack{$\Delta F \;\;\; [{\mathcal C}^2/{\mathcal L}^2]$}}%
\special{ps: currentpoint grestore moveto}%
}
\put(5145,151){\makebox(0,0){3}}
\put(4388,151){\makebox(0,0){2.5}}
\put(3630,151){\makebox(0,0){2}}
\put(2873,151){\makebox(0,0){1.5}}
\put(2115,151){\makebox(0,0){1}}
\put(1358,151){\makebox(0,0){0.5}}
\put(600,151){\makebox(0,0){0}}
\put(540,3145){\makebox(0,0)[r]{$10^{0}$}}
\put(540,2732){\makebox(0,0)[r]{$10^{-1}$}}
\put(540,2318){\makebox(0,0)[r]{$10^{-2}$}}
\put(540,1905){\makebox(0,0)[r]{$10^{-3}$}}
\put(540,1491){\makebox(0,0)[r]{$10^{-4}$}}
\put(540,1078){\makebox(0,0)[r]{$10^{-5}$}}
\put(540,664){\makebox(0,0)[r]{$10^{-6}$}}
\put(540,251){\makebox(0,0)[r]{$10^{-7}$}}
\end{picture}

%% file: f3.tex
\setlength{\unitlength}{0.1bp}
\special{!
/gnudict 40 dict def
gnudict begin
/Color false def
/Solid false def
/gnulinewidth 5.000 def
/vshift -33 def
/dl {10 mul} def
/hpt 31.5 def
/vpt 31.5 def
/M {moveto} bind def
/L {lineto} bind def
/R {rmoveto} bind def
/V {rlineto} bind def
/vpt2 vpt 2 mul def
/hpt2 hpt 2 mul def
/Lshow { currentpoint stroke M
  0 vshift R show } def
/Rshow { currentpoint stroke M
  dup stringwidth pop neg vshift R show } def
/Cshow { currentpoint stroke M
  dup stringwidth pop -2 div vshift R show } def
/DL { Color {setrgbcolor Solid {pop []} if 0 setdash }
 {pop pop pop Solid {pop []} if 0 setdash} ifelse } def
/BL { stroke gnulinewidth 2 mul setlinewidth } def
/AL { stroke gnulinewidth 2 div setlinewidth } def
/PL { stroke gnulinewidth setlinewidth } def
/LTb { BL [] 0 0 0 DL } def
/LTa { AL [1 dl 2 dl] 0 setdash 0 0 0 setrgbcolor } def
/LT0 { PL [] 0 1 0 DL } def
/LT1 { PL [4 dl 2 dl] 0 0 1 DL } def
/LT2 { PL [2 dl 3 dl] 1 0 0 DL } def
/LT3 { PL [1 dl 1.5 dl] 1 0 1 DL } def
/LT4 { PL [5 dl 2 dl 1 dl 2 dl] 0 1 1 DL } def
/LT5 { PL [4 dl 3 dl 1 dl 3 dl] 1 1 0 DL } def
/LT6 { PL [2 dl 2 dl 2 dl 4 dl] 0 0 0 DL } def
/LT7 { PL [2 dl 2 dl 2 dl 2 dl 2 dl 4 dl] 1 0.3 0 DL } def
/LT8 { PL [2 dl 2 dl 2 dl 2 dl 2 dl 2 dl 2 dl 4 dl] 0.5 0.5 0.5 DL } def
/P { stroke [] 0 setdash
  currentlinewidth 2 div sub M
  0 currentlinewidth V stroke } def
/D { stroke [] 0 setdash 2 copy vpt add M
  hpt neg vpt neg V hpt vpt neg V
  hpt vpt V hpt neg vpt V closepath stroke
  P } def
/A { stroke [] 0 setdash vpt sub M 0 vpt2 V
  currentpoint stroke M
  hpt neg vpt neg R hpt2 0 V stroke
  } def
/B { stroke [] 0 setdash 2 copy exch hpt sub exch vpt add M
  0 vpt2 neg V hpt2 0 V 0 vpt2 V
  hpt2 neg 0 V closepath stroke
  P } def
/C { stroke [] 0 setdash exch hpt sub exch vpt add M
  hpt2 vpt2 neg V currentpoint stroke M
  hpt2 neg 0 R hpt2 vpt2 V stroke } def
/T { stroke [] 0 setdash 2 copy vpt 1.12 mul add M
  hpt neg vpt -1.62 mul V
  hpt 2 mul 0 V
  hpt neg vpt 1.62 mul V closepath stroke
  P  } def
/S { 2 copy A C} def
end
}
\begin{picture}(2789,1836)(0,0)
\special{"
gnudict begin
gsave
50 50 translate
0.100 0.100 scale
0 setgray
/Helvetica findfont 100 scalefont setfont
newpath
-500.000000 -500.000000 translate
LTa
600 251 M
0 1534 V
LTb
600 251 M
63 0 V
1943 0 R
-63 0 V
600 470 M
63 0 V
1943 0 R
-63 0 V
600 689 M
63 0 V
1943 0 R
-63 0 V
600 908 M
63 0 V
1943 0 R
-63 0 V
600 1128 M
63 0 V
1943 0 R
-63 0 V
600 1347 M
63 0 V
1943 0 R
-63 0 V
600 1566 M
63 0 V
1943 0 R
-63 0 V
600 1785 M
63 0 V
1943 0 R
-63 0 V
600 251 M
0 63 V
0 1471 R
0 -63 V
934 251 M
0 63 V
0 1471 R
0 -63 V
1269 251 M
0 63 V
0 1471 R
0 -63 V
1603 251 M
0 63 V
0 1471 R
0 -63 V
1937 251 M
0 63 V
0 1471 R
0 -63 V
2272 251 M
0 63 V
0 1471 R
0 -63 V
2606 251 M
0 63 V
0 1471 R
0 -63 V
600 251 M
2006 0 V
0 1534 V
-2006 0 V
600 251 L
LT0
667 1617 M
7 -1 V
6 -1 V
7 -2 V
7 -2 V
6 -2 V
7 -2 V
7 -3 V
6 -4 V
7 -4 V
7 -5 V
6 -5 V
7 -5 V
7 -6 V
6 -7 V
7 -6 V
7 -7 V
7 -8 V
6 -7 V
7 -8 V
7 -8 V
6 -9 V
7 -8 V
7 -9 V
6 -8 V
7 -9 V
7 -7 V
6 -7 V
7 -6 V
7 -5 V
6 -3 V
7 -2 V
7 0 V
7 1 V
6 2 V
7 2 V
7 4 V
6 3 V
7 4 V
7 3 V
6 4 V
7 4 V
7 4 V
6 3 V
7 4 V
7 3 V
6 3 V
7 4 V
7 3 V
7 3 V
6 3 V
7 3 V
7 3 V
6 3 V
7 2 V
7 3 V
6 3 V
7 2 V
7 3 V
6 2 V
7 2 V
7 3 V
6 2 V
7 2 V
7 2 V
7 3 V
6 2 V
7 2 V
7 2 V
6 2 V
7 2 V
7 2 V
6 2 V
7 2 V
7 1 V
6 2 V
7 2 V
7 2 V
6 2 V
7 1 V
7 2 V
6 2 V
7 1 V
7 2 V
7 1 V
6 2 V
7 2 V
7 1 V
6 2 V
7 1 V
7 2 V
6 1 V
7 2 V
7 1 V
6 2 V
7 1 V
7 1 V
6 2 V
7 1 V
7 1 V
7 2 V
6 1 V
7 1 V
7 2 V
6 1 V
7 1 V
7 1 V
6 2 V
7 1 V
7 1 V
6 1 V
7 2 V
7 1 V
6 1 V
7 1 V
7 1 V
7 1 V
6 1 V
7 2 V
7 1 V
6 1 V
7 1 V
7 1 V
6 1 V
7 1 V
7 1 V
6 1 V
7 1 V
7 1 V
6 1 V
7 1 V
7 1 V
7 1 V
6 1 V
7 1 V
7 1 V
6 1 V
7 1 V
7 1 V
6 1 V
7 0 V
7 1 V
6 1 V
7 1 V
7 1 V
6 1 V
7 1 V
7 1 V
6 0 V
7 1 V
7 1 V
7 1 V
6 1 V
7 1 V
7 0 V
6 1 V
7 1 V
7 1 V
6 1 V
7 0 V
7 1 V
6 1 V
7 1 V
7 0 V
6 1 V
7 1 V
7 1 V
7 0 V
6 1 V
7 1 V
7 1 V
6 0 V
7 1 V
7 1 V
6 0 V
7 1 V
7 1 V
6 0 V
7 1 V
7 1 V
6 0 V
7 1 V
7 1 V
7 0 V
6 1 V
7 1 V
7 0 V
6 1 V
7 1 V
7 0 V
6 1 V
7 0 V
7 1 V
6 1 V
7 0 V
7 1 V
6 1 V
7 0 V
7 1 V
7 0 V
6 1 V
7 0 V
7 1 V
6 1 V
7 0 V
7 1 V
6 0 V
7 1 V
7 1 V
6 0 V
7 1 V
7 0 V
6 1 V
7 0 V
7 1 V
7 0 V
6 1 V
7 1 V
7 0 V
6 1 V
7 0 V
7 1 V
6 0 V
7 1 V
7 0 V
6 1 V
7 0 V
7 1 V
6 0 V
7 1 V
7 0 V
6 1 V
7 0 V
7 1 V
7 0 V
6 1 V
7 1 V
7 0 V
6 1 V
7 0 V
7 1 V
6 0 V
7 1 V
7 0 V
6 1 V
7 0 V
7 1 V
6 0 V
7 1 V
7 0 V
7 1 V
6 0 V
7 0 V
7 1 V
6 0 V
7 1 V
7 0 V
6 1 V
7 0 V
7 1 V
6 0 V
7 1 V
7 0 V
6 1 V
7 0 V
7 1 V
7 0 V
6 1 V
7 0 V
7 1 V
6 0 V
7 1 V
7 0 V
6 0 V
7 1 V
7 0 V
6 1 V
7 0 V
7 1 V
6 0 V
7 1 V
7 0 V
7 1 V
6 0 V
7 0 V
7 1 V
6 0 V
7 1 V
7 0 V
6 1 V
7 0 V
LT1
667 1617 M
7 -1 V
6 -1 V
7 -2 V
7 -2 V
6 -2 V
7 -2 V
7 -3 V
6 -4 V
7 -4 V
7 -5 V
6 -5 V
7 -5 V
7 -6 V
6 -7 V
7 -6 V
7 -8 V
7 -7 V
6 -8 V
7 -8 V
7 -8 V
6 -9 V
7 -9 V
7 -10 V
6 -10 V
7 -10 V
7 -10 V
6 -11 V
7 -11 V
7 -12 V
6 -12 V
7 -12 V
7 -12 V
7 -13 V
6 -13 V
7 -14 V
7 -13 V
6 -15 V
7 -14 V
7 -14 V
6 -15 V
7 -15 V
7 -14 V
6 -14 V
7 -14 V
7 -12 V
6 -10 V
7 -9 V
7 -6 V
7 -3 V
6 -2 V
7 1 V
7 1 V
6 3 V
7 2 V
7 4 V
6 3 V
7 4 V
7 4 V
6 3 V
7 4 V
7 4 V
6 3 V
7 4 V
7 3 V
7 4 V
6 4 V
7 3 V
7 4 V
6 3 V
7 4 V
7 4 V
6 3 V
7 4 V
7 3 V
6 4 V
7 3 V
7 4 V
6 3 V
7 3 V
7 4 V
6 3 V
7 4 V
7 3 V
7 3 V
6 4 V
7 3 V
7 3 V
6 3 V
7 4 V
7 3 V
6 3 V
7 3 V
7 3 V
6 3 V
7 3 V
7 3 V
6 3 V
7 3 V
7 3 V
7 3 V
6 2 V
7 3 V
7 3 V
6 3 V
7 2 V
7 3 V
6 3 V
7 2 V
7 3 V
6 2 V
7 3 V
7 2 V
6 3 V
7 2 V
7 2 V
7 3 V
6 2 V
7 2 V
7 3 V
6 2 V
7 2 V
7 2 V
6 2 V
7 2 V
7 2 V
6 2 V
7 2 V
7 2 V
6 2 V
7 2 V
7 2 V
7 2 V
6 2 V
7 2 V
7 2 V
6 1 V
7 2 V
7 2 V
6 2 V
7 1 V
7 2 V
6 2 V
7 1 V
7 2 V
6 1 V
7 2 V
7 2 V
6 1 V
7 2 V
7 1 V
7 2 V
6 1 V
7 2 V
7 1 V
6 2 V
7 1 V
7 1 V
6 2 V
7 1 V
7 1 V
6 2 V
7 1 V
7 1 V
6 2 V
7 1 V
7 1 V
7 2 V
6 1 V
7 1 V
7 1 V
6 2 V
7 1 V
7 1 V
6 1 V
7 1 V
7 1 V
6 2 V
7 1 V
7 1 V
6 1 V
7 1 V
7 1 V
7 1 V
6 2 V
7 1 V
7 1 V
6 1 V
7 1 V
7 1 V
6 1 V
7 1 V
7 1 V
6 1 V
7 1 V
7 1 V
6 1 V
7 1 V
7 1 V
7 1 V
6 1 V
7 1 V
7 1 V
6 1 V
7 1 V
7 1 V
6 1 V
7 1 V
7 1 V
6 1 V
7 1 V
7 1 V
6 1 V
7 1 V
7 1 V
7 1 V
6 1 V
7 1 V
7 1 V
6 1 V
7 1 V
7 1 V
6 0 V
7 1 V
7 1 V
6 1 V
7 1 V
7 1 V
6 1 V
7 1 V
7 1 V
6 1 V
7 1 V
7 1 V
7 1 V
6 0 V
7 1 V
7 1 V
6 1 V
7 1 V
7 1 V
6 1 V
7 1 V
7 1 V
6 1 V
7 1 V
7 0 V
6 1 V
7 1 V
7 1 V
7 1 V
6 1 V
7 1 V
7 1 V
6 1 V
7 0 V
7 1 V
6 1 V
7 1 V
7 1 V
6 1 V
7 1 V
7 1 V
6 0 V
7 1 V
7 1 V
7 1 V
6 1 V
7 1 V
7 1 V
6 1 V
7 0 V
7 1 V
6 1 V
7 1 V
7 1 V
6 1 V
7 1 V
7 0 V
6 1 V
7 1 V
7 1 V
7 1 V
6 1 V
7 0 V
7 1 V
6 1 V
7 1 V
7 1 V
6 0 V
7 1 V
LT2
667 1617 M
7 -1 V
6 -1 V
7 -2 V
7 -2 V
6 -2 V
7 -2 V
7 -3 V
6 -4 V
7 -4 V
7 -5 V
6 -5 V
7 -5 V
7 -6 V
6 -7 V
7 -6 V
7 -7 V
7 -8 V
6 -8 V
7 -8 V
7 -8 V
6 -9 V
7 -9 V
7 -10 V
6 -9 V
7 -10 V
7 -11 V
6 -11 V
7 -11 V
7 -11 V
6 -12 V
7 -12 V
7 -13 V
7 -12 V
6 -13 V
7 -14 V
7 -13 V
6 -14 V
7 -15 V
7 -14 V
6 -15 V
7 -16 V
7 -15 V
6 -16 V
7 -16 V
7 -16 V
6 -17 V
7 -17 V
7 -17 V
7 -18 V
6 -17 V
7 -18 V
7 -18 V
6 -17 V
7 -16 V
7 -16 V
6 -14 V
7 -11 V
7 -8 V
6 -5 V
7 -3 V
7 0 V
6 2 V
7 3 V
7 3 V
7 4 V
6 4 V
7 4 V
7 4 V
6 5 V
7 4 V
7 4 V
6 4 V
7 4 V
7 4 V
6 4 V
7 4 V
7 4 V
6 4 V
7 4 V
7 4 V
6 4 V
7 3 V
7 4 V
7 4 V
6 3 V
7 4 V
7 4 V
6 3 V
7 4 V
7 3 V
6 4 V
7 3 V
7 4 V
6 3 V
7 4 V
7 3 V
6 3 V
7 4 V
7 3 V
7 3 V
6 3 V
7 4 V
7 3 V
6 3 V
7 3 V
7 3 V
6 3 V
7 4 V
7 3 V
6 3 V
7 3 V
7 3 V
6 3 V
7 3 V
7 3 V
7 3 V
6 3 V
7 3 V
7 3 V
6 3 V
7 3 V
7 2 V
6 3 V
7 3 V
7 3 V
6 3 V
7 3 V
7 2 V
6 3 V
7 3 V
7 3 V
7 2 V
6 3 V
7 3 V
7 2 V
6 3 V
7 3 V
7 2 V
6 3 V
7 2 V
7 3 V
6 3 V
7 2 V
7 3 V
6 2 V
7 3 V
7 2 V
6 3 V
7 2 V
7 3 V
7 2 V
6 2 V
7 3 V
7 2 V
6 3 V
7 2 V
7 2 V
6 3 V
7 2 V
7 3 V
6 2 V
7 2 V
7 2 V
6 3 V
7 2 V
7 2 V
7 3 V
6 2 V
7 2 V
7 2 V
6 3 V
7 2 V
7 2 V
6 2 V
7 3 V
7 2 V
6 2 V
7 2 V
7 2 V
6 3 V
7 2 V
7 2 V
7 2 V
6 2 V
7 2 V
7 2 V
6 3 V
7 2 V
7 2 V
6 2 V
7 2 V
7 2 V
6 2 V
7 2 V
7 2 V
6 3 V
7 2 V
7 2 V
7 2 V
6 2 V
7 2 V
7 2 V
6 2 V
7 2 V
7 2 V
6 2 V
7 2 V
7 2 V
6 2 V
7 2 V
7 2 V
6 2 V
7 2 V
7 2 V
7 2 V
6 2 V
7 1 V
7 2 V
6 2 V
7 2 V
7 2 V
6 2 V
7 2 V
7 2 V
6 2 V
7 1 V
7 2 V
6 2 V
7 2 V
7 2 V
6 2 V
7 1 V
7 2 V
7 2 V
6 2 V
7 1 V
7 2 V
6 2 V
7 2 V
7 1 V
6 2 V
7 2 V
7 1 V
6 2 V
7 2 V
7 1 V
6 2 V
7 2 V
7 1 V
7 2 V
6 2 V
7 1 V
7 2 V
6 1 V
7 2 V
7 2 V
6 1 V
7 2 V
7 1 V
6 2 V
7 1 V
7 2 V
6 1 V
7 2 V
7 1 V
7 2 V
6 1 V
7 2 V
7 1 V
6 2 V
7 1 V
7 1 V
6 2 V
7 1 V
7 2 V
6 1 V
7 1 V
7 2 V
6 1 V
7 1 V
7 2 V
7 1 V
6 1 V
7 2 V
7 1 V
6 1 V
7 1 V
7 2 V
6 1 V
7 1 V
LT3
667 1617 M
7 -1 V
6 -1 V
7 -2 V
7 -2 V
6 -2 V
7 -2 V
7 -3 V
6 -4 V
7 -4 V
7 -5 V
6 -5 V
7 -5 V
7 -6 V
6 -7 V
7 -6 V
7 -7 V
7 -8 V
6 -8 V
7 -8 V
7 -8 V
6 -9 V
7 -9 V
7 -10 V
6 -9 V
7 -10 V
7 -11 V
6 -11 V
7 -11 V
7 -11 V
6 -12 V
7 -12 V
7 -13 V
7 -12 V
6 -13 V
7 -14 V
7 -13 V
6 -14 V
7 -15 V
7 -14 V
6 -15 V
7 -16 V
7 -15 V
6 -16 V
7 -16 V
7 -17 V
6 -16 V
7 -18 V
7 -17 V
7 -18 V
6 -18 V
7 -18 V
7 -19 V
6 -18 V
7 -20 V
7 -19 V
6 -20 V
7 -20 V
7 -20 V
6 -21 V
7 -20 V
7 -21 V
6 -20 V
7 -20 V
7 -18 V
7 -15 V
6 -12 V
7 -7 V
7 -3 V
6 0 V
7 4 V
7 4 V
6 6 V
7 6 V
7 6 V
6 6 V
7 6 V
7 6 V
6 6 V
7 6 V
7 6 V
6 6 V
7 6 V
7 5 V
7 6 V
6 5 V
7 6 V
7 5 V
6 5 V
7 5 V
7 6 V
6 5 V
7 5 V
7 4 V
6 5 V
7 5 V
7 5 V
6 4 V
7 5 V
7 4 V
7 5 V
6 4 V
7 5 V
7 4 V
6 4 V
7 4 V
7 4 V
6 4 V
7 4 V
7 4 V
6 4 V
7 4 V
7 4 V
6 4 V
7 4 V
7 3 V
7 4 V
6 4 V
7 3 V
7 4 V
6 3 V
7 4 V
7 3 V
6 4 V
7 3 V
7 4 V
6 3 V
7 3 V
7 4 V
6 3 V
7 3 V
7 3 V
7 4 V
6 3 V
7 3 V
7 3 V
6 3 V
7 4 V
7 3 V
6 3 V
7 3 V
7 3 V
6 3 V
7 4 V
7 3 V
6 3 V
7 3 V
7 3 V
6 3 V
7 3 V
7 4 V
7 3 V
6 3 V
7 3 V
7 3 V
6 3 V
7 3 V
7 3 V
6 4 V
7 3 V
7 3 V
6 3 V
7 3 V
7 3 V
6 3 V
7 3 V
7 4 V
7 3 V
6 3 V
7 3 V
7 3 V
6 3 V
7 3 V
7 3 V
6 3 V
7 3 V
7 4 V
6 3 V
7 3 V
7 3 V
6 3 V
7 3 V
7 3 V
7 3 V
6 3 V
7 3 V
7 3 V
6 3 V
7 3 V
7 3 V
6 3 V
7 3 V
7 3 V
6 3 V
7 3 V
7 3 V
6 3 V
7 3 V
7 3 V
7 3 V
6 3 V
7 3 V
7 2 V
6 3 V
7 3 V
7 3 V
6 3 V
7 3 V
7 3 V
6 2 V
7 3 V
7 3 V
6 3 V
7 2 V
7 3 V
7 3 V
6 2 V
7 3 V
7 3 V
6 2 V
7 3 V
7 3 V
6 2 V
7 3 V
7 2 V
6 3 V
7 3 V
7 2 V
6 3 V
7 2 V
7 3 V
6 2 V
7 2 V
7 3 V
7 2 V
6 3 V
7 2 V
7 2 V
6 3 V
7 2 V
7 2 V
6 3 V
7 2 V
7 2 V
6 2 V
7 3 V
7 2 V
6 2 V
7 2 V
7 2 V
7 2 V
6 3 V
7 2 V
7 2 V
6 2 V
7 2 V
7 2 V
6 2 V
7 2 V
7 2 V
6 2 V
7 2 V
7 2 V
6 2 V
7 2 V
7 1 V
7 2 V
6 2 V
7 2 V
7 2 V
6 2 V
7 1 V
7 2 V
6 2 V
7 2 V
7 2 V
6 1 V
7 2 V
7 2 V
6 1 V
7 2 V
7 2 V
7 1 V
6 2 V
7 2 V
7 1 V
6 2 V
7 1 V
7 2 V
6 1 V
7 2 V
LT4
667 1617 M
7 -1 V
6 -1 V
7 -2 V
7 -2 V
6 -2 V
7 -2 V
7 -3 V
6 -4 V
7 -4 V
7 -5 V
6 -5 V
7 -5 V
7 -6 V
6 -7 V
7 -6 V
7 -7 V
7 -8 V
6 -8 V
7 -8 V
7 -8 V
6 -9 V
7 -9 V
7 -10 V
6 -9 V
7 -10 V
7 -11 V
6 -11 V
7 -11 V
7 -11 V
6 -12 V
7 -12 V
7 -13 V
7 -12 V
6 -13 V
7 -14 V
7 -13 V
6 -14 V
7 -15 V
7 -14 V
6 -15 V
7 -16 V
7 -15 V
6 -16 V
7 -16 V
7 -17 V
6 -16 V
7 -18 V
7 -17 V
7 -18 V
6 -18 V
7 -18 V
7 -19 V
6 -18 V
7 -20 V
7 -19 V
6 -20 V
7 -20 V
7 -21 V
6 -20 V
7 -21 V
7 -22 V
6 -21 V
7 -22 V
7 -22 V
7 -23 V
6 -22 V
7 -23 V
7 -22 V
6 -22 V
7 -20 V
7 -18 V
6 -13 V
7 -9 V
7 -4 V
6 0 V
7 3 V
7 5 V
6 5 V
7 7 V
7 6 V
6 6 V
7 7 V
7 6 V
7 7 V
6 6 V
7 6 V
7 7 V
6 6 V
7 6 V
7 6 V
6 6 V
7 6 V
7 6 V
6 6 V
7 6 V
7 6 V
6 5 V
7 6 V
7 6 V
7 5 V
6 6 V
7 5 V
7 6 V
6 5 V
7 6 V
7 5 V
6 5 V
7 5 V
7 6 V
6 5 V
7 5 V
7 5 V
6 5 V
7 5 V
7 5 V
7 5 V
6 5 V
7 5 V
7 4 V
6 5 V
7 5 V
7 5 V
6 4 V
7 5 V
7 5 V
6 5 V
7 4 V
7 5 V
6 4 V
7 5 V
7 5 V
7 4 V
6 5 V
7 4 V
7 5 V
6 5 V
7 4 V
7 5 V
6 4 V
7 5 V
7 4 V
6 5 V
7 4 V
7 5 V
6 4 V
7 5 V
7 4 V
6 4 V
7 5 V
7 4 V
7 5 V
6 4 V
7 5 V
7 4 V
6 4 V
7 5 V
7 4 V
6 4 V
7 5 V
7 4 V
6 4 V
7 5 V
7 4 V
6 4 V
7 4 V
7 5 V
7 4 V
6 4 V
7 4 V
7 4 V
6 5 V
7 4 V
7 4 V
6 4 V
7 4 V
7 4 V
6 4 V
7 4 V
7 4 V
6 4 V
7 4 V
7 4 V
7 4 V
6 4 V
7 4 V
7 4 V
6 4 V
7 4 V
7 3 V
6 4 V
7 4 V
7 4 V
6 4 V
7 3 V
7 4 V
6 4 V
7 3 V
7 4 V
7 4 V
6 3 V
7 4 V
7 4 V
6 3 V
7 4 V
7 3 V
6 4 V
7 3 V
7 3 V
6 4 V
7 3 V
7 4 V
6 3 V
7 3 V
7 4 V
7 3 V
6 3 V
7 3 V
7 3 V
6 4 V
7 3 V
7 3 V
6 3 V
7 3 V
7 3 V
6 3 V
7 3 V
7 3 V
6 3 V
7 3 V
7 3 V
6 2 V
7 3 V
7 3 V
7 3 V
6 3 V
7 2 V
7 3 V
6 3 V
7 2 V
7 3 V
6 3 V
7 2 V
7 3 V
6 2 V
7 3 V
7 2 V
6 3 V
7 2 V
7 3 V
7 2 V
6 3 V
7 2 V
7 2 V
6 3 V
7 2 V
7 2 V
6 2 V
7 3 V
7 2 V
6 2 V
7 2 V
7 2 V
6 3 V
7 2 V
7 2 V
7 2 V
6 2 V
7 2 V
7 2 V
6 2 V
7 2 V
7 2 V
6 2 V
7 2 V
7 2 V
6 2 V
7 2 V
7 1 V
6 2 V
7 2 V
7 2 V
7 2 V
6 1 V
7 2 V
7 2 V
6 2 V
7 1 V
7 2 V
6 2 V
7 1 V
LT5
667 1617 M
7 -1 V
6 -1 V
7 -2 V
7 -2 V
6 -2 V
7 -2 V
7 -3 V
6 -4 V
7 -4 V
7 -5 V
6 -5 V
7 -5 V
7 -6 V
6 -7 V
7 -6 V
7 -7 V
7 -8 V
6 -8 V
7 -8 V
7 -8 V
6 -9 V
7 -9 V
7 -10 V
6 -9 V
7 -10 V
7 -11 V
6 -11 V
7 -11 V
7 -11 V
6 -12 V
7 -12 V
7 -13 V
7 -12 V
6 -13 V
7 -14 V
7 -13 V
6 -14 V
7 -15 V
7 -14 V
6 -15 V
7 -16 V
7 -15 V
6 -16 V
7 -16 V
7 -17 V
6 -16 V
7 -18 V
7 -17 V
7 -18 V
6 -18 V
7 -18 V
7 -19 V
6 -18 V
7 -20 V
7 -19 V
6 -20 V
7 -20 V
7 -21 V
6 -20 V
7 -21 V
7 -22 V
6 -21 V
7 -22 V
7 -23 V
7 -22 V
6 -23 V
7 -23 V
7 -24 V
6 -23 V
7 -24 V
7 -24 V
6 -24 V
7 -24 V
7 -23 V
6 -22 V
7 -19 V
7 -14 V
6 -9 V
7 -3 V
7 1 V
6 4 V
7 5 V
7 7 V
7 7 V
6 7 V
7 7 V
7 7 V
6 8 V
7 7 V
7 7 V
6 7 V
7 7 V
7 6 V
6 7 V
7 7 V
7 7 V
6 6 V
7 7 V
7 6 V
7 7 V
6 6 V
7 6 V
7 7 V
6 6 V
7 6 V
7 6 V
6 7 V
7 6 V
7 6 V
6 6 V
7 6 V
7 6 V
6 6 V
7 6 V
7 6 V
7 6 V
6 6 V
7 6 V
7 6 V
6 6 V
7 5 V
7 6 V
6 6 V
7 6 V
7 6 V
6 6 V
7 6 V
7 6 V
6 6 V
7 6 V
7 6 V
7 6 V
6 5 V
7 6 V
7 6 V
6 6 V
7 6 V
7 6 V
6 6 V
7 5 V
7 6 V
6 6 V
7 6 V
7 5 V
6 6 V
7 6 V
7 6 V
6 5 V
7 6 V
7 5 V
7 6 V
6 6 V
7 5 V
7 6 V
6 5 V
7 6 V
7 5 V
6 5 V
7 6 V
7 5 V
6 6 V
7 5 V
7 5 V
6 5 V
7 6 V
7 5 V
7 5 V
6 5 V
7 6 V
7 5 V
6 5 V
7 5 V
7 5 V
6 5 V
7 5 V
7 5 V
6 5 V
7 5 V
7 5 V
6 5 V
7 5 V
7 5 V
7 5 V
6 4 V
7 5 V
7 5 V
6 4 V
7 5 V
7 5 V
6 4 V
7 5 V
7 4 V
6 5 V
7 4 V
7 5 V
6 4 V
7 5 V
7 4 V
7 4 V
6 4 V
7 5 V
7 4 V
6 4 V
7 4 V
7 4 V
6 4 V
7 4 V
7 4 V
6 4 V
7 4 V
7 4 V
6 4 V
7 3 V
7 4 V
7 4 V
6 3 V
7 4 V
7 4 V
6 3 V
7 4 V
7 3 V
6 4 V
7 3 V
7 4 V
6 3 V
7 3 V
7 4 V
6 3 V
7 3 V
7 3 V
6 3 V
7 4 V
7 3 V
7 3 V
6 3 V
7 3 V
7 3 V
6 3 V
7 3 V
7 2 V
6 3 V
7 3 V
7 3 V
6 3 V
7 2 V
7 3 V
6 3 V
7 3 V
7 2 V
7 3 V
6 2 V
7 3 V
7 2 V
6 3 V
7 2 V
7 3 V
6 2 V
7 3 V
7 2 V
6 2 V
7 3 V
7 2 V
6 2 V
7 3 V
7 2 V
7 2 V
6 2 V
7 3 V
7 2 V
6 2 V
7 2 V
7 2 V
6 2 V
7 2 V
7 2 V
6 2 V
7 2 V
7 2 V
6 2 V
7 2 V
7 2 V
7 2 V
6 2 V
7 2 V
7 2 V
6 1 V
7 2 V
7 2 V
6 2 V
7 2 V
LT6
667 1617 M
7 -1 V
6 -1 V
7 -2 V
7 -2 V
6 -2 V
7 -2 V
7 -3 V
6 -4 V
7 -4 V
7 -5 V
6 -5 V
7 -5 V
7 -6 V
6 -7 V
7 -6 V
7 -7 V
7 -8 V
6 -8 V
7 -8 V
7 -8 V
6 -9 V
7 -9 V
7 -10 V
6 -9 V
7 -10 V
7 -11 V
6 -11 V
7 -11 V
7 -11 V
6 -12 V
7 -12 V
7 -13 V
7 -12 V
6 -13 V
7 -14 V
7 -13 V
6 -14 V
7 -15 V
7 -14 V
6 -15 V
7 -16 V
7 -15 V
6 -16 V
7 -16 V
7 -17 V
6 -16 V
7 -18 V
7 -17 V
7 -18 V
6 -18 V
7 -18 V
7 -19 V
6 -18 V
7 -20 V
7 -19 V
6 -20 V
7 -20 V
7 -21 V
6 -20 V
7 -21 V
7 -22 V
6 -21 V
7 -22 V
7 -23 V
7 -22 V
6 -23 V
7 -23 V
7 -24 V
6 -23 V
7 -25 V
7 -24 V
6 -24 V
7 -25 V
7 -25 V
6 -25 V
7 -25 V
7 -25 V
6 -23 V
7 -22 V
7 -19 V
6 -14 V
7 -9 V
7 -4 V
7 0 V
6 3 V
7 4 V
7 5 V
6 7 V
7 6 V
7 7 V
6 7 V
7 7 V
7 8 V
6 7 V
7 8 V
7 7 V
6 8 V
7 8 V
7 7 V
7 8 V
6 7 V
7 8 V
7 8 V
6 7 V
7 8 V
7 7 V
6 8 V
7 8 V
7 7 V
6 8 V
7 7 V
7 8 V
6 7 V
7 8 V
7 7 V
7 8 V
6 7 V
7 8 V
7 7 V
6 8 V
7 7 V
7 8 V
6 7 V
7 8 V
7 7 V
6 7 V
7 8 V
7 7 V
6 7 V
7 8 V
7 7 V
7 7 V
6 7 V
7 8 V
7 7 V
6 7 V
7 7 V
7 7 V
6 7 V
7 7 V
7 7 V
6 7 V
7 6 V
7 7 V
6 7 V
7 7 V
7 6 V
6 7 V
7 7 V
7 6 V
7 7 V
6 7 V
7 6 V
7 7 V
6 6 V
7 7 V
7 6 V
6 7 V
7 6 V
7 6 V
6 7 V
7 6 V
7 6 V
6 7 V
7 6 V
7 6 V
7 6 V
6 6 V
7 7 V
7 6 V
6 6 V
7 6 V
7 6 V
6 5 V
7 6 V
7 6 V
6 6 V
7 6 V
7 5 V
6 6 V
7 6 V
7 5 V
7 6 V
6 5 V
7 6 V
7 5 V
6 5 V
7 6 V
7 5 V
6 5 V
7 5 V
7 5 V
6 5 V
7 5 V
7 5 V
6 5 V
7 5 V
7 5 V
7 4 V
6 5 V
7 5 V
7 4 V
6 5 V
7 4 V
7 5 V
6 4 V
7 4 V
7 5 V
6 4 V
7 4 V
7 4 V
6 4 V
7 5 V
7 4 V
7 4 V
6 3 V
7 4 V
7 4 V
6 4 V
7 4 V
7 3 V
6 4 V
7 4 V
7 3 V
6 4 V
7 4 V
7 3 V
6 3 V
7 4 V
7 3 V
6 4 V
7 3 V
7 3 V
7 3 V
6 4 V
7 3 V
7 3 V
6 3 V
7 3 V
7 3 V
6 3 V
7 3 V
7 3 V
6 3 V
7 3 V
7 3 V
6 2 V
7 3 V
7 3 V
7 3 V
6 2 V
7 3 V
7 3 V
6 2 V
7 3 V
7 2 V
6 3 V
7 3 V
7 2 V
6 2 V
7 3 V
7 2 V
6 3 V
7 2 V
7 2 V
7 3 V
6 2 V
7 2 V
7 3 V
6 2 V
7 2 V
7 2 V
6 2 V
7 3 V
7 2 V
6 2 V
7 2 V
7 2 V
6 2 V
7 2 V
7 2 V
7 2 V
6 2 V
7 2 V
7 2 V
6 2 V
7 2 V
7 1 V
6 2 V
7 2 V
stroke
grestore
end
showpage
}
\put(1603,0){\makebox(0,0){$\alpha \;\;\; [{\mathcal L}^{-1}]$}}
\put(240,1018){%
\special{ps: gsave currentpoint currentpoint translate
270 rotate neg exch neg exch translate}%
\makebox(0,0)[b]{\shortstack{$\Delta F \;\;\; [{\mathcal C}^2/{\mathcal L}^2]$}}%
\special{ps: currentpoint grestore moveto}%
}
\put(2606,151){\makebox(0,0){3}}
\put(2272,151){\makebox(0,0){2.5}}
\put(1937,151){\makebox(0,0){2}}
\put(1603,151){\makebox(0,0){1.5}}
\put(1269,151){\makebox(0,0){1}}
\put(934,151){\makebox(0,0){0.5}}
\put(600,151){\makebox(0,0){0}}
\put(540,1785){\makebox(0,0)[r]{$10^{0}$}}
\put(540,1566){\makebox(0,0)[r]{$10^{-1}$}}
\put(540,1347){\makebox(0,0)[r]{$10^{-2}$}}
\put(540,1128){\makebox(0,0)[r]{$10^{-3}$}}
\put(540,908){\makebox(0,0)[r]{$10^{-4}$}}
\put(540,689){\makebox(0,0)[r]{$10^{-5}$}}
\put(540,470){\makebox(0,0)[r]{$10^{-6}$}}
\put(540,251){\makebox(0,0)[r]{$10^{-7}$}}
\end{picture}

%% file: f4.tex
\setlength{\unitlength}{0.1bp}
\special{!
/gnudict 40 dict def
gnudict begin
/Color false def
/Solid false def
/gnulinewidth 5.000 def
/vshift -33 def
/dl {10 mul} def
/hpt 31.5 def
/vpt 31.5 def
/M {moveto} bind def
/L {lineto} bind def
/R {rmoveto} bind def
/V {rlineto} bind def
/vpt2 vpt 2 mul def
/hpt2 hpt 2 mul def
/Lshow { currentpoint stroke M
  0 vshift R show } def
/Rshow { currentpoint stroke M
  dup stringwidth pop neg vshift R show } def
/Cshow { currentpoint stroke M
  dup stringwidth pop -2 div vshift R show } def
/DL { Color {setrgbcolor Solid {pop []} if 0 setdash }
 {pop pop pop Solid {pop []} if 0 setdash} ifelse } def
/BL { stroke gnulinewidth 2 mul setlinewidth } def
/AL { stroke gnulinewidth 2 div setlinewidth } def
/PL { stroke gnulinewidth setlinewidth } def
/LTb { BL [] 0 0 0 DL } def
/LTa { AL [1 dl 2 dl] 0 setdash 0 0 0 setrgbcolor } def
/LT0 { PL [] 0 1 0 DL } def
/LT1 { PL [4 dl 2 dl] 0 0 1 DL } def
/LT2 { PL [2 dl 3 dl] 1 0 0 DL } def
/LT3 { PL [1 dl 1.5 dl] 1 0 1 DL } def
/LT4 { PL [5 dl 2 dl 1 dl 2 dl] 0 1 1 DL } def
/LT5 { PL [4 dl 3 dl 1 dl 3 dl] 1 1 0 DL } def
/LT6 { PL [2 dl 2 dl 2 dl 4 dl] 0 0 0 DL } def
/LT7 { PL [2 dl 2 dl 2 dl 2 dl 2 dl 4 dl] 1 0.3 0 DL } def
/LT8 { PL [2 dl 2 dl 2 dl 2 dl 2 dl 2 dl 2 dl 4 dl] 0.5 0.5 0.5 DL } def
/P { stroke [] 0 setdash
  currentlinewidth 2 div sub M
  0 currentlinewidth V stroke } def
/D { stroke [] 0 setdash 2 copy vpt add M
  hpt neg vpt neg V hpt vpt neg V
  hpt vpt V hpt neg vpt V closepath stroke
  P } def
/A { stroke [] 0 setdash vpt sub M 0 vpt2 V
  currentpoint stroke M
  hpt neg vpt neg R hpt2 0 V stroke
  } def
/B { stroke [] 0 setdash 2 copy exch hpt sub exch vpt add M
  0 vpt2 neg V hpt2 0 V 0 vpt2 V
  hpt2 neg 0 V closepath stroke
  P } def
/C { stroke [] 0 setdash exch hpt sub exch vpt add M
  hpt2 vpt2 neg V currentpoint stroke M
  hpt2 neg 0 R hpt2 vpt2 V stroke } def
/T { stroke [] 0 setdash 2 copy vpt 1.12 mul add M
  hpt neg vpt -1.62 mul V
  hpt 2 mul 0 V
  hpt neg vpt 1.62 mul V closepath stroke
  P  } def
/S { 2 copy A C} def
end
}
\begin{picture}(2789,1836)(0,0)
\special{"
gnudict begin
gsave
50 50 translate
0.100 0.100 scale
0 setgray
/Helvetica findfont 100 scalefont setfont
newpath
-500.000000 -500.000000 translate
LTa
600 251 M
0 1534 V
LTb
600 385 M
63 0 V
1943 0 R
-63 0 V
600 640 M
63 0 V
1943 0 R
-63 0 V
600 896 M
63 0 V
1943 0 R
-63 0 V
600 1152 M
63 0 V
1943 0 R
-63 0 V
600 1407 M
63 0 V
1943 0 R
-63 0 V
600 1663 M
63 0 V
1943 0 R
-63 0 V
600 251 M
0 63 V
0 1471 R
0 -63 V
934 251 M
0 63 V
0 1471 R
0 -63 V
1269 251 M
0 63 V
0 1471 R
0 -63 V
1603 251 M
0 63 V
0 1471 R
0 -63 V
1937 251 M
0 63 V
0 1471 R
0 -63 V
2272 251 M
0 63 V
0 1471 R
0 -63 V
2606 251 M
0 63 V
0 1471 R
0 -63 V
600 251 M
2006 0 V
0 1534 V
-2006 0 V
600 251 L
LT0
667 1467 M
7 -1 V
6 -2 V
7 -1 V
7 -2 V
6 -3 V
7 -3 V
7 -4 V
6 -4 V
7 -4 V
7 -6 V
6 -5 V
7 -7 V
7 -6 V
6 -8 V
7 -7 V
7 -8 V
7 -8 V
6 -9 V
7 -9 V
7 -10 V
6 -9 V
7 -10 V
7 -11 V
6 -10 V
7 -9 V
7 -9 V
6 -7 V
7 -5 V
7 -1 V
6 2 V
7 5 V
7 8 V
7 9 V
6 10 V
7 10 V
7 10 V
6 10 V
7 10 V
7 10 V
6 8 V
7 9 V
7 8 V
6 8 V
7 7 V
7 7 V
6 7 V
7 7 V
7 6 V
7 6 V
6 5 V
7 6 V
7 5 V
6 5 V
7 5 V
7 5 V
6 4 V
7 5 V
7 4 V
6 4 V
7 4 V
7 4 V
6 3 V
7 4 V
7 3 V
7 4 V
6 3 V
7 3 V
7 3 V
6 3 V
7 3 V
7 3 V
6 3 V
7 2 V
7 3 V
6 3 V
7 2 V
7 2 V
6 3 V
7 2 V
7 2 V
6 3 V
7 2 V
7 2 V
7 2 V
6 2 V
7 2 V
7 2 V
6 2 V
7 2 V
7 1 V
6 2 V
7 2 V
7 2 V
6 1 V
7 2 V
7 2 V
6 1 V
7 2 V
7 1 V
7 2 V
6 1 V
7 2 V
7 1 V
6 2 V
7 1 V
7 1 V
6 2 V
7 1 V
7 1 V
6 2 V
7 1 V
7 1 V
6 1 V
7 2 V
7 1 V
7 1 V
6 1 V
7 1 V
7 1 V
6 1 V
7 2 V
7 1 V
6 1 V
7 1 V
7 1 V
6 1 V
7 1 V
7 1 V
6 1 V
7 1 V
7 1 V
7 1 V
6 1 V
7 0 V
7 1 V
6 1 V
7 1 V
7 1 V
6 1 V
7 1 V
7 1 V
6 0 V
7 1 V
7 1 V
6 1 V
7 1 V
7 0 V
6 1 V
7 1 V
7 1 V
7 1 V
6 0 V
7 1 V
7 1 V
6 0 V
7 1 V
7 1 V
6 1 V
7 0 V
7 1 V
6 1 V
7 0 V
7 1 V
6 1 V
7 0 V
7 1 V
7 1 V
6 0 V
7 1 V
7 1 V
6 0 V
7 1 V
7 0 V
6 1 V
7 1 V
7 0 V
6 1 V
7 0 V
7 1 V
6 1 V
7 0 V
7 1 V
7 0 V
6 1 V
7 0 V
7 1 V
6 1 V
7 0 V
7 1 V
6 0 V
7 1 V
7 0 V
6 1 V
7 0 V
7 1 V
6 0 V
7 1 V
7 0 V
7 1 V
6 0 V
7 1 V
7 0 V
6 1 V
7 0 V
7 1 V
6 0 V
7 1 V
7 0 V
6 1 V
7 0 V
7 0 V
6 1 V
7 0 V
7 1 V
7 0 V
6 1 V
7 0 V
7 1 V
6 0 V
7 0 V
7 1 V
6 0 V
7 1 V
7 0 V
6 1 V
7 0 V
7 0 V
6 1 V
7 0 V
7 1 V
6 0 V
7 0 V
7 1 V
7 0 V
6 1 V
7 0 V
7 0 V
6 1 V
7 0 V
7 1 V
6 0 V
7 0 V
7 1 V
6 0 V
7 0 V
7 1 V
6 0 V
7 1 V
7 0 V
7 0 V
6 1 V
7 0 V
7 0 V
6 1 V
7 0 V
7 0 V
6 1 V
7 0 V
7 0 V
6 1 V
7 0 V
7 0 V
6 1 V
7 0 V
7 0 V
7 1 V
6 0 V
7 0 V
7 1 V
6 0 V
7 0 V
7 1 V
6 0 V
7 0 V
7 1 V
6 0 V
7 0 V
7 1 V
6 0 V
7 0 V
7 0 V
7 1 V
6 0 V
7 0 V
7 1 V
6 0 V
7 0 V
7 0 V
6 1 V
7 0 V
LT1
667 1467 M
7 -1 V
6 -2 V
7 -1 V
7 -2 V
6 -3 V
7 -3 V
7 -4 V
6 -4 V
7 -5 V
7 -5 V
6 -6 V
7 -7 V
7 -6 V
6 -8 V
7 -8 V
7 -8 V
7 -9 V
6 -9 V
7 -9 V
7 -10 V
6 -11 V
7 -10 V
7 -11 V
6 -12 V
7 -12 V
7 -12 V
6 -13 V
7 -13 V
7 -14 V
6 -14 V
7 -14 V
7 -15 V
7 -15 V
6 -14 V
7 -15 V
7 -14 V
6 -13 V
7 -10 V
7 -8 V
6 -4 V
7 0 V
7 4 V
6 6 V
7 7 V
7 8 V
6 9 V
7 10 V
7 9 V
7 9 V
6 9 V
7 9 V
7 9 V
6 8 V
7 9 V
7 8 V
6 9 V
7 8 V
7 7 V
6 8 V
7 8 V
7 7 V
6 8 V
7 7 V
7 7 V
7 7 V
6 6 V
7 7 V
7 6 V
6 7 V
7 6 V
7 6 V
6 6 V
7 6 V
7 5 V
6 6 V
7 5 V
7 5 V
6 6 V
7 5 V
7 4 V
6 5 V
7 5 V
7 5 V
7 4 V
6 4 V
7 5 V
7 4 V
6 4 V
7 4 V
7 4 V
6 4 V
7 4 V
7 3 V
6 4 V
7 3 V
7 4 V
6 3 V
7 4 V
7 3 V
7 3 V
6 3 V
7 3 V
7 3 V
6 3 V
7 3 V
7 3 V
6 3 V
7 3 V
7 2 V
6 3 V
7 2 V
7 3 V
6 3 V
7 2 V
7 2 V
7 3 V
6 2 V
7 2 V
7 3 V
6 2 V
7 2 V
7 2 V
6 2 V
7 2 V
7 2 V
6 2 V
7 2 V
7 2 V
6 2 V
7 2 V
7 2 V
7 2 V
6 2 V
7 1 V
7 2 V
6 2 V
7 2 V
7 1 V
6 2 V
7 2 V
7 1 V
6 2 V
7 1 V
7 2 V
6 1 V
7 2 V
7 1 V
6 2 V
7 1 V
7 2 V
7 1 V
6 1 V
7 2 V
7 1 V
6 2 V
7 1 V
7 1 V
6 1 V
7 2 V
7 1 V
6 1 V
7 1 V
7 2 V
6 1 V
7 1 V
7 1 V
7 1 V
6 1 V
7 2 V
7 1 V
6 1 V
7 1 V
7 1 V
6 1 V
7 1 V
7 1 V
6 1 V
7 1 V
7 1 V
6 1 V
7 1 V
7 1 V
7 1 V
6 1 V
7 1 V
7 1 V
6 1 V
7 1 V
7 1 V
6 1 V
7 0 V
7 1 V
6 1 V
7 1 V
7 1 V
6 1 V
7 1 V
7 0 V
7 1 V
6 1 V
7 1 V
7 1 V
6 1 V
7 0 V
7 1 V
6 1 V
7 1 V
7 0 V
6 1 V
7 1 V
7 1 V
6 0 V
7 1 V
7 1 V
7 0 V
6 1 V
7 1 V
7 1 V
6 0 V
7 1 V
7 1 V
6 0 V
7 1 V
7 1 V
6 0 V
7 1 V
7 1 V
6 0 V
7 1 V
7 0 V
6 1 V
7 1 V
7 0 V
7 1 V
6 1 V
7 0 V
7 1 V
6 0 V
7 1 V
7 1 V
6 0 V
7 1 V
7 0 V
6 1 V
7 0 V
7 1 V
6 0 V
7 1 V
7 1 V
7 0 V
6 1 V
7 0 V
7 1 V
6 0 V
7 1 V
7 0 V
6 1 V
7 0 V
7 1 V
6 0 V
7 1 V
7 0 V
6 1 V
7 0 V
7 1 V
7 0 V
6 1 V
7 0 V
7 1 V
6 0 V
7 1 V
7 0 V
6 1 V
7 0 V
7 0 V
6 1 V
7 0 V
7 1 V
6 0 V
7 1 V
7 0 V
7 1 V
6 0 V
7 0 V
7 1 V
6 0 V
7 1 V
7 0 V
6 1 V
7 0 V
LT2
667 1467 M
7 -1 V
6 -2 V
7 -1 V
7 -2 V
6 -3 V
7 -3 V
7 -4 V
6 -4 V
7 -5 V
7 -5 V
6 -6 V
7 -6 V
7 -7 V
6 -8 V
7 -8 V
7 -8 V
7 -9 V
6 -9 V
7 -9 V
7 -10 V
6 -10 V
7 -11 V
7 -11 V
6 -11 V
7 -12 V
7 -12 V
6 -13 V
7 -13 V
7 -13 V
6 -14 V
7 -14 V
7 -15 V
7 -14 V
6 -16 V
7 -15 V
7 -16 V
6 -17 V
7 -16 V
7 -17 V
6 -17 V
7 -18 V
7 -17 V
6 -18 V
7 -17 V
7 -17 V
6 -16 V
7 -14 V
7 -12 V
7 -9 V
6 -5 V
7 -2 V
7 1 V
6 3 V
7 4 V
7 5 V
6 6 V
7 6 V
7 7 V
6 6 V
7 6 V
7 7 V
6 6 V
7 6 V
7 7 V
7 6 V
6 6 V
7 6 V
7 6 V
6 6 V
7 6 V
7 6 V
6 5 V
7 6 V
7 6 V
6 5 V
7 6 V
7 5 V
6 5 V
7 6 V
7 5 V
6 5 V
7 5 V
7 6 V
7 5 V
6 5 V
7 5 V
7 5 V
6 4 V
7 5 V
7 5 V
6 5 V
7 4 V
7 5 V
6 5 V
7 4 V
7 5 V
6 4 V
7 5 V
7 4 V
7 4 V
6 5 V
7 4 V
7 4 V
6 4 V
7 5 V
7 4 V
6 4 V
7 4 V
7 4 V
6 4 V
7 4 V
7 4 V
6 4 V
7 4 V
7 4 V
7 4 V
6 4 V
7 4 V
7 4 V
6 3 V
7 4 V
7 4 V
6 4 V
7 3 V
7 4 V
6 4 V
7 3 V
7 4 V
6 4 V
7 3 V
7 4 V
7 3 V
6 4 V
7 3 V
7 4 V
6 3 V
7 3 V
7 4 V
6 3 V
7 3 V
7 4 V
6 3 V
7 3 V
7 3 V
6 3 V
7 4 V
7 3 V
6 3 V
7 3 V
7 3 V
7 3 V
6 3 V
7 3 V
7 3 V
6 2 V
7 3 V
7 3 V
6 3 V
7 3 V
7 2 V
6 3 V
7 3 V
7 3 V
6 2 V
7 3 V
7 2 V
7 3 V
6 2 V
7 3 V
7 2 V
6 3 V
7 2 V
7 3 V
6 2 V
7 3 V
7 2 V
6 2 V
7 2 V
7 3 V
6 2 V
7 2 V
7 2 V
7 3 V
6 2 V
7 2 V
7 2 V
6 2 V
7 2 V
7 2 V
6 2 V
7 2 V
7 2 V
6 2 V
7 2 V
7 2 V
6 2 V
7 2 V
7 2 V
7 1 V
6 2 V
7 2 V
7 2 V
6 2 V
7 1 V
7 2 V
6 2 V
7 2 V
7 1 V
6 2 V
7 2 V
7 1 V
6 2 V
7 1 V
7 2 V
7 2 V
6 1 V
7 2 V
7 1 V
6 2 V
7 1 V
7 2 V
6 1 V
7 2 V
7 1 V
6 2 V
7 1 V
7 1 V
6 2 V
7 1 V
7 2 V
6 1 V
7 1 V
7 2 V
7 1 V
6 1 V
7 2 V
7 1 V
6 1 V
7 1 V
7 2 V
6 1 V
7 1 V
7 1 V
6 1 V
7 2 V
7 1 V
6 1 V
7 1 V
7 1 V
7 1 V
6 2 V
7 1 V
7 1 V
6 1 V
7 1 V
7 1 V
6 1 V
7 1 V
7 1 V
6 1 V
7 1 V
7 1 V
6 1 V
7 1 V
7 1 V
7 1 V
6 1 V
7 1 V
7 1 V
6 1 V
7 1 V
7 1 V
6 1 V
7 1 V
7 1 V
6 1 V
7 1 V
7 1 V
6 1 V
7 0 V
7 1 V
7 1 V
6 1 V
7 1 V
7 1 V
6 1 V
7 0 V
7 1 V
6 1 V
7 1 V
LT3
667 1467 M
7 -1 V
6 -2 V
7 -1 V
7 -2 V
6 -3 V
7 -3 V
7 -4 V
6 -4 V
7 -5 V
7 -5 V
6 -6 V
7 -6 V
7 -7 V
6 -8 V
7 -8 V
7 -8 V
7 -8 V
6 -10 V
7 -9 V
7 -10 V
6 -10 V
7 -11 V
7 -11 V
6 -11 V
7 -12 V
7 -12 V
6 -13 V
7 -13 V
7 -13 V
6 -14 V
7 -14 V
7 -14 V
7 -15 V
6 -16 V
7 -15 V
7 -16 V
6 -16 V
7 -17 V
7 -17 V
6 -18 V
7 -17 V
7 -19 V
6 -18 V
7 -19 V
7 -19 V
6 -20 V
7 -19 V
7 -20 V
7 -21 V
6 -20 V
7 -21 V
7 -20 V
6 -20 V
7 -20 V
7 -18 V
6 -16 V
7 -13 V
7 -10 V
6 -6 V
7 -2 V
7 0 V
6 2 V
7 3 V
7 4 V
7 4 V
6 5 V
7 5 V
7 5 V
6 4 V
7 5 V
7 5 V
6 5 V
7 5 V
7 5 V
6 4 V
7 5 V
7 4 V
6 5 V
7 5 V
7 4 V
6 4 V
7 5 V
7 4 V
7 5 V
6 4 V
7 4 V
7 4 V
6 4 V
7 5 V
7 4 V
6 4 V
7 4 V
7 4 V
6 4 V
7 4 V
7 4 V
6 4 V
7 4 V
7 3 V
7 4 V
6 4 V
7 4 V
7 4 V
6 3 V
7 4 V
7 4 V
6 3 V
7 4 V
7 4 V
6 3 V
7 4 V
7 3 V
6 4 V
7 3 V
7 4 V
7 3 V
6 4 V
7 3 V
7 4 V
6 3 V
7 3 V
7 4 V
6 3 V
7 3 V
7 4 V
6 3 V
7 3 V
7 3 V
6 4 V
7 3 V
7 3 V
7 3 V
6 3 V
7 3 V
7 4 V
6 3 V
7 3 V
7 3 V
6 3 V
7 3 V
7 3 V
6 3 V
7 3 V
7 3 V
6 3 V
7 3 V
7 3 V
6 2 V
7 3 V
7 3 V
7 3 V
6 3 V
7 3 V
7 3 V
6 2 V
7 3 V
7 3 V
6 3 V
7 2 V
7 3 V
6 3 V
7 3 V
7 2 V
6 3 V
7 3 V
7 2 V
7 3 V
6 3 V
7 2 V
7 3 V
6 2 V
7 3 V
7 3 V
6 2 V
7 3 V
7 2 V
6 3 V
7 3 V
7 2 V
6 3 V
7 2 V
7 3 V
7 2 V
6 3 V
7 2 V
7 3 V
6 2 V
7 3 V
7 2 V
6 3 V
7 2 V
7 3 V
6 2 V
7 2 V
7 3 V
6 2 V
7 3 V
7 2 V
7 3 V
6 2 V
7 2 V
7 3 V
6 2 V
7 2 V
7 3 V
6 2 V
7 2 V
7 3 V
6 2 V
7 2 V
7 3 V
6 2 V
7 2 V
7 2 V
7 3 V
6 2 V
7 2 V
7 2 V
6 3 V
7 2 V
7 2 V
6 2 V
7 2 V
7 3 V
6 2 V
7 2 V
7 2 V
6 2 V
7 2 V
7 3 V
6 2 V
7 2 V
7 2 V
7 2 V
6 2 V
7 2 V
7 2 V
6 2 V
7 2 V
7 2 V
6 2 V
7 2 V
7 2 V
6 2 V
7 2 V
7 2 V
6 2 V
7 2 V
7 2 V
7 2 V
6 1 V
7 2 V
7 2 V
6 2 V
7 2 V
7 2 V
6 2 V
7 1 V
7 2 V
6 2 V
7 2 V
7 1 V
6 2 V
7 2 V
7 2 V
7 1 V
6 2 V
7 2 V
7 1 V
6 2 V
7 2 V
7 1 V
6 2 V
7 2 V
7 1 V
6 2 V
7 2 V
7 1 V
6 2 V
7 1 V
7 2 V
7 1 V
6 2 V
7 1 V
7 2 V
6 1 V
7 2 V
7 1 V
6 2 V
7 1 V
LT4
667 1467 M
7 -1 V
6 -2 V
7 -1 V
7 -2 V
6 -3 V
7 -3 V
7 -4 V
6 -4 V
7 -5 V
7 -5 V
6 -6 V
7 -6 V
7 -7 V
6 -8 V
7 -8 V
7 -8 V
7 -8 V
6 -10 V
7 -9 V
7 -10 V
6 -10 V
7 -11 V
7 -11 V
6 -11 V
7 -12 V
7 -12 V
6 -13 V
7 -13 V
7 -13 V
6 -14 V
7 -14 V
7 -14 V
7 -15 V
6 -16 V
7 -15 V
7 -16 V
6 -16 V
7 -17 V
7 -17 V
6 -18 V
7 -17 V
7 -19 V
6 -18 V
7 -19 V
7 -19 V
6 -20 V
7 -20 V
7 -20 V
7 -21 V
6 -21 V
7 -21 V
7 -22 V
6 -22 V
7 -22 V
7 -23 V
6 -23 V
7 -23 V
7 -23 V
6 -24 V
7 -23 V
7 -23 V
6 -22 V
7 -20 V
7 -18 V
7 -14 V
6 -10 V
7 -6 V
7 -2 V
6 0 V
7 3 V
7 3 V
6 4 V
7 4 V
7 4 V
6 4 V
7 5 V
7 4 V
6 4 V
7 5 V
7 4 V
6 4 V
7 4 V
7 4 V
7 4 V
6 4 V
7 4 V
7 4 V
6 3 V
7 4 V
7 4 V
6 4 V
7 3 V
7 4 V
6 4 V
7 3 V
7 4 V
6 3 V
7 4 V
7 3 V
7 4 V
6 3 V
7 4 V
7 3 V
6 4 V
7 3 V
7 3 V
6 4 V
7 3 V
7 4 V
6 3 V
7 3 V
7 4 V
6 3 V
7 3 V
7 3 V
7 4 V
6 3 V
7 3 V
7 3 V
6 4 V
7 3 V
7 3 V
6 3 V
7 3 V
7 3 V
6 4 V
7 3 V
7 3 V
6 3 V
7 3 V
7 3 V
7 3 V
6 3 V
7 3 V
7 3 V
6 3 V
7 3 V
7 3 V
6 2 V
7 3 V
7 3 V
6 3 V
7 3 V
7 3 V
6 2 V
7 3 V
7 3 V
6 2 V
7 3 V
7 3 V
7 2 V
6 3 V
7 3 V
7 2 V
6 3 V
7 2 V
7 3 V
6 2 V
7 3 V
7 2 V
6 3 V
7 2 V
7 2 V
6 3 V
7 2 V
7 2 V
7 3 V
6 2 V
7 2 V
7 3 V
6 2 V
7 2 V
7 2 V
6 2 V
7 3 V
7 2 V
6 2 V
7 2 V
7 2 V
6 2 V
7 2 V
7 3 V
7 2 V
6 2 V
7 2 V
7 2 V
6 2 V
7 2 V
7 2 V
6 2 V
7 2 V
7 2 V
6 2 V
7 2 V
7 2 V
6 2 V
7 1 V
7 2 V
7 2 V
6 2 V
7 2 V
7 2 V
6 2 V
7 1 V
7 2 V
6 2 V
7 2 V
7 2 V
6 1 V
7 2 V
7 2 V
6 2 V
7 1 V
7 2 V
7 2 V
6 2 V
7 1 V
7 2 V
6 2 V
7 1 V
7 2 V
6 2 V
7 1 V
7 2 V
6 2 V
7 1 V
7 2 V
6 1 V
7 2 V
7 2 V
6 1 V
7 2 V
7 1 V
7 2 V
6 1 V
7 2 V
7 1 V
6 2 V
7 1 V
7 2 V
6 1 V
7 2 V
7 1 V
6 2 V
7 1 V
7 2 V
6 1 V
7 2 V
7 1 V
7 2 V
6 1 V
7 1 V
7 2 V
6 1 V
7 2 V
7 1 V
6 2 V
7 1 V
7 1 V
6 2 V
7 1 V
7 1 V
6 2 V
7 1 V
7 2 V
7 1 V
6 1 V
7 2 V
7 1 V
6 1 V
7 2 V
7 1 V
6 1 V
7 1 V
7 2 V
6 1 V
7 1 V
7 2 V
6 1 V
7 1 V
7 2 V
7 1 V
6 1 V
7 1 V
7 2 V
6 1 V
7 1 V
7 2 V
6 1 V
7 1 V
LT5
667 1467 M
7 -1 V
6 -2 V
7 -1 V
7 -2 V
6 -3 V
7 -3 V
7 -4 V
6 -4 V
7 -5 V
7 -5 V
6 -6 V
7 -6 V
7 -7 V
6 -8 V
7 -8 V
7 -8 V
7 -8 V
6 -10 V
7 -9 V
7 -10 V
6 -10 V
7 -11 V
7 -11 V
6 -11 V
7 -12 V
7 -12 V
6 -13 V
7 -13 V
7 -13 V
6 -14 V
7 -14 V
7 -14 V
7 -15 V
6 -16 V
7 -15 V
7 -16 V
6 -16 V
7 -17 V
7 -17 V
6 -18 V
7 -17 V
7 -19 V
6 -18 V
7 -19 V
7 -19 V
6 -20 V
7 -20 V
7 -20 V
7 -21 V
6 -21 V
7 -21 V
7 -22 V
6 -22 V
7 -22 V
7 -23 V
6 -23 V
7 -24 V
7 -23 V
6 -25 V
7 -24 V
7 -25 V
6 -25 V
7 -25 V
7 -26 V
7 -26 V
6 -26 V
7 -26 V
7 -26 V
6 -25 V
7 -23 V
7 -21 V
6 -18 V
7 -13 V
7 -8 V
6 -4 V
7 -1 V
7 2 V
6 2 V
7 4 V
7 3 V
6 4 V
7 4 V
7 5 V
7 4 V
6 4 V
7 4 V
7 4 V
6 4 V
7 4 V
7 4 V
6 4 V
7 4 V
7 4 V
6 3 V
7 4 V
7 4 V
6 4 V
7 3 V
7 4 V
7 4 V
6 3 V
7 4 V
7 4 V
6 3 V
7 4 V
7 3 V
6 4 V
7 3 V
7 3 V
6 4 V
7 3 V
7 4 V
6 3 V
7 3 V
7 3 V
7 4 V
6 3 V
7 3 V
7 3 V
6 3 V
7 4 V
7 3 V
6 3 V
7 3 V
7 3 V
6 3 V
7 3 V
7 3 V
6 3 V
7 3 V
7 2 V
7 3 V
6 3 V
7 3 V
7 3 V
6 2 V
7 3 V
7 3 V
6 3 V
7 2 V
7 3 V
6 2 V
7 3 V
7 3 V
6 2 V
7 3 V
7 2 V
6 3 V
7 2 V
7 3 V
7 2 V
6 2 V
7 3 V
7 2 V
6 2 V
7 3 V
7 2 V
6 2 V
7 3 V
7 2 V
6 2 V
7 2 V
7 2 V
6 2 V
7 3 V
7 2 V
7 2 V
6 2 V
7 2 V
7 2 V
6 2 V
7 2 V
7 2 V
6 2 V
7 2 V
7 2 V
6 2 V
7 2 V
7 2 V
6 2 V
7 2 V
7 1 V
7 2 V
6 2 V
7 2 V
7 2 V
6 2 V
7 1 V
7 2 V
6 2 V
7 2 V
7 1 V
6 2 V
7 2 V
7 1 V
6 2 V
7 2 V
7 1 V
7 2 V
6 2 V
7 1 V
7 2 V
6 1 V
7 2 V
7 2 V
6 1 V
7 2 V
7 1 V
6 2 V
7 1 V
7 2 V
6 1 V
7 2 V
7 1 V
7 2 V
6 1 V
7 2 V
7 1 V
6 2 V
7 1 V
7 2 V
6 1 V
7 1 V
7 2 V
6 1 V
7 2 V
7 1 V
6 1 V
7 2 V
7 1 V
6 2 V
7 1 V
7 1 V
7 2 V
6 1 V
7 1 V
7 2 V
6 1 V
7 2 V
7 1 V
6 1 V
7 2 V
7 1 V
6 1 V
7 2 V
7 1 V
6 1 V
7 2 V
7 1 V
7 1 V
6 2 V
7 1 V
7 1 V
6 2 V
7 1 V
7 1 V
6 2 V
7 1 V
7 1 V
6 2 V
7 1 V
7 1 V
6 2 V
7 1 V
7 1 V
7 2 V
6 1 V
7 1 V
7 2 V
6 1 V
7 1 V
7 2 V
6 1 V
7 1 V
7 2 V
6 1 V
7 1 V
7 2 V
6 1 V
7 1 V
7 2 V
7 1 V
6 1 V
7 2 V
7 1 V
6 1 V
7 2 V
7 1 V
6 1 V
7 2 V
stroke
grestore
end
showpage
}
\put(1603,0){\makebox(0,0){$\alpha \;\;\; [{\mathcal L}^{-1}]$}}
\put(240,1018){%
\special{ps: gsave currentpoint currentpoint translate
270 rotate neg exch neg exch translate}%
\makebox(0,0)[b]{\shortstack{$\Delta F \;\;\; [{\mathcal C}^2/{\mathcal L}^2]$}}%
\special{ps: currentpoint grestore moveto}%
}
\put(2606,151){\makebox(0,0){3}}
\put(2272,151){\makebox(0,0){2.5}}
\put(1937,151){\makebox(0,0){2}}
\put(1603,151){\makebox(0,0){1.5}}
\put(1269,151){\makebox(0,0){1}}
\put(934,151){\makebox(0,0){0.5}}
\put(600,151){\makebox(0,0){0}}
\put(540,1663){\makebox(0,0)[r]{$10^{0}$}}
\put(540,1407){\makebox(0,0)[r]{$10^{-1}$}}
\put(540,1152){\makebox(0,0)[r]{$10^{-2}$}}
\put(540,896){\makebox(0,0)[r]{$10^{-3}$}}
\put(540,640){\makebox(0,0)[r]{$10^{-4}$}}
\put(540,385){\makebox(0,0)[r]{$10^{-5}$}}
\end{picture}

%% file: f5.tex
\setlength{\unitlength}{0.1bp}
\special{!
/gnudict 40 dict def
gnudict begin
/Color false def
/Solid false def
/gnulinewidth 5.000 def
/vshift -33 def
/dl {10 mul} def
/hpt 31.5 def
/vpt 31.5 def
/M {moveto} bind def
/L {lineto} bind def
/R {rmoveto} bind def
/V {rlineto} bind def
/vpt2 vpt 2 mul def
/hpt2 hpt 2 mul def
/Lshow { currentpoint stroke M
  0 vshift R show } def
/Rshow { currentpoint stroke M
  dup stringwidth pop neg vshift R show } def
/Cshow { currentpoint stroke M
  dup stringwidth pop -2 div vshift R show } def
/DL { Color {setrgbcolor Solid {pop []} if 0 setdash }
 {pop pop pop Solid {pop []} if 0 setdash} ifelse } def
/BL { stroke gnulinewidth 2 mul setlinewidth } def
/AL { stroke gnulinewidth 2 div setlinewidth } def
/PL { stroke gnulinewidth setlinewidth } def
/LTb { BL [] 0 0 0 DL } def
/LTa { AL [1 dl 2 dl] 0 setdash 0 0 0 setrgbcolor } def
/LT0 { PL [] 0 1 0 DL } def
/LT1 { PL [4 dl 2 dl] 0 0 1 DL } def
/LT2 { PL [2 dl 3 dl] 1 0 0 DL } def
/LT3 { PL [1 dl 1.5 dl] 1 0 1 DL } def
/LT4 { PL [5 dl 2 dl 1 dl 2 dl] 0 1 1 DL } def
/LT5 { PL [4 dl 3 dl 1 dl 3 dl] 1 1 0 DL } def
/LT6 { PL [2 dl 2 dl 2 dl 4 dl] 0 0 0 DL } def
/LT7 { PL [2 dl 2 dl 2 dl 2 dl 2 dl 4 dl] 1 0.3 0 DL } def
/LT8 { PL [2 dl 2 dl 2 dl 2 dl 2 dl 2 dl 2 dl 4 dl] 0.5 0.5 0.5 DL } def
/P { stroke [] 0 setdash
  currentlinewidth 2 div sub M
  0 currentlinewidth V stroke } def
/D { stroke [] 0 setdash 2 copy vpt add M
  hpt neg vpt neg V hpt vpt neg V
  hpt vpt V hpt neg vpt V closepath stroke
  P } def
/A { stroke [] 0 setdash vpt sub M 0 vpt2 V
  currentpoint stroke M
  hpt neg vpt neg R hpt2 0 V stroke
  } def
/B { stroke [] 0 setdash 2 copy exch hpt sub exch vpt add M
  0 vpt2 neg V hpt2 0 V 0 vpt2 V
  hpt2 neg 0 V closepath stroke
  P } def
/C { stroke [] 0 setdash exch hpt sub exch vpt add M
  hpt2 vpt2 neg V currentpoint stroke M
  hpt2 neg 0 R hpt2 vpt2 V stroke } def
/T { stroke [] 0 setdash 2 copy vpt 1.12 mul add M
  hpt neg vpt -1.62 mul V
  hpt 2 mul 0 V
  hpt neg vpt 1.62 mul V closepath stroke
  P  } def
/S { 2 copy A C} def
end
}
\begin{picture}(2789,1836)(0,0)
\special{"
gnudict begin
gsave
50 50 translate
0.100 0.100 scale
0 setgray
/Helvetica findfont 100 scalefont setfont
newpath
-500.000000 -500.000000 translate
LTa
600 251 M
0 1534 V
LTb
600 251 M
63 0 V
1943 0 R
-63 0 V
600 470 M
63 0 V
1943 0 R
-63 0 V
600 689 M
63 0 V
1943 0 R
-63 0 V
600 908 M
63 0 V
1943 0 R
-63 0 V
600 1128 M
63 0 V
1943 0 R
-63 0 V
600 1347 M
63 0 V
1943 0 R
-63 0 V
600 1566 M
63 0 V
1943 0 R
-63 0 V
600 1785 M
63 0 V
1943 0 R
-63 0 V
600 251 M
0 63 V
0 1471 R
0 -63 V
934 251 M
0 63 V
0 1471 R
0 -63 V
1269 251 M
0 63 V
0 1471 R
0 -63 V
1603 251 M
0 63 V
0 1471 R
0 -63 V
1937 251 M
0 63 V
0 1471 R
0 -63 V
2272 251 M
0 63 V
0 1471 R
0 -63 V
2606 251 M
0 63 V
0 1471 R
0 -63 V
600 251 M
2006 0 V
0 1534 V
-2006 0 V
600 251 L
LT0
667 1617 M
7 -1 V
6 -1 V
7 -2 V
7 -2 V
6 -2 V
7 -2 V
7 -3 V
6 -4 V
7 -4 V
7 -5 V
6 -5 V
7 -5 V
7 -6 V
6 -7 V
7 -6 V
7 -7 V
7 -8 V
6 -8 V
7 -8 V
7 -8 V
6 -9 V
7 -9 V
7 -10 V
6 -9 V
7 -10 V
7 -11 V
6 -11 V
7 -11 V
7 -11 V
6 -12 V
7 -12 V
7 -13 V
7 -12 V
6 -13 V
7 -14 V
7 -13 V
6 -14 V
7 -15 V
7 -14 V
6 -15 V
7 -16 V
7 -15 V
6 -16 V
7 -16 V
7 -17 V
6 -16 V
7 -18 V
7 -17 V
7 -18 V
6 -18 V
7 -18 V
7 -19 V
6 -18 V
7 -20 V
7 -19 V
6 -20 V
7 -20 V
7 -21 V
6 -20 V
7 -21 V
7 -22 V
6 -21 V
7 -22 V
7 -23 V
7 -22 V
6 -23 V
7 -23 V
7 -24 V
6 -23 V
7 -25 V
7 -24 V
6 -24 V
7 -25 V
7 -25 V
6 -25 V
7 -25 V
7 -25 V
6 -23 V
7 -22 V
7 -19 V
6 -14 V
7 -9 V
7 -4 V
7 0 V
6 3 V
7 4 V
7 5 V
6 7 V
7 6 V
7 7 V
6 7 V
7 7 V
7 8 V
6 7 V
7 8 V
7 7 V
6 8 V
7 8 V
7 7 V
7 8 V
6 7 V
7 8 V
7 8 V
6 7 V
7 8 V
7 7 V
6 8 V
7 8 V
7 7 V
6 8 V
7 7 V
7 8 V
6 7 V
7 8 V
7 7 V
7 8 V
6 7 V
7 8 V
7 7 V
6 8 V
7 7 V
7 8 V
6 7 V
7 8 V
7 7 V
6 7 V
7 8 V
7 7 V
6 7 V
7 8 V
7 7 V
7 7 V
6 7 V
7 8 V
7 7 V
6 7 V
7 7 V
7 7 V
6 7 V
7 7 V
7 7 V
6 7 V
7 6 V
7 7 V
6 7 V
7 7 V
7 6 V
6 7 V
7 7 V
7 6 V
7 7 V
6 7 V
7 6 V
7 7 V
6 6 V
7 7 V
7 6 V
6 7 V
7 6 V
7 6 V
6 7 V
7 6 V
7 6 V
6 7 V
7 6 V
7 6 V
7 6 V
6 6 V
7 7 V
7 6 V
6 6 V
7 6 V
7 6 V
6 5 V
7 6 V
7 6 V
6 6 V
7 6 V
7 5 V
6 6 V
7 6 V
7 5 V
7 6 V
6 5 V
7 6 V
7 5 V
6 5 V
7 6 V
7 5 V
6 5 V
7 5 V
7 5 V
6 5 V
7 5 V
7 5 V
6 5 V
7 5 V
7 5 V
7 4 V
6 5 V
7 5 V
7 4 V
6 5 V
7 4 V
7 5 V
6 4 V
7 4 V
7 5 V
6 4 V
7 4 V
7 4 V
6 4 V
7 5 V
7 4 V
7 4 V
6 3 V
7 4 V
7 4 V
6 4 V
7 4 V
7 3 V
6 4 V
7 4 V
7 3 V
6 4 V
7 4 V
7 3 V
6 3 V
7 4 V
7 3 V
6 4 V
7 3 V
7 3 V
7 3 V
6 4 V
7 3 V
7 3 V
6 3 V
7 3 V
7 3 V
6 3 V
7 3 V
7 3 V
6 3 V
7 3 V
7 3 V
6 2 V
7 3 V
7 3 V
7 3 V
6 2 V
7 3 V
7 3 V
6 2 V
7 3 V
7 2 V
6 3 V
7 3 V
7 2 V
6 2 V
7 3 V
7 2 V
6 3 V
7 2 V
7 2 V
7 3 V
6 2 V
7 2 V
7 3 V
6 2 V
7 2 V
7 2 V
6 2 V
7 3 V
7 2 V
6 2 V
7 2 V
7 2 V
6 2 V
7 2 V
7 2 V
7 2 V
6 2 V
7 2 V
7 2 V
6 2 V
7 2 V
7 1 V
6 2 V
7 2 V
LT1
667 1617 M
7 -1 V
6 -1 V
7 -2 V
7 -2 V
6 -2 V
7 -2 V
7 -3 V
6 -4 V
7 -4 V
7 -5 V
6 -5 V
7 -5 V
7 -6 V
6 -7 V
7 -6 V
7 -7 V
7 -8 V
6 -8 V
7 -8 V
7 -8 V
6 -9 V
7 -9 V
7 -10 V
6 -9 V
7 -10 V
7 -11 V
6 -11 V
7 -11 V
7 -11 V
6 -12 V
7 -12 V
7 -13 V
7 -12 V
6 -13 V
7 -14 V
7 -13 V
6 -14 V
7 -14 V
7 -15 V
6 -15 V
7 -15 V
7 -15 V
6 -14 V
7 -15 V
7 -14 V
6 -14 V
7 -12 V
7 -10 V
7 -7 V
6 -5 V
7 -3 V
7 0 V
6 1 V
7 3 V
7 3 V
6 4 V
7 4 V
7 4 V
6 5 V
7 4 V
7 4 V
6 4 V
7 5 V
7 4 V
7 4 V
6 3 V
7 4 V
7 4 V
6 4 V
7 3 V
7 4 V
6 4 V
7 3 V
7 3 V
6 4 V
7 3 V
7 3 V
6 4 V
7 3 V
7 3 V
6 3 V
7 3 V
7 3 V
7 3 V
6 3 V
7 3 V
7 3 V
6 3 V
7 3 V
7 2 V
6 3 V
7 3 V
7 3 V
6 2 V
7 3 V
7 3 V
6 2 V
7 3 V
7 2 V
7 3 V
6 3 V
7 2 V
7 2 V
6 3 V
7 2 V
7 3 V
6 2 V
7 2 V
7 3 V
6 2 V
7 2 V
7 3 V
6 2 V
7 2 V
7 2 V
7 2 V
6 3 V
7 2 V
7 2 V
6 2 V
7 2 V
7 2 V
6 2 V
7 2 V
7 2 V
6 2 V
7 2 V
7 2 V
6 2 V
7 2 V
7 2 V
7 2 V
6 2 V
7 2 V
7 2 V
6 2 V
7 1 V
7 2 V
6 2 V
7 2 V
7 2 V
6 1 V
7 2 V
7 2 V
6 2 V
7 1 V
7 2 V
6 2 V
7 1 V
7 2 V
7 2 V
6 1 V
7 2 V
7 2 V
6 1 V
7 2 V
7 2 V
6 1 V
7 2 V
7 1 V
6 2 V
7 2 V
7 1 V
6 2 V
7 1 V
7 2 V
7 1 V
6 2 V
7 1 V
7 2 V
6 1 V
7 2 V
7 1 V
6 2 V
7 1 V
7 1 V
6 2 V
7 1 V
7 2 V
6 1 V
7 2 V
7 1 V
7 1 V
6 2 V
7 1 V
7 1 V
6 2 V
7 1 V
7 2 V
6 1 V
7 1 V
7 2 V
6 1 V
7 1 V
7 1 V
6 2 V
7 1 V
7 1 V
7 2 V
6 1 V
7 1 V
7 1 V
6 2 V
7 1 V
7 1 V
6 1 V
7 2 V
7 1 V
6 1 V
7 1 V
7 2 V
6 1 V
7 1 V
7 1 V
7 1 V
6 2 V
7 1 V
7 1 V
6 1 V
7 1 V
7 2 V
6 1 V
7 1 V
7 1 V
6 1 V
7 1 V
7 2 V
6 1 V
7 1 V
7 1 V
6 1 V
7 1 V
7 1 V
7 2 V
6 1 V
7 1 V
7 1 V
6 1 V
7 1 V
7 1 V
6 1 V
7 2 V
7 1 V
6 1 V
7 1 V
7 1 V
6 1 V
7 1 V
7 1 V
7 1 V
6 1 V
7 1 V
7 1 V
6 2 V
7 1 V
7 1 V
6 1 V
7 1 V
7 1 V
6 1 V
7 1 V
7 1 V
6 1 V
7 1 V
7 1 V
7 1 V
6 1 V
7 1 V
7 1 V
6 1 V
7 1 V
7 1 V
6 1 V
7 1 V
7 1 V
6 1 V
7 1 V
7 1 V
6 1 V
7 1 V
7 1 V
7 1 V
6 1 V
7 1 V
7 1 V
6 1 V
7 1 V
7 1 V
6 1 V
7 1 V
LT2
667 1617 M
7 -1 V
6 -1 V
7 -2 V
7 -2 V
6 -2 V
7 -2 V
7 -3 V
6 -4 V
7 -4 V
7 -5 V
6 -5 V
7 -5 V
7 -6 V
6 -7 V
7 -6 V
7 -7 V
7 -8 V
6 -8 V
7 -8 V
7 -8 V
6 -9 V
7 -9 V
7 -10 V
6 -9 V
7 -10 V
7 -11 V
6 -11 V
7 -11 V
7 -11 V
6 -12 V
7 -12 V
7 -13 V
7 -12 V
6 -13 V
7 -14 V
7 -13 V
6 -14 V
7 -15 V
7 -14 V
6 -15 V
7 -16 V
7 -15 V
6 -16 V
7 -16 V
7 -17 V
6 -16 V
7 -17 V
7 -18 V
7 -17 V
6 -18 V
7 -19 V
7 -18 V
6 -18 V
7 -19 V
7 -18 V
6 -18 V
7 -18 V
7 -15 V
6 -13 V
7 -10 V
7 -6 V
6 -3 V
7 0 V
7 3 V
7 4 V
6 5 V
7 5 V
7 5 V
6 6 V
7 5 V
7 6 V
6 5 V
7 6 V
7 5 V
6 5 V
7 5 V
7 6 V
6 5 V
7 5 V
7 5 V
6 5 V
7 5 V
7 5 V
7 4 V
6 5 V
7 5 V
7 4 V
6 5 V
7 4 V
7 5 V
6 4 V
7 5 V
7 4 V
6 4 V
7 5 V
7 4 V
6 4 V
7 4 V
7 4 V
7 4 V
6 4 V
7 4 V
7 4 V
6 4 V
7 4 V
7 3 V
6 4 V
7 4 V
7 3 V
6 4 V
7 4 V
7 3 V
6 4 V
7 3 V
7 4 V
7 3 V
6 3 V
7 4 V
7 3 V
6 3 V
7 4 V
7 3 V
6 3 V
7 3 V
7 3 V
6 4 V
7 3 V
7 3 V
6 3 V
7 3 V
7 3 V
7 3 V
6 3 V
7 3 V
7 3 V
6 3 V
7 2 V
7 3 V
6 3 V
7 3 V
7 3 V
6 3 V
7 2 V
7 3 V
6 3 V
7 3 V
7 2 V
6 3 V
7 2 V
7 3 V
7 3 V
6 2 V
7 3 V
7 2 V
6 3 V
7 2 V
7 3 V
6 2 V
7 3 V
7 2 V
6 3 V
7 2 V
7 3 V
6 2 V
7 2 V
7 3 V
7 2 V
6 2 V
7 3 V
7 2 V
6 2 V
7 2 V
7 3 V
6 2 V
7 2 V
7 2 V
6 2 V
7 3 V
7 2 V
6 2 V
7 2 V
7 2 V
7 2 V
6 2 V
7 3 V
7 2 V
6 2 V
7 2 V
7 2 V
6 2 V
7 2 V
7 2 V
6 2 V
7 2 V
7 2 V
6 2 V
7 2 V
7 2 V
7 2 V
6 2 V
7 2 V
7 2 V
6 2 V
7 2 V
7 2 V
6 2 V
7 2 V
7 2 V
6 2 V
7 2 V
7 1 V
6 2 V
7 2 V
7 2 V
7 2 V
6 2 V
7 2 V
7 2 V
6 2 V
7 2 V
7 1 V
6 2 V
7 2 V
7 2 V
6 2 V
7 2 V
7 1 V
6 2 V
7 2 V
7 2 V
6 2 V
7 2 V
7 1 V
7 2 V
6 2 V
7 2 V
7 2 V
6 1 V
7 2 V
7 2 V
6 2 V
7 1 V
7 2 V
6 2 V
7 2 V
7 1 V
6 2 V
7 2 V
7 1 V
7 2 V
6 2 V
7 1 V
7 2 V
6 2 V
7 1 V
7 2 V
6 2 V
7 1 V
7 2 V
6 1 V
7 2 V
7 2 V
6 1 V
7 2 V
7 1 V
7 2 V
6 1 V
7 2 V
7 1 V
6 2 V
7 1 V
7 2 V
6 1 V
7 2 V
7 1 V
6 2 V
7 1 V
7 2 V
6 1 V
7 1 V
7 2 V
7 1 V
6 2 V
7 1 V
7 1 V
6 2 V
7 1 V
7 1 V
6 2 V
7 1 V
LT3
667 1617 M
7 -1 V
6 -1 V
7 -2 V
7 -2 V
6 -2 V
7 -2 V
7 -3 V
6 -4 V
7 -4 V
7 -5 V
6 -5 V
7 -5 V
7 -6 V
6 -7 V
7 -6 V
7 -7 V
7 -8 V
6 -8 V
7 -8 V
7 -8 V
6 -9 V
7 -9 V
7 -10 V
6 -9 V
7 -10 V
7 -11 V
6 -11 V
7 -11 V
7 -11 V
6 -12 V
7 -12 V
7 -13 V
7 -12 V
6 -13 V
7 -14 V
7 -13 V
6 -14 V
7 -15 V
7 -14 V
6 -15 V
7 -16 V
7 -15 V
6 -16 V
7 -16 V
7 -17 V
6 -16 V
7 -18 V
7 -17 V
7 -18 V
6 -18 V
7 -18 V
7 -19 V
6 -18 V
7 -20 V
7 -19 V
6 -20 V
7 -20 V
7 -20 V
6 -21 V
7 -21 V
7 -21 V
6 -21 V
7 -21 V
7 -20 V
7 -19 V
6 -17 V
7 -14 V
7 -9 V
6 -5 V
7 0 V
7 3 V
6 5 V
7 6 V
7 7 V
6 7 V
7 7 V
7 7 V
6 7 V
7 7 V
7 7 V
6 6 V
7 7 V
7 7 V
7 6 V
6 7 V
7 6 V
7 7 V
6 6 V
7 6 V
7 6 V
6 6 V
7 6 V
7 6 V
6 6 V
7 5 V
7 6 V
6 6 V
7 5 V
7 6 V
7 5 V
6 5 V
7 6 V
7 5 V
6 5 V
7 5 V
7 5 V
6 5 V
7 5 V
7 5 V
6 5 V
7 5 V
7 5 V
6 4 V
7 5 V
7 5 V
7 4 V
6 5 V
7 5 V
7 4 V
6 5 V
7 4 V
7 4 V
6 5 V
7 4 V
7 4 V
6 5 V
7 4 V
7 4 V
6 4 V
7 4 V
7 4 V
7 4 V
6 4 V
7 4 V
7 4 V
6 4 V
7 4 V
7 4 V
6 4 V
7 3 V
7 4 V
6 4 V
7 4 V
7 3 V
6 4 V
7 4 V
7 3 V
6 4 V
7 3 V
7 4 V
7 3 V
6 3 V
7 4 V
7 3 V
6 4 V
7 3 V
7 3 V
6 3 V
7 4 V
7 3 V
6 3 V
7 3 V
7 3 V
6 4 V
7 3 V
7 3 V
7 3 V
6 3 V
7 3 V
7 3 V
6 3 V
7 3 V
7 3 V
6 3 V
7 3 V
7 3 V
6 3 V
7 3 V
7 2 V
6 3 V
7 3 V
7 3 V
7 3 V
6 3 V
7 3 V
7 3 V
6 2 V
7 3 V
7 3 V
6 3 V
7 3 V
7 2 V
6 3 V
7 3 V
7 3 V
6 2 V
7 3 V
7 3 V
7 3 V
6 2 V
7 3 V
7 3 V
6 3 V
7 2 V
7 3 V
6 3 V
7 2 V
7 3 V
6 3 V
7 2 V
7 3 V
6 3 V
7 2 V
7 3 V
7 2 V
6 3 V
7 3 V
7 2 V
6 3 V
7 2 V
7 3 V
6 2 V
7 3 V
7 2 V
6 3 V
7 2 V
7 3 V
6 2 V
7 3 V
7 2 V
6 2 V
7 3 V
7 2 V
7 2 V
6 3 V
7 2 V
7 2 V
6 3 V
7 2 V
7 2 V
6 2 V
7 3 V
7 2 V
6 2 V
7 2 V
7 2 V
6 3 V
7 2 V
7 2 V
7 2 V
6 2 V
7 2 V
7 2 V
6 2 V
7 2 V
7 2 V
6 2 V
7 2 V
7 2 V
6 2 V
7 2 V
7 2 V
6 2 V
7 2 V
7 2 V
7 2 V
6 1 V
7 2 V
7 2 V
6 2 V
7 2 V
7 1 V
6 2 V
7 2 V
7 2 V
6 1 V
7 2 V
7 2 V
6 1 V
7 2 V
7 2 V
7 1 V
6 2 V
7 1 V
7 2 V
6 2 V
7 1 V
7 2 V
6 1 V
7 2 V
LT4
667 1617 M
7 -1 V
6 -1 V
7 -2 V
7 -2 V
6 -2 V
7 -2 V
7 -3 V
6 -4 V
7 -4 V
7 -5 V
6 -5 V
7 -5 V
7 -6 V
6 -7 V
7 -6 V
7 -7 V
7 -8 V
6 -8 V
7 -8 V
7 -8 V
6 -9 V
7 -9 V
7 -10 V
6 -9 V
7 -10 V
7 -11 V
6 -11 V
7 -11 V
7 -11 V
6 -12 V
7 -12 V
7 -13 V
7 -12 V
6 -13 V
7 -14 V
7 -13 V
6 -14 V
7 -15 V
7 -14 V
6 -15 V
7 -16 V
7 -15 V
6 -16 V
7 -16 V
7 -17 V
6 -16 V
7 -18 V
7 -17 V
7 -18 V
6 -18 V
7 -18 V
7 -19 V
6 -18 V
7 -20 V
7 -19 V
6 -20 V
7 -20 V
7 -21 V
6 -20 V
7 -21 V
7 -22 V
6 -21 V
7 -22 V
7 -22 V
7 -23 V
6 -23 V
7 -22 V
7 -23 V
6 -22 V
7 -22 V
7 -20 V
6 -16 V
7 -12 V
7 -6 V
6 -1 V
7 3 V
7 5 V
6 7 V
7 8 V
7 8 V
6 8 V
7 8 V
7 8 V
7 9 V
6 8 V
7 8 V
7 7 V
6 8 V
7 8 V
7 7 V
6 8 V
7 7 V
7 7 V
6 8 V
7 7 V
7 7 V
6 7 V
7 7 V
7 7 V
7 6 V
6 7 V
7 7 V
7 6 V
6 7 V
7 6 V
7 6 V
6 7 V
7 6 V
7 6 V
6 6 V
7 6 V
7 6 V
6 6 V
7 6 V
7 6 V
7 6 V
6 6 V
7 5 V
7 6 V
6 6 V
7 5 V
7 6 V
6 5 V
7 5 V
7 6 V
6 5 V
7 5 V
7 6 V
6 5 V
7 5 V
7 5 V
7 5 V
6 5 V
7 5 V
7 4 V
6 5 V
7 5 V
7 5 V
6 4 V
7 5 V
7 5 V
6 4 V
7 5 V
7 4 V
6 5 V
7 4 V
7 4 V
6 5 V
7 4 V
7 4 V
7 4 V
6 5 V
7 4 V
7 4 V
6 4 V
7 4 V
7 4 V
6 4 V
7 4 V
7 4 V
6 4 V
7 4 V
7 4 V
6 4 V
7 4 V
7 3 V
7 4 V
6 4 V
7 4 V
7 4 V
6 3 V
7 4 V
7 4 V
6 4 V
7 3 V
7 4 V
6 4 V
7 4 V
7 3 V
6 4 V
7 4 V
7 3 V
7 4 V
6 4 V
7 3 V
7 4 V
6 4 V
7 3 V
7 4 V
6 3 V
7 4 V
7 3 V
6 4 V
7 3 V
7 4 V
6 3 V
7 4 V
7 3 V
7 4 V
6 3 V
7 4 V
7 3 V
6 3 V
7 4 V
7 3 V
6 3 V
7 4 V
7 3 V
6 3 V
7 4 V
7 3 V
6 3 V
7 3 V
7 3 V
7 3 V
6 4 V
7 3 V
7 3 V
6 3 V
7 3 V
7 3 V
6 3 V
7 3 V
7 3 V
6 2 V
7 3 V
7 3 V
6 3 V
7 3 V
7 3 V
6 2 V
7 3 V
7 3 V
7 3 V
6 2 V
7 3 V
7 3 V
6 2 V
7 3 V
7 2 V
6 3 V
7 2 V
7 3 V
6 2 V
7 3 V
7 2 V
6 3 V
7 2 V
7 2 V
7 3 V
6 2 V
7 2 V
7 3 V
6 2 V
7 2 V
7 2 V
6 2 V
7 3 V
7 2 V
6 2 V
7 2 V
7 2 V
6 2 V
7 2 V
7 2 V
7 2 V
6 2 V
7 2 V
7 2 V
6 2 V
7 2 V
7 2 V
6 2 V
7 2 V
7 2 V
6 1 V
7 2 V
7 2 V
6 2 V
7 2 V
7 1 V
7 2 V
6 2 V
7 1 V
7 2 V
6 2 V
7 2 V
7 1 V
6 2 V
7 1 V
LT5
667 1617 M
7 -1 V
6 -1 V
7 -2 V
7 -2 V
6 -2 V
7 -2 V
7 -3 V
6 -4 V
7 -4 V
7 -5 V
6 -5 V
7 -5 V
7 -6 V
6 -7 V
7 -6 V
7 -7 V
7 -8 V
6 -8 V
7 -8 V
7 -8 V
6 -9 V
7 -9 V
7 -10 V
6 -9 V
7 -10 V
7 -11 V
6 -11 V
7 -11 V
7 -11 V
6 -12 V
7 -12 V
7 -13 V
7 -12 V
6 -13 V
7 -14 V
7 -13 V
6 -14 V
7 -15 V
7 -14 V
6 -15 V
7 -16 V
7 -15 V
6 -16 V
7 -16 V
7 -17 V
6 -16 V
7 -18 V
7 -17 V
7 -18 V
6 -18 V
7 -18 V
7 -19 V
6 -18 V
7 -20 V
7 -19 V
6 -20 V
7 -20 V
7 -21 V
6 -20 V
7 -21 V
7 -22 V
6 -21 V
7 -22 V
7 -23 V
7 -22 V
6 -23 V
7 -23 V
7 -24 V
6 -23 V
7 -24 V
7 -24 V
6 -24 V
7 -23 V
7 -23 V
6 -21 V
7 -18 V
7 -13 V
6 -7 V
7 -1 V
7 3 V
6 6 V
7 8 V
7 9 V
7 9 V
6 9 V
7 9 V
7 9 V
6 9 V
7 9 V
7 9 V
6 9 V
7 9 V
7 9 V
6 8 V
7 9 V
7 8 V
6 9 V
7 8 V
7 8 V
7 8 V
6 8 V
7 8 V
7 8 V
6 8 V
7 7 V
7 8 V
6 8 V
7 7 V
7 7 V
6 8 V
7 7 V
7 7 V
6 7 V
7 7 V
7 7 V
7 7 V
6 7 V
7 7 V
7 7 V
6 6 V
7 7 V
7 6 V
6 7 V
7 6 V
7 6 V
6 6 V
7 7 V
7 6 V
6 6 V
7 6 V
7 6 V
7 5 V
6 6 V
7 6 V
7 6 V
6 5 V
7 6 V
7 5 V
6 6 V
7 5 V
7 6 V
6 5 V
7 5 V
7 5 V
6 6 V
7 5 V
7 5 V
6 5 V
7 5 V
7 5 V
7 5 V
6 5 V
7 5 V
7 5 V
6 5 V
7 4 V
7 5 V
6 5 V
7 5 V
7 5 V
6 4 V
7 5 V
7 5 V
6 5 V
7 4 V
7 5 V
7 5 V
6 4 V
7 5 V
7 4 V
6 5 V
7 5 V
7 4 V
6 5 V
7 4 V
7 5 V
6 4 V
7 5 V
7 4 V
6 5 V
7 4 V
7 4 V
7 5 V
6 4 V
7 4 V
7 5 V
6 4 V
7 4 V
7 5 V
6 4 V
7 4 V
7 4 V
6 4 V
7 4 V
7 4 V
6 4 V
7 4 V
7 4 V
7 4 V
6 4 V
7 4 V
7 4 V
6 4 V
7 4 V
7 3 V
6 4 V
7 4 V
7 4 V
6 3 V
7 4 V
7 3 V
6 4 V
7 3 V
7 4 V
7 3 V
6 4 V
7 3 V
7 4 V
6 3 V
7 3 V
7 3 V
6 4 V
7 3 V
7 3 V
6 3 V
7 3 V
7 3 V
6 4 V
7 3 V
7 3 V
6 3 V
7 2 V
7 3 V
7 3 V
6 3 V
7 3 V
7 3 V
6 3 V
7 2 V
7 3 V
6 3 V
7 2 V
7 3 V
6 3 V
7 2 V
7 3 V
6 2 V
7 3 V
7 2 V
7 3 V
6 2 V
7 3 V
7 2 V
6 2 V
7 3 V
7 2 V
6 2 V
7 3 V
7 2 V
6 2 V
7 2 V
7 3 V
6 2 V
7 2 V
7 2 V
7 2 V
6 2 V
7 2 V
7 2 V
6 2 V
7 2 V
7 2 V
6 2 V
7 2 V
7 2 V
6 2 V
7 2 V
7 2 V
6 2 V
7 1 V
7 2 V
7 2 V
6 2 V
7 2 V
7 1 V
6 2 V
7 2 V
7 2 V
6 1 V
7 2 V
LT6
667 1617 M
7 -1 V
6 -1 V
7 -2 V
7 -2 V
6 -2 V
7 -2 V
7 -3 V
6 -4 V
7 -4 V
7 -5 V
6 -5 V
7 -5 V
7 -6 V
6 -7 V
7 -6 V
7 -7 V
7 -8 V
6 -8 V
7 -8 V
7 -8 V
6 -9 V
7 -9 V
7 -10 V
6 -9 V
7 -10 V
7 -11 V
6 -11 V
7 -11 V
7 -11 V
6 -12 V
7 -12 V
7 -13 V
7 -12 V
6 -13 V
7 -14 V
7 -13 V
6 -14 V
7 -15 V
7 -14 V
6 -15 V
7 -16 V
7 -15 V
6 -16 V
7 -16 V
7 -17 V
6 -16 V
7 -18 V
7 -17 V
7 -18 V
6 -18 V
7 -18 V
7 -19 V
6 -18 V
7 -20 V
7 -19 V
6 -20 V
7 -20 V
7 -21 V
6 -20 V
7 -21 V
7 -22 V
6 -21 V
7 -22 V
7 -23 V
7 -22 V
6 -23 V
7 -23 V
7 -24 V
6 -23 V
7 -24 V
7 -25 V
6 -24 V
7 -25 V
7 -25 V
6 -24 V
7 -24 V
7 -23 V
6 -21 V
7 -17 V
7 -13 V
6 -6 V
7 -1 V
7 4 V
7 6 V
6 8 V
7 8 V
7 10 V
6 9 V
7 10 V
7 10 V
6 9 V
7 10 V
7 10 V
6 9 V
7 10 V
7 9 V
6 9 V
7 10 V
7 9 V
7 9 V
6 9 V
7 9 V
7 9 V
6 9 V
7 9 V
7 8 V
6 9 V
7 8 V
7 9 V
6 8 V
7 8 V
7 8 V
6 8 V
7 8 V
7 8 V
7 8 V
6 8 V
7 7 V
7 8 V
6 8 V
7 7 V
7 7 V
6 8 V
7 7 V
7 7 V
6 7 V
7 7 V
7 7 V
6 7 V
7 6 V
7 7 V
7 7 V
6 6 V
7 7 V
7 6 V
6 7 V
7 6 V
7 6 V
6 6 V
7 7 V
7 6 V
6 6 V
7 6 V
7 6 V
6 6 V
7 6 V
7 6 V
6 5 V
7 6 V
7 6 V
7 6 V
6 6 V
7 5 V
7 6 V
6 6 V
7 5 V
7 6 V
6 6 V
7 5 V
7 6 V
6 5 V
7 6 V
7 5 V
6 6 V
7 5 V
7 5 V
7 6 V
6 5 V
7 5 V
7 6 V
6 5 V
7 5 V
7 5 V
6 6 V
7 5 V
7 5 V
6 5 V
7 5 V
7 5 V
6 5 V
7 5 V
7 5 V
7 5 V
6 4 V
7 5 V
7 5 V
6 5 V
7 4 V
7 5 V
6 5 V
7 4 V
7 5 V
6 4 V
7 5 V
7 4 V
6 5 V
7 4 V
7 4 V
7 5 V
6 4 V
7 4 V
7 4 V
6 4 V
7 4 V
7 4 V
6 4 V
7 4 V
7 4 V
6 4 V
7 4 V
7 4 V
6 3 V
7 4 V
7 4 V
7 3 V
6 4 V
7 4 V
7 3 V
6 4 V
7 3 V
7 4 V
6 3 V
7 3 V
7 4 V
6 3 V
7 3 V
7 3 V
6 4 V
7 3 V
7 3 V
6 3 V
7 3 V
7 3 V
7 3 V
6 3 V
7 3 V
7 3 V
6 3 V
7 3 V
7 2 V
6 3 V
7 3 V
7 3 V
6 2 V
7 3 V
7 3 V
6 2 V
7 3 V
7 2 V
7 3 V
6 3 V
7 2 V
7 2 V
6 3 V
7 2 V
7 3 V
6 2 V
7 2 V
7 3 V
6 2 V
7 2 V
7 3 V
6 2 V
7 2 V
7 2 V
7 2 V
6 2 V
7 3 V
7 2 V
6 2 V
7 2 V
7 2 V
6 2 V
7 2 V
7 2 V
6 2 V
7 2 V
7 2 V
6 1 V
7 2 V
7 2 V
7 2 V
6 2 V
7 2 V
7 2 V
6 1 V
7 2 V
7 2 V
6 2 V
7 1 V
stroke
grestore
end
showpage
}
\put(1603,0){\makebox(0,0){$\alpha \;\;\; [{\mathcal L}^{-1}]$}}
\put(240,1018){%
\special{ps: gsave currentpoint currentpoint translate
270 rotate neg exch neg exch translate}%
\makebox(0,0)[b]{\shortstack{$\Delta F \;\;\; [{\mathcal C}^2/{\mathcal L}^2]$}}%
\special{ps: currentpoint grestore moveto}%
}
\put(2606,151){\makebox(0,0){3}}
\put(2272,151){\makebox(0,0){2.5}}
\put(1937,151){\makebox(0,0){2}}
\put(1603,151){\makebox(0,0){1.5}}
\put(1269,151){\makebox(0,0){1}}
\put(934,151){\makebox(0,0){0.5}}
\put(600,151){\makebox(0,0){0}}
\put(540,1785){\makebox(0,0)[r]{$10^{0}$}}
\put(540,1566){\makebox(0,0)[r]{$10^{-1}$}}
\put(540,1347){\makebox(0,0)[r]{$10^{-2}$}}
\put(540,1128){\makebox(0,0)[r]{$10^{-3}$}}
\put(540,908){\makebox(0,0)[r]{$10^{-4}$}}
\put(540,689){\makebox(0,0)[r]{$10^{-5}$}}
\put(540,470){\makebox(0,0)[r]{$10^{-6}$}}
\put(540,251){\makebox(0,0)[r]{$10^{-7}$}}
\end{picture}

%% file: f6.tex
\setlength{\unitlength}{0.1bp}
\special{!
/gnudict 40 dict def
gnudict begin
/Color false def
/Solid false def
/gnulinewidth 5.000 def
/vshift -33 def
/dl {10 mul} def
/hpt 31.5 def
/vpt 31.5 def
/M {moveto} bind def
/L {lineto} bind def
/R {rmoveto} bind def
/V {rlineto} bind def
/vpt2 vpt 2 mul def
/hpt2 hpt 2 mul def
/Lshow { currentpoint stroke M
  0 vshift R show } def
/Rshow { currentpoint stroke M
  dup stringwidth pop neg vshift R show } def
/Cshow { currentpoint stroke M
  dup stringwidth pop -2 div vshift R show } def
/DL { Color {setrgbcolor Solid {pop []} if 0 setdash }
 {pop pop pop Solid {pop []} if 0 setdash} ifelse } def
/BL { stroke gnulinewidth 2 mul setlinewidth } def
/AL { stroke gnulinewidth 2 div setlinewidth } def
/PL { stroke gnulinewidth setlinewidth } def
/LTb { BL [] 0 0 0 DL } def
/LTa { AL [1 dl 2 dl] 0 setdash 0 0 0 setrgbcolor } def
/LT0 { PL [] 0 1 0 DL } def
/LT1 { PL [4 dl 2 dl] 0 0 1 DL } def
/LT2 { PL [2 dl 3 dl] 1 0 0 DL } def
/LT3 { PL [1 dl 1.5 dl] 1 0 1 DL } def
/LT4 { PL [5 dl 2 dl 1 dl 2 dl] 0 1 1 DL } def
/LT5 { PL [4 dl 3 dl 1 dl 3 dl] 1 1 0 DL } def
/LT6 { PL [2 dl 2 dl 2 dl 4 dl] 0 0 0 DL } def
/LT7 { PL [2 dl 2 dl 2 dl 2 dl 2 dl 4 dl] 1 0.3 0 DL } def
/LT8 { PL [2 dl 2 dl 2 dl 2 dl 2 dl 2 dl 2 dl 4 dl] 0.5 0.5 0.5 DL } def
/P { stroke [] 0 setdash
  currentlinewidth 2 div sub M
  0 currentlinewidth V stroke } def
/D { stroke [] 0 setdash 2 copy vpt add M
  hpt neg vpt neg V hpt vpt neg V
  hpt vpt V hpt neg vpt V closepath stroke
  P } def
/A { stroke [] 0 setdash vpt sub M 0 vpt2 V
  currentpoint stroke M
  hpt neg vpt neg R hpt2 0 V stroke
  } def
/B { stroke [] 0 setdash 2 copy exch hpt sub exch vpt add M
  0 vpt2 neg V hpt2 0 V 0 vpt2 V
  hpt2 neg 0 V closepath stroke
  P } def
/C { stroke [] 0 setdash exch hpt sub exch vpt add M
  hpt2 vpt2 neg V currentpoint stroke M
  hpt2 neg 0 R hpt2 vpt2 V stroke } def
/T { stroke [] 0 setdash 2 copy vpt 1.12 mul add M
  hpt neg vpt -1.62 mul V
  hpt 2 mul 0 V
  hpt neg vpt 1.62 mul V closepath stroke
  P  } def
/S { 2 copy A C} def
end
}
\begin{picture}(2789,1836)(0,0)
\special{"
gnudict begin
gsave
50 50 translate
0.100 0.100 scale
0 setgray
/Helvetica findfont 100 scalefont setfont
newpath
-500.000000 -500.000000 translate
LTa
600 251 M
0 1534 V
LTb
600 251 M
63 0 V
1943 0 R
-63 0 V
600 507 M
63 0 V
1943 0 R
-63 0 V
600 762 M
63 0 V
1943 0 R
-63 0 V
600 1018 M
63 0 V
1943 0 R
-63 0 V
600 1274 M
63 0 V
1943 0 R
-63 0 V
600 1529 M
63 0 V
1943 0 R
-63 0 V
600 1785 M
63 0 V
1943 0 R
-63 0 V
600 251 M
0 63 V
0 1471 R
0 -63 V
934 251 M
0 63 V
0 1471 R
0 -63 V
1269 251 M
0 63 V
0 1471 R
0 -63 V
1603 251 M
0 63 V
0 1471 R
0 -63 V
1937 251 M
0 63 V
0 1471 R
0 -63 V
2272 251 M
0 63 V
0 1471 R
0 -63 V
2606 251 M
0 63 V
0 1471 R
0 -63 V
600 251 M
2006 0 V
0 1534 V
-2006 0 V
600 251 L
LT0
2265 890 M
180 0 V
667 1687 M
7 -1 V
6 0 V
7 -1 V
7 -1 V
6 -1 V
7 -2 V
7 -2 V
6 -2 V
7 -2 V
7 -3 V
6 -3 V
7 -3 V
7 -4 V
6 -3 V
7 -4 V
7 -4 V
7 -5 V
6 -4 V
7 -5 V
7 -5 V
6 -5 V
7 -5 V
7 -6 V
6 -6 V
7 -5 V
7 -7 V
6 -6 V
7 -6 V
7 -7 V
6 -7 V
7 -7 V
7 -7 V
7 -8 V
6 -7 V
7 -8 V
7 -8 V
6 -8 V
7 -9 V
7 -8 V
6 -9 V
7 -9 V
7 -9 V
6 -9 V
7 -10 V
7 -9 V
6 -10 V
7 -10 V
7 -10 V
7 -10 V
6 -11 V
7 -11 V
7 -10 V
6 -11 V
7 -12 V
7 -11 V
6 -12 V
7 -11 V
7 -12 V
6 -12 V
7 -13 V
7 -12 V
6 -13 V
7 -13 V
7 -13 V
7 -13 V
6 -13 V
7 -14 V
7 -13 V
6 -14 V
7 -14 V
7 -15 V
6 -14 V
7 -15 V
7 -14 V
6 -15 V
7 -15 V
7 -16 V
6 -15 V
7 -16 V
7 -16 V
6 -16 V
7 -16 V
7 -17 V
7 -16 V
6 -17 V
7 -17 V
7 -17 V
6 -17 V
7 -18 V
7 -17 V
6 -18 V
7 -18 V
7 -18 V
6 -19 V
7 -18 V
7 -19 V
6 -19 V
7 -19 V
7 -19 V
7 -20 V
6 -20 V
7 -19 V
7 -20 V
6 -21 V
7 -20 V
7 -20 V
6 -21 V
7 -21 V
7 -21 V
6 -21 V
7 -22 V
7 -21 V
6 -22 V
7 -19 V
7 -11 V
7 7 V
6 17 V
7 21 V
7 21 V
6 20 V
7 21 V
7 19 V
6 19 V
7 19 V
7 18 V
6 18 V
7 18 V
7 17 V
6 17 V
7 16 V
7 17 V
7 15 V
6 16 V
7 15 V
7 15 V
6 14 V
7 15 V
7 13 V
6 14 V
7 14 V
7 13 V
6 13 V
7 12 V
7 13 V
6 12 V
7 12 V
7 12 V
6 11 V
7 11 V
7 11 V
7 11 V
6 11 V
7 10 V
7 11 V
6 10 V
7 10 V
7 10 V
6 9 V
7 10 V
7 9 V
6 9 V
7 9 V
7 9 V
6 8 V
7 9 V
7 8 V
7 8 V
6 9 V
7 7 V
7 8 V
6 8 V
7 8 V
7 7 V
6 7 V
7 8 V
7 7 V
6 7 V
7 7 V
7 6 V
6 7 V
7 7 V
7 6 V
7 6 V
6 7 V
7 6 V
7 6 V
6 6 V
7 6 V
7 6 V
6 6 V
7 5 V
7 6 V
6 5 V
7 6 V
7 5 V
6 5 V
7 6 V
7 5 V
7 5 V
6 5 V
7 5 V
7 4 V
6 5 V
7 5 V
7 5 V
6 4 V
7 5 V
7 4 V
6 5 V
7 4 V
7 4 V
6 4 V
7 5 V
7 4 V
7 4 V
6 4 V
7 4 V
7 4 V
6 4 V
7 3 V
7 4 V
6 4 V
7 4 V
7 3 V
6 4 V
7 3 V
7 4 V
6 3 V
7 4 V
7 3 V
6 3 V
7 4 V
7 3 V
7 3 V
6 3 V
7 3 V
7 4 V
6 3 V
7 3 V
7 3 V
6 3 V
7 3 V
7 2 V
6 3 V
7 3 V
7 3 V
6 3 V
7 2 V
7 3 V
7 3 V
6 2 V
7 3 V
7 3 V
6 2 V
7 3 V
7 2 V
6 3 V
7 2 V
7 3 V
6 2 V
7 2 V
7 3 V
6 2 V
7 2 V
7 3 V
7 2 V
6 2 V
7 2 V
7 2 V
6 3 V
7 2 V
7 2 V
6 2 V
7 2 V
7 2 V
6 2 V
7 2 V
7 2 V
6 2 V
7 2 V
7 2 V
7 2 V
6 2 V
7 2 V
7 2 V
6 1 V
7 2 V
7 2 V
6 2 V
7 2 V
LT1
2265 790 M
180 0 V
667 1687 M
7 -1 V
6 0 V
7 -1 V
7 -1 V
6 -1 V
7 -2 V
7 -2 V
6 -2 V
7 -2 V
7 -3 V
6 -3 V
7 -3 V
7 -4 V
6 -3 V
7 -4 V
7 -4 V
7 -5 V
6 -4 V
7 -5 V
7 -5 V
6 -5 V
7 -5 V
7 -6 V
6 -6 V
7 -5 V
7 -7 V
6 -6 V
7 -6 V
7 -7 V
6 -7 V
7 -7 V
7 -7 V
7 -8 V
6 -7 V
7 -8 V
7 -8 V
6 -8 V
7 -9 V
7 -8 V
6 -9 V
7 -9 V
7 -9 V
6 -9 V
7 -10 V
7 -9 V
6 -10 V
7 -10 V
7 -10 V
7 -10 V
6 -11 V
7 -11 V
7 -10 V
6 -11 V
7 -12 V
7 -11 V
6 -12 V
7 -11 V
7 -12 V
6 -12 V
7 -13 V
7 -12 V
6 -12 V
7 -13 V
7 -12 V
7 -12 V
6 -11 V
7 -9 V
7 -7 V
6 -4 V
7 -1 V
7 2 V
6 3 V
7 4 V
7 4 V
6 4 V
7 5 V
7 5 V
6 4 V
7 5 V
7 4 V
6 5 V
7 4 V
7 4 V
7 5 V
6 4 V
7 4 V
7 4 V
6 5 V
7 4 V
7 4 V
6 4 V
7 4 V
7 4 V
6 4 V
7 4 V
7 4 V
6 3 V
7 4 V
7 4 V
7 3 V
6 4 V
7 4 V
7 3 V
6 4 V
7 3 V
7 4 V
6 3 V
7 3 V
7 4 V
6 3 V
7 3 V
7 3 V
6 4 V
7 3 V
7 3 V
7 3 V
6 3 V
7 3 V
7 3 V
6 3 V
7 2 V
7 3 V
6 3 V
7 3 V
7 3 V
6 2 V
7 3 V
7 3 V
6 2 V
7 3 V
7 3 V
7 2 V
6 3 V
7 2 V
7 3 V
6 2 V
7 3 V
7 2 V
6 3 V
7 2 V
7 3 V
6 2 V
7 2 V
7 3 V
6 2 V
7 2 V
7 3 V
6 2 V
7 2 V
7 2 V
7 3 V
6 2 V
7 2 V
7 2 V
6 2 V
7 3 V
7 2 V
6 2 V
7 2 V
7 2 V
6 2 V
7 2 V
7 3 V
6 2 V
7 2 V
7 2 V
7 2 V
6 2 V
7 2 V
7 2 V
6 2 V
7 2 V
7 2 V
6 2 V
7 1 V
7 2 V
6 2 V
7 2 V
7 2 V
6 2 V
7 2 V
7 2 V
7 1 V
6 2 V
7 2 V
7 2 V
6 2 V
7 1 V
7 2 V
6 2 V
7 1 V
7 2 V
6 2 V
7 1 V
7 2 V
6 2 V
7 1 V
7 2 V
7 2 V
6 1 V
7 2 V
7 1 V
6 2 V
7 2 V
7 1 V
6 2 V
7 1 V
7 2 V
6 1 V
7 2 V
7 1 V
6 1 V
7 2 V
7 1 V
7 2 V
6 1 V
7 2 V
7 1 V
6 1 V
7 2 V
7 1 V
6 1 V
7 2 V
7 1 V
6 1 V
7 2 V
7 1 V
6 1 V
7 1 V
7 2 V
6 1 V
7 1 V
7 1 V
7 2 V
6 1 V
7 1 V
7 1 V
6 1 V
7 1 V
7 2 V
6 1 V
7 1 V
7 1 V
6 1 V
7 1 V
7 1 V
6 1 V
7 2 V
7 1 V
7 1 V
6 1 V
7 1 V
7 1 V
6 1 V
7 1 V
7 1 V
6 1 V
7 1 V
7 1 V
6 1 V
7 1 V
7 1 V
6 1 V
7 1 V
7 0 V
7 1 V
6 1 V
7 1 V
7 1 V
6 1 V
7 1 V
7 1 V
6 1 V
7 0 V
7 1 V
6 1 V
7 1 V
7 1 V
6 1 V
7 0 V
7 1 V
7 1 V
6 1 V
7 1 V
7 0 V
6 1 V
7 1 V
7 1 V
6 0 V
7 1 V
LT2
2265 690 M
180 0 V
600 1721 M
20 -1 V
21 -2 V
20 -4 V
20 -6 V
20 -7 V
21 -9 V
20 -11 V
20 -12 V
20 -14 V
21 -15 V
20 -18 V
20 -18 V
20 -21 V
21 -22 V
20 -23 V
20 -26 V
20 -27 V
21 -28 V
20 -30 V
20 -32 V
21 -34 V
20 -35 V
20 -36 V
20 -39 V
21 -40 V
20 -41 V
20 -44 V
20 -44 V
21 -47 V
20 -48 V
20 -50 V
20 -51 V
21 -53 V
20 -55 V
20 -56 V
20 -58 V
21 -60 V
20 -61 V
20 -63 V
21 -64 V
20 -66 V
20 -68 V
9 -31 V
LT3
2265 590 M
180 0 V
1417 251 M
14 50 V
20 68 V
20 63 V
21 60 V
20 55 V
20 52 V
20 49 V
21 45 V
20 43 V
20 41 V
20 38 V
21 36 V
20 34 V
20 32 V
20 31 V
21 29 V
20 27 V
20 26 V
20 25 V
21 24 V
20 22 V
20 22 V
21 20 V
20 20 V
20 18 V
20 18 V
21 17 V
20 17 V
20 15 V
20 15 V
21 15 V
20 13 V
20 14 V
20 13 V
21 12 V
20 12 V
20 11 V
20 11 V
21 11 V
20 10 V
20 10 V
21 9 V
20 9 V
20 9 V
20 9 V
21 8 V
20 8 V
20 8 V
20 7 V
21 8 V
20 7 V
20 6 V
20 7 V
21 7 V
20 6 V
20 6 V
20 6 V
21 5 V
20 6 V
LT4
2265 490 M
180 0 V
659 251 M
2 12 V
20 120 V
20 93 V
21 76 V
20 64 V
20 55 V
20 49 V
21 44 V
20 40 V
20 36 V
20 34 V
21 30 V
20 29 V
20 27 V
20 25 V
21 24 V
20 23 V
20 21 V
21 20 V
20 20 V
20 18 V
20 18 V
21 17 V
20 16 V
20 16 V
20 15 V
21 15 V
20 14 V
20 13 V
20 14 V
21 12 V
20 13 V
20 12 V
20 12 V
21 11 V
20 11 V
20 11 V
21 11 V
20 10 V
20 10 V
20 10 V
21 9 V
20 10 V
20 9 V
20 9 V
21 8 V
20 9 V
20 8 V
20 9 V
21 8 V
20 8 V
20 8 V
20 7 V
21 8 V
20 7 V
20 7 V
20 7 V
21 7 V
20 7 V
20 7 V
21 7 V
20 6 V
20 7 V
20 6 V
21 6 V
20 7 V
20 6 V
20 6 V
21 5 V
20 6 V
20 6 V
20 6 V
21 5 V
20 6 V
20 5 V
20 6 V
21 5 V
20 5 V
20 5 V
21 5 V
20 6 V
20 4 V
20 5 V
21 5 V
20 5 V
20 5 V
20 5 V
21 4 V
20 5 V
20 4 V
20 5 V
21 4 V
20 5 V
20 4 V
20 4 V
21 5 V
20 4 V
stroke
grestore
end
showpage
}
\put(2205,490){\makebox(0,0)[r]{5}}
\put(2205,590){\makebox(0,0)[r]{4}}
\put(2205,690){\makebox(0,0)[r]{3}}
\put(2205,790){\makebox(0,0)[r]{2}}
\put(2205,890){\makebox(0,0)[r]{1}}
\put(1603,0){\makebox(0,0){$\alpha \;\;\; [{\mathcal L}^{-1}]$}}
\put(240,1018){%
\special{ps: gsave currentpoint currentpoint translate
270 rotate neg exch neg exch translate}%
\makebox(0,0)[b]{\shortstack{$\Delta F \;\;\; [{\mathcal C}^2/{\mathcal L}^2]$}}%
\special{ps: currentpoint grestore moveto}%
}
\put(2606,151){\makebox(0,0){3}}
\put(2272,151){\makebox(0,0){2.5}}
\put(1937,151){\makebox(0,0){2}}
\put(1603,151){\makebox(0,0){1.5}}
\put(1269,151){\makebox(0,0){1}}
\put(934,151){\makebox(0,0){0.5}}
\put(600,151){\makebox(0,0){0}}
\put(540,1785){\makebox(0,0)[r]{$10^{0}$}}
\put(540,1529){\makebox(0,0)[r]{$10^{-2}$}}
\put(540,1274){\makebox(0,0)[r]{$10^{-4}$}}
\put(540,1018){\makebox(0,0)[r]{$10^{-6}$}}
\put(540,762){\makebox(0,0)[r]{$10^{-8}$}}
\put(540,507){\makebox(0,0)[r]{$10^{-10}$}}
\put(540,251){\makebox(0,0)[r]{$10^{-12}$}}
\end{picture}

%% file: f7.tex
\setlength{\unitlength}{0.1bp}
\special{!
/gnudict 40 dict def
gnudict begin
/Color false def
/Solid false def
/gnulinewidth 5.000 def
/vshift -33 def
/dl {10 mul} def
/hpt 31.5 def
/vpt 31.5 def
/M {moveto} bind def
/L {lineto} bind def
/R {rmoveto} bind def
/V {rlineto} bind def
/vpt2 vpt 2 mul def
/hpt2 hpt 2 mul def
/Lshow { currentpoint stroke M
  0 vshift R show } def
/Rshow { currentpoint stroke M
  dup stringwidth pop neg vshift R show } def
/Cshow { currentpoint stroke M
  dup stringwidth pop -2 div vshift R show } def
/DL { Color {setrgbcolor Solid {pop []} if 0 setdash }
 {pop pop pop Solid {pop []} if 0 setdash} ifelse } def
/BL { stroke gnulinewidth 2 mul setlinewidth } def
/AL { stroke gnulinewidth 2 div setlinewidth } def
/PL { stroke gnulinewidth setlinewidth } def
/LTb { BL [] 0 0 0 DL } def
/LTa { AL [1 dl 2 dl] 0 setdash 0 0 0 setrgbcolor } def
/LT0 { PL [] 0 1 0 DL } def
/LT1 { PL [4 dl 2 dl] 0 0 1 DL } def
/LT2 { PL [2 dl 3 dl] 1 0 0 DL } def
/LT3 { PL [1 dl 1.5 dl] 1 0 1 DL } def
/LT4 { PL [5 dl 2 dl 1 dl 2 dl] 0 1 1 DL } def
/LT5 { PL [4 dl 3 dl 1 dl 3 dl] 1 1 0 DL } def
/LT6 { PL [2 dl 2 dl 2 dl 4 dl] 0 0 0 DL } def
/LT7 { PL [2 dl 2 dl 2 dl 2 dl 2 dl 4 dl] 1 0.3 0 DL } def
/LT8 { PL [2 dl 2 dl 2 dl 2 dl 2 dl 2 dl 2 dl 4 dl] 0.5 0.5 0.5 DL } def
/P { stroke [] 0 setdash
  currentlinewidth 2 div sub M
  0 currentlinewidth V stroke } def
/D { stroke [] 0 setdash 2 copy vpt add M
  hpt neg vpt neg V hpt vpt neg V
  hpt vpt V hpt neg vpt V closepath stroke
  P } def
/A { stroke [] 0 setdash vpt sub M 0 vpt2 V
  currentpoint stroke M
  hpt neg vpt neg R hpt2 0 V stroke
  } def
/B { stroke [] 0 setdash 2 copy exch hpt sub exch vpt add M
  0 vpt2 neg V hpt2 0 V 0 vpt2 V
  hpt2 neg 0 V closepath stroke
  P } def
/C { stroke [] 0 setdash exch hpt sub exch vpt add M
  hpt2 vpt2 neg V currentpoint stroke M
  hpt2 neg 0 R hpt2 vpt2 V stroke } def
/T { stroke [] 0 setdash 2 copy vpt 1.12 mul add M
  hpt neg vpt -1.62 mul V
  hpt 2 mul 0 V
  hpt neg vpt 1.62 mul V closepath stroke
  P  } def
/S { 2 copy A C} def
end
}
\begin{picture}(2789,1836)(0,0)
\special{"
gnudict begin
gsave
50 50 translate
0.100 0.100 scale
0 setgray
/Helvetica findfont 100 scalefont setfont
newpath
-500.000000 -500.000000 translate
LTa
600 251 M
0 1534 V
LTb
600 452 M
63 0 V
1943 0 R
-63 0 V
600 835 M
63 0 V
1943 0 R
-63 0 V
600 1219 M
63 0 V
1943 0 R
-63 0 V
600 1602 M
63 0 V
1943 0 R
-63 0 V
600 251 M
0 63 V
0 1471 R
0 -63 V
832 251 M
0 63 V
0 1471 R
0 -63 V
1063 251 M
0 63 V
0 1471 R
0 -63 V
1295 251 M
0 63 V
0 1471 R
0 -63 V
1527 251 M
0 63 V
0 1471 R
0 -63 V
1758 251 M
0 63 V
0 1471 R
0 -63 V
1990 251 M
0 63 V
0 1471 R
0 -63 V
2221 251 M
0 63 V
0 1471 R
0 -63 V
2453 251 M
0 63 V
0 1471 R
0 -63 V
600 251 M
2006 0 V
0 1534 V
-2006 0 V
600 251 L
LT0
868 750 M
180 0 V
601 1664 M
2 4 V
2 4 V
2 5 V
2 4 V
2 4 V
2 4 V
2 3 V
2 4 V
2 4 V
2 3 V
2 4 V
2 3 V
2 3 V
2 3 V
2 3 V
2 3 V
2 2 V
2 3 V
2 2 V
2 2 V
2 1 V
2 2 V
2 1 V
2 1 V
2 1 V
2 1 V
2 0 V
2 0 V
2 0 V
2 0 V
2 0 V
2 -1 V
2 -1 V
2 0 V
2 -1 V
2 -1 V
2 -2 V
2 -1 V
2 -1 V
2 -2 V
2 -1 V
2 -2 V
2 -1 V
2 -2 V
2 -2 V
2 -1 V
2 -2 V
2 -2 V
2 -2 V
2 -2 V
2 -1 V
2 -2 V
2 -2 V
2 -2 V
2 -1 V
2 -2 V
2 -2 V
2 -2 V
2 -1 V
2 -2 V
2 -2 V
2 -2 V
2 -1 V
2 -2 V
2 -1 V
2 -2 V
2 -1 V
2 -2 V
2 -1 V
2 -2 V
2 -1 V
2 -2 V
2 -1 V
2 -1 V
2 -1 V
2 -2 V
2 -1 V
2 -1 V
2 -1 V
2 -1 V
2 -1 V
2 0 V
2 -1 V
2 -1 V
2 -1 V
2 -1 V
2 0 V
2 -1 V
2 -1 V
2 -1 V
2 0 V
2 -1 V
2 -1 V
2 -1 V
2 0 V
2 -1 V
2 -1 V
2 -1 V
2 -1 V
2 -1 V
2 -1 V
2 -1 V
2 0 V
2 -1 V
2 -2 V
2 -1 V
2 -1 V
2 -1 V
2 -1 V
2 -1 V
2 -1 V
2 -1 V
2 -1 V
2 -1 V
2 -1 V
2 -1 V
2 0 V
2 -1 V
2 -1 V
2 -1 V
2 -1 V
2 0 V
2 -1 V
2 -1 V
2 -1 V
2 0 V
2 -1 V
2 -1 V
2 0 V
2 -1 V
2 -1 V
2 -1 V
2 0 V
2 -1 V
2 -1 V
2 -1 V
2 -1 V
2 -1 V
2 -1 V
2 -1 V
2 -1 V
2 -1 V
2 -2 V
2 -1 V
2 -1 V
2 -2 V
2 -1 V
2 -2 V
2 -1 V
2 -2 V
2 -1 V
2 -2 V
2 -2 V
2 -1 V
2 -2 V
2 -1 V
2 -2 V
2 -2 V
2 -1 V
2 -2 V
2 -2 V
2 -2 V
2 -1 V
2 -2 V
2 -2 V
2 -1 V
2 -2 V
2 -2 V
2 -2 V
2 -2 V
2 -1 V
2 -2 V
2 -2 V
2 -2 V
2 -2 V
2 -2 V
2 -2 V
2 -2 V
2 -2 V
2 -2 V
2 -2 V
2 -2 V
2 -2 V
2 -3 V
2 -2 V
2 -2 V
2 -2 V
2 -2 V
2 -2 V
2 -2 V
2 -3 V
2 -2 V
2 -2 V
2 -2 V
2 -2 V
2 -2 V
2 -2 V
2 -2 V
2 -3 V
2 -2 V
2 -2 V
2 -2 V
2 -2 V
2 -2 V
2 -2 V
2 -2 V
2 -3 V
2 -2 V
2 -2 V
2 -2 V
2 -2 V
2 -2 V
2 -2 V
2 -2 V
2 -2 V
2 -2 V
2 -2 V
2 -3 V
2 -2 V
2 -2 V
2 -2 V
2 -2 V
2 -2 V
2 -2 V
2 -2 V
2 -2 V
2 -2 V
2 -2 V
2 -2 V
2 -2 V
2 -2 V
2 -2 V
2 -2 V
2 -2 V
3 -2 V
2 -2 V
2 -2 V
2 -2 V
2 -2 V
2 -2 V
2 -2 V
2 -1 V
2 -2 V
2 -2 V
2 -2 V
2 -2 V
2 -2 V
2 -2 V
2 -2 V
2 -2 V
2 -2 V
2 -2 V
2 -2 V
2 -2 V
2 -1 V
2 -2 V
2 -2 V
2 -2 V
2 -1 V
2 -2 V
2 -2 V
2 -2 V
2 -1 V
2 -2 V
2 -2 V
2 -1 V
2 -2 V
2 -2 V
2 -1 V
2 -2 V
2 -2 V
2 -2 V
2 -2 V
2 -2 V
2 -2 V
2 -2 V
2 -2 V
2 -2 V
2 -2 V
2 -3 V
2 -2 V
2 -3 V
2 -2 V
2 -2 V
2 -3 V
2 -3 V
2 -2 V
2 -3 V
2 -2 V
2 -3 V
2 -3 V
2 -2 V
2 -3 V
2 -3 V
2 -2 V
2 -3 V
2 -3 V
2 -3 V
2 -2 V
2 -3 V
2 -3 V
2 -2 V
2 -3 V
2 -3 V
2 -3 V
2 -3 V
2 -3 V
2 -2 V
2 -3 V
2 -3 V
2 -3 V
2 -3 V
2 -3 V
2 -4 V
2 -3 V
2 -3 V
2 -3 V
2 -3 V
2 -4 V
2 -3 V
2 -3 V
2 -3 V
2 -4 V
2 -3 V
2 -4 V
2 -3 V
2 -3 V
2 -4 V
2 -3 V
2 -4 V
2 -3 V
2 -3 V
2 -4 V
2 -3 V
2 -4 V
2 -3 V
2 -3 V
2 -4 V
2 -3 V
2 -3 V
2 -4 V
2 -3 V
2 -3 V
2 -4 V
2 -3 V
2 -3 V
2 -3 V
2 -4 V
2 -3 V
2 -3 V
2 -4 V
2 -3 V
2 -3 V
2 -3 V
2 -4 V
2 -3 V
2 -3 V
2 -3 V
2 -4 V
2 -3 V
2 -3 V
2 -3 V
2 -3 V
2 -3 V
2 -3 V
2 -3 V
2 -3 V
2 -3 V
2 -3 V
2 -3 V
2 -3 V
2 -3 V
2 -2 V
2 -3 V
2 -3 V
2 -3 V
2 -2 V
2 -3 V
2 -2 V
2 -3 V
2 -2 V
2 -3 V
2 -3 V
2 -2 V
2 -3 V
2 -2 V
2 -3 V
2 -2 V
2 -3 V
2 -2 V
2 -3 V
2 -3 V
2 -2 V
2 -3 V
2 -3 V
2 -3 V
2 -2 V
currentpoint stroke M
2 -3 V
2 -3 V
2 -3 V
2 -3 V
2 -3 V
2 -3 V
2 -3 V
2 -2 V
2 -3 V
2 -3 V
2 -3 V
2 -3 V
2 -3 V
2 -3 V
2 -2 V
2 -3 V
2 -3 V
2 -3 V
2 -2 V
2 -3 V
2 -3 V
2 -2 V
2 -3 V
2 -3 V
2 -2 V
2 -3 V
2 -3 V
2 -3 V
2 -2 V
2 -3 V
2 -3 V
2 -3 V
2 -3 V
2 -3 V
2 -3 V
2 -3 V
2 -3 V
2 -3 V
2 -3 V
2 -4 V
2 -3 V
2 -3 V
2 -3 V
2 -3 V
2 -3 V
2 -3 V
2 -3 V
2 -2 V
2 -3 V
2 -3 V
2 -3 V
2 -3 V
2 -2 V
2 -3 V
2 -3 V
2 -3 V
2 -2 V
2 -3 V
2 -3 V
2 -2 V
2 -3 V
2 -3 V
2 -3 V
2 -3 V
2 -3 V
2 -3 V
2 -3 V
2 -3 V
2 -3 V
2 -3 V
2 -4 V
2 -3 V
2 -3 V
2 -4 V
2 -3 V
2 -4 V
2 -3 V
2 -4 V
2 -3 V
2 -4 V
2 -3 V
2 -4 V
2 -4 V
2 -3 V
2 -4 V
2 -3 V
2 -4 V
2 -3 V
2 -4 V
2 -3 V
2 -4 V
2 -3 V
2 -3 V
2 -3 V
2 -4 V
2 -3 V
2 -3 V
2 -3 V
2 -3 V
2 -3 V
2 -3 V
2 -3 V
2 -2 V
2 -3 V
2 -3 V
2 -3 V
2 -2 V
2 -3 V
2 -3 V
2 -3 V
2 -2 V
2 -3 V
2 -3 V
2 -3 V
2 -2 V
2 -3 V
2 -3 V
2 -3 V
2 -2 V
2 -3 V
2 -3 V
2 -3 V
2 -3 V
2 -3 V
2 -2 V
2 -3 V
2 -3 V
2 -3 V
2 -2 V
2 -3 V
2 -3 V
2 -2 V
2 -3 V
2 -3 V
2 -2 V
2 -3 V
2 -2 V
2 -2 V
2 -3 V
2 -2 V
2 -2 V
2 -2 V
2 -2 V
2 -3 V
2 -2 V
2 -2 V
2 -2 V
2 -2 V
2 -2 V
2 -2 V
2 -2 V
2 -1 V
2 -2 V
2 -2 V
2 -2 V
2 -2 V
2 -2 V
2 -2 V
2 -1 V
2 -2 V
2 -2 V
2 -2 V
2 -1 V
2 -2 V
2 -2 V
2 -2 V
2 -1 V
2 -2 V
2 -2 V
2 -1 V
2 -2 V
2 -1 V
2 -2 V
2 -1 V
2 -2 V
2 -1 V
2 -2 V
2 -1 V
2 -2 V
2 -1 V
2 -2 V
2 -1 V
2 -1 V
2 -2 V
2 -1 V
2 -2 V
2 -1 V
2 -1 V
2 -2 V
2 -1 V
2 -1 V
2 -1 V
2 -2 V
2 -1 V
2 -1 V
2 -1 V
2 -2 V
2 -1 V
2 -1 V
2 -1 V
2 -2 V
2 -1 V
2 -1 V
2 -1 V
2 -2 V
2 -1 V
2 -1 V
2 -2 V
2 -1 V
2 -1 V
2 -2 V
2 -1 V
2 -1 V
2 -2 V
2 -1 V
2 -1 V
2 -2 V
2 -1 V
2 -2 V
2 -1 V
2 -1 V
2 -1 V
2 -2 V
2 -1 V
2 -1 V
2 -2 V
2 -1 V
2 -1 V
2 -1 V
2 -1 V
2 -2 V
2 -1 V
2 -1 V
2 -1 V
2 -1 V
2 -1 V
2 -1 V
2 -1 V
2 -1 V
2 -1 V
2 -1 V
2 -1 V
2 -1 V
2 -1 V
2 -2 V
2 -1 V
2 -1 V
2 -1 V
2 -1 V
2 -1 V
2 -1 V
2 -1 V
2 -1 V
2 -1 V
2 -1 V
2 -1 V
2 -1 V
2 -1 V
2 -1 V
2 -1 V
2 -1 V
2 -1 V
2 -1 V
2 -1 V
2 -2 V
2 -1 V
2 -1 V
2 -1 V
2 -1 V
2 -1 V
2 0 V
2 -1 V
2 -1 V
2 -1 V
2 -1 V
2 -1 V
2 -1 V
2 0 V
2 -1 V
2 -1 V
2 -1 V
2 0 V
2 -1 V
2 0 V
2 -1 V
2 -1 V
2 0 V
2 -1 V
2 0 V
2 -1 V
2 0 V
2 -1 V
2 0 V
2 -1 V
2 0 V
2 -1 V
2 0 V
2 -1 V
2 0 V
2 -1 V
2 0 V
2 -1 V
2 0 V
2 -1 V
2 0 V
2 -1 V
2 0 V
2 -1 V
2 0 V
2 -1 V
2 0 V
2 -1 V
2 0 V
2 -1 V
2 -1 V
2 0 V
2 -1 V
2 0 V
2 -1 V
2 -1 V
2 0 V
2 -1 V
2 0 V
2 -1 V
2 -1 V
3 0 V
2 -1 V
2 -1 V
2 0 V
2 -1 V
2 0 V
2 -1 V
2 -1 V
2 0 V
2 -1 V
2 0 V
2 -1 V
2 0 V
2 -1 V
2 0 V
2 -1 V
2 0 V
2 -1 V
2 0 V
2 -1 V
2 0 V
2 0 V
2 -1 V
2 0 V
2 -1 V
2 0 V
2 -1 V
2 0 V
2 -1 V
2 0 V
2 -1 V
2 0 V
2 -1 V
2 0 V
2 -1 V
2 0 V
2 -1 V
2 0 V
2 -1 V
2 0 V
2 -1 V
2 0 V
2 -1 V
2 0 V
2 -1 V
2 0 V
2 0 V
2 -1 V
2 0 V
2 -1 V
2 0 V
2 0 V
2 -1 V
2 0 V
2 0 V
2 -1 V
2 0 V
2 0 V
2 -1 V
2 0 V
2 0 V
2 -1 V
2 0 V
2 0 V
2 0 V
2 -1 V
2 0 V
2 0 V
2 -1 V
2 0 V
2 0 V
2 -1 V
2 0 V
2 0 V
2 -1 V
currentpoint stroke M
2 0 V
2 -1 V
2 0 V
2 -1 V
2 0 V
2 -1 V
2 0 V
2 -1 V
2 0 V
2 -1 V
2 0 V
2 -1 V
2 0 V
2 -1 V
2 -1 V
2 0 V
2 -1 V
2 0 V
2 -1 V
2 0 V
2 -1 V
2 0 V
2 -1 V
2 0 V
2 -1 V
2 0 V
2 -1 V
2 0 V
2 0 V
2 -1 V
2 0 V
2 -1 V
2 0 V
2 -1 V
2 0 V
2 0 V
2 -1 V
2 0 V
2 -1 V
2 0 V
2 -1 V
2 0 V
2 -1 V
2 0 V
2 -1 V
2 0 V
2 -1 V
2 0 V
2 -1 V
2 0 V
2 -1 V
2 0 V
2 -1 V
2 0 V
2 0 V
2 -1 V
2 0 V
2 0 V
2 -1 V
2 0 V
2 0 V
2 -1 V
2 0 V
2 0 V
2 0 V
2 0 V
2 0 V
2 -1 V
2 0 V
2 0 V
2 0 V
2 0 V
2 0 V
2 0 V
2 0 V
2 0 V
2 0 V
2 0 V
2 0 V
2 0 V
2 0 V
2 0 V
2 0 V
2 0 V
2 0 V
2 0 V
2 0 V
2 0 V
2 0 V
2 0 V
2 0 V
2 0 V
2 0 V
2 0 V
2 0 V
2 0 V
2 0 V
2 0 V
2 0 V
2 -1 V
2 0 V
2 0 V
2 0 V
2 0 V
2 0 V
2 0 V
2 -1 V
2 0 V
2 0 V
2 0 V
2 0 V
2 0 V
2 -1 V
2 0 V
2 0 V
2 0 V
2 0 V
2 0 V
2 -1 V
2 0 V
2 0 V
2 0 V
2 0 V
2 0 V
2 0 V
2 0 V
2 0 V
2 -1 V
2 0 V
2 0 V
2 0 V
2 0 V
2 0 V
2 0 V
2 0 V
2 0 V
2 0 V
2 0 V
2 0 V
2 0 V
2 0 V
2 -1 V
2 0 V
2 0 V
2 0 V
2 0 V
2 0 V
2 0 V
2 0 V
2 0 V
2 0 V
2 0 V
2 0 V
2 0 V
2 0 V
2 0 V
2 0 V
2 0 V
2 0 V
2 0 V
2 0 V
2 0 V
2 1 V
2 0 V
2 0 V
2 0 V
2 0 V
2 0 V
2 0 V
2 0 V
2 0 V
2 0 V
2 0 V
2 1 V
2 0 V
2 0 V
2 0 V
2 0 V
2 1 V
2 0 V
2 0 V
2 0 V
2 0 V
2 1 V
2 0 V
2 0 V
2 0 V
2 0 V
2 1 V
2 0 V
2 0 V
2 0 V
2 0 V
2 0 V
2 1 V
2 0 V
2 0 V
2 0 V
2 0 V
2 0 V
2 0 V
2 0 V
LT1
868 650 M
180 0 V
601 1244 M
2 0 V
2 0 V
2 1 V
2 0 V
2 0 V
2 1 V
2 0 V
2 0 V
2 1 V
2 0 V
2 0 V
2 0 V
2 1 V
2 0 V
2 0 V
2 0 V
2 0 V
2 0 V
2 0 V
2 1 V
2 0 V
2 0 V
2 0 V
2 0 V
2 0 V
2 1 V
2 0 V
2 0 V
2 0 V
2 1 V
2 0 V
2 0 V
2 1 V
2 0 V
2 1 V
2 0 V
2 1 V
2 1 V
2 0 V
2 1 V
2 0 V
2 1 V
2 1 V
2 0 V
2 1 V
2 0 V
2 1 V
2 0 V
2 1 V
2 0 V
2 1 V
2 0 V
2 0 V
2 0 V
2 0 V
2 1 V
2 0 V
2 -1 V
2 0 V
2 0 V
2 0 V
2 0 V
2 0 V
2 -1 V
2 0 V
2 0 V
2 0 V
2 -1 V
2 0 V
2 0 V
2 -1 V
2 0 V
2 0 V
2 0 V
2 0 V
2 -1 V
2 0 V
2 0 V
2 0 V
2 0 V
2 0 V
2 0 V
2 0 V
2 0 V
2 0 V
2 0 V
2 0 V
2 -1 V
2 0 V
2 0 V
2 0 V
2 0 V
2 -1 V
2 0 V
2 0 V
2 -1 V
2 0 V
2 -1 V
2 0 V
2 -1 V
2 -1 V
2 0 V
2 -1 V
2 -1 V
2 -1 V
2 0 V
2 -1 V
2 -1 V
2 -1 V
2 0 V
2 -1 V
2 -1 V
2 0 V
2 -1 V
2 0 V
2 -1 V
2 0 V
2 -1 V
2 0 V
2 0 V
2 -1 V
2 0 V
2 0 V
2 0 V
2 0 V
2 -1 V
2 0 V
2 0 V
2 0 V
2 0 V
2 0 V
2 0 V
2 0 V
2 -1 V
2 0 V
2 0 V
2 0 V
2 0 V
2 0 V
2 -1 V
2 0 V
2 0 V
2 0 V
2 -1 V
2 0 V
2 0 V
2 -1 V
2 0 V
2 -1 V
2 0 V
2 0 V
2 -1 V
2 0 V
2 -1 V
2 0 V
2 0 V
2 -1 V
2 0 V
2 -1 V
2 0 V
2 0 V
2 -1 V
2 0 V
2 0 V
2 -1 V
2 0 V
2 0 V
2 -1 V
2 0 V
2 0 V
2 -1 V
2 0 V
2 0 V
2 -1 V
2 0 V
2 -1 V
2 0 V
2 0 V
2 -1 V
2 0 V
2 -1 V
2 0 V
2 0 V
2 -1 V
2 0 V
2 -1 V
2 0 V
2 0 V
2 -1 V
2 0 V
2 -1 V
2 0 V
2 0 V
2 0 V
2 -1 V
2 0 V
2 0 V
2 0 V
2 -1 V
2 0 V
2 0 V
2 0 V
2 0 V
2 -1 V
2 0 V
2 0 V
2 0 V
2 0 V
2 -1 V
2 0 V
2 0 V
2 0 V
2 -1 V
2 0 V
2 0 V
2 0 V
2 -1 V
2 0 V
2 0 V
2 -1 V
2 0 V
2 -1 V
2 0 V
2 0 V
2 -1 V
2 0 V
2 0 V
2 -1 V
2 0 V
2 0 V
2 -1 V
2 0 V
2 -1 V
2 0 V
3 0 V
2 0 V
2 -1 V
2 0 V
2 0 V
2 0 V
2 -1 V
2 0 V
2 0 V
2 0 V
2 0 V
2 0 V
2 0 V
2 0 V
2 0 V
2 0 V
2 0 V
2 0 V
2 0 V
2 0 V
2 0 V
2 0 V
2 0 V
2 0 V
2 0 V
2 0 V
2 0 V
2 1 V
2 0 V
2 0 V
2 0 V
2 0 V
2 1 V
2 0 V
2 0 V
2 0 V
2 0 V
2 0 V
2 0 V
2 0 V
2 0 V
2 0 V
2 0 V
2 0 V
2 0 V
2 0 V
2 0 V
2 0 V
2 0 V
2 0 V
2 -1 V
2 0 V
2 0 V
2 0 V
2 0 V
2 0 V
2 0 V
2 0 V
2 0 V
2 -1 V
2 0 V
2 0 V
2 0 V
2 0 V
2 0 V
2 0 V
2 0 V
2 0 V
2 0 V
2 0 V
2 0 V
2 0 V
2 0 V
2 0 V
2 0 V
2 0 V
2 0 V
2 -1 V
2 0 V
2 0 V
2 0 V
2 0 V
2 0 V
2 0 V
2 -1 V
2 0 V
2 0 V
2 0 V
2 0 V
2 0 V
2 0 V
2 -1 V
2 0 V
2 0 V
2 0 V
2 0 V
2 0 V
2 0 V
2 0 V
2 0 V
2 0 V
2 0 V
2 0 V
2 0 V
2 0 V
2 0 V
2 0 V
2 0 V
2 1 V
2 0 V
2 0 V
2 0 V
2 0 V
2 0 V
2 0 V
2 0 V
2 0 V
2 1 V
2 0 V
2 0 V
2 0 V
2 0 V
2 0 V
2 0 V
2 0 V
2 0 V
2 0 V
2 0 V
2 0 V
2 0 V
2 0 V
2 0 V
2 0 V
2 0 V
2 -1 V
2 0 V
2 0 V
2 0 V
2 0 V
2 0 V
2 0 V
2 0 V
2 -1 V
2 0 V
2 0 V
2 0 V
2 0 V
2 0 V
2 0 V
2 0 V
2 0 V
2 0 V
2 0 V
2 0 V
2 0 V
2 0 V
2 0 V
2 0 V
2 0 V
2 0 V
2 0 V
2 0 V
2 0 V
currentpoint stroke M
2 0 V
2 0 V
2 0 V
2 0 V
2 0 V
2 0 V
2 0 V
2 0 V
2 0 V
2 0 V
2 0 V
2 0 V
2 0 V
2 0 V
2 0 V
2 0 V
2 0 V
2 0 V
2 0 V
2 0 V
2 0 V
2 0 V
2 0 V
2 0 V
2 0 V
2 0 V
2 0 V
2 0 V
2 0 V
2 0 V
2 0 V
2 0 V
2 0 V
2 0 V
2 0 V
2 0 V
2 0 V
2 0 V
2 0 V
2 0 V
2 0 V
2 0 V
2 0 V
2 0 V
2 0 V
2 0 V
2 0 V
2 0 V
2 0 V
2 0 V
2 0 V
2 0 V
2 1 V
2 0 V
2 0 V
2 0 V
2 0 V
2 0 V
2 0 V
2 1 V
2 0 V
2 0 V
2 0 V
2 0 V
2 0 V
2 1 V
2 0 V
2 0 V
2 0 V
2 0 V
2 0 V
2 0 V
2 0 V
2 1 V
2 0 V
2 0 V
2 0 V
2 0 V
2 0 V
2 0 V
2 0 V
2 -1 V
2 0 V
2 0 V
2 0 V
2 0 V
2 0 V
2 0 V
2 0 V
2 0 V
2 0 V
2 0 V
2 -1 V
2 0 V
2 0 V
2 0 V
2 0 V
2 0 V
2 0 V
2 -1 V
2 0 V
2 0 V
2 0 V
2 0 V
2 0 V
2 -1 V
2 0 V
2 0 V
2 0 V
2 0 V
2 0 V
2 -1 V
2 0 V
2 0 V
2 0 V
2 0 V
2 0 V
2 0 V
2 -1 V
2 0 V
2 0 V
2 0 V
2 0 V
2 0 V
2 0 V
2 0 V
2 0 V
2 0 V
2 1 V
2 0 V
2 0 V
2 0 V
2 0 V
2 0 V
2 0 V
2 0 V
2 0 V
2 1 V
2 0 V
2 0 V
2 0 V
2 0 V
2 0 V
2 0 V
2 0 V
2 0 V
2 0 V
2 0 V
2 0 V
2 0 V
2 0 V
2 0 V
2 0 V
2 -1 V
2 0 V
2 0 V
2 0 V
2 0 V
2 0 V
2 0 V
2 0 V
2 0 V
2 0 V
2 0 V
2 0 V
2 0 V
2 0 V
2 0 V
2 0 V
2 0 V
2 0 V
2 1 V
2 0 V
2 0 V
2 0 V
2 0 V
2 0 V
2 1 V
2 0 V
2 0 V
2 0 V
2 0 V
2 0 V
2 1 V
2 0 V
2 0 V
2 0 V
2 0 V
2 0 V
2 0 V
2 0 V
2 0 V
2 0 V
2 0 V
2 0 V
2 0 V
2 1 V
2 0 V
2 0 V
2 0 V
2 0 V
2 0 V
2 0 V
2 0 V
2 0 V
2 0 V
2 0 V
2 0 V
2 0 V
2 0 V
2 0 V
2 -1 V
2 0 V
2 0 V
2 0 V
2 0 V
2 0 V
2 0 V
2 0 V
2 0 V
2 0 V
2 0 V
2 -1 V
2 0 V
2 0 V
2 0 V
2 0 V
2 0 V
2 0 V
2 0 V
2 0 V
2 0 V
2 0 V
2 0 V
2 0 V
2 0 V
2 0 V
2 0 V
2 0 V
2 0 V
2 0 V
2 0 V
2 0 V
2 0 V
2 0 V
2 0 V
2 1 V
2 0 V
2 0 V
2 0 V
2 0 V
2 0 V
2 0 V
2 0 V
2 0 V
2 0 V
2 0 V
2 0 V
2 0 V
2 0 V
2 0 V
2 0 V
2 0 V
2 0 V
2 0 V
2 -1 V
2 0 V
2 0 V
2 0 V
2 0 V
2 0 V
2 0 V
2 0 V
2 0 V
2 0 V
2 0 V
2 0 V
2 -1 V
2 0 V
2 0 V
2 0 V
2 0 V
2 0 V
2 0 V
2 0 V
2 0 V
2 0 V
2 0 V
2 0 V
2 0 V
2 0 V
2 0 V
2 0 V
2 0 V
2 0 V
2 0 V
2 0 V
2 0 V
2 0 V
2 0 V
2 0 V
2 0 V
2 0 V
2 0 V
2 0 V
2 0 V
2 0 V
2 0 V
2 1 V
2 0 V
2 0 V
2 0 V
2 0 V
2 0 V
2 0 V
2 1 V
2 0 V
2 0 V
2 0 V
2 0 V
2 1 V
2 0 V
2 0 V
2 0 V
2 0 V
3 0 V
2 0 V
2 1 V
2 0 V
2 0 V
2 0 V
2 0 V
2 0 V
2 0 V
2 0 V
2 0 V
2 0 V
2 0 V
2 0 V
2 0 V
2 0 V
2 0 V
2 0 V
2 0 V
2 0 V
2 0 V
2 0 V
2 0 V
2 0 V
2 0 V
2 0 V
2 0 V
2 0 V
2 0 V
2 -1 V
2 0 V
2 0 V
2 0 V
2 0 V
2 0 V
2 0 V
2 0 V
2 0 V
2 0 V
2 -1 V
2 0 V
2 0 V
2 0 V
2 0 V
2 0 V
2 0 V
2 0 V
2 0 V
2 0 V
2 0 V
2 0 V
2 0 V
2 0 V
2 0 V
2 0 V
2 0 V
2 0 V
2 0 V
2 0 V
2 0 V
2 0 V
2 0 V
2 0 V
2 0 V
2 0 V
2 0 V
2 0 V
2 0 V
2 0 V
2 0 V
2 0 V
2 0 V
2 0 V
2 0 V
2 0 V
currentpoint stroke M
2 -1 V
2 0 V
2 0 V
2 0 V
2 0 V
2 0 V
2 0 V
2 0 V
2 0 V
2 0 V
2 0 V
2 0 V
2 0 V
2 0 V
2 0 V
2 0 V
2 0 V
2 0 V
2 0 V
2 0 V
2 1 V
2 0 V
2 0 V
2 -1 V
2 0 V
2 0 V
2 0 V
2 0 V
2 0 V
2 0 V
2 0 V
2 0 V
2 0 V
2 0 V
2 -1 V
2 0 V
2 0 V
2 0 V
2 0 V
2 0 V
2 0 V
2 -1 V
2 0 V
2 0 V
2 0 V
2 0 V
2 0 V
2 0 V
2 0 V
2 0 V
2 0 V
2 0 V
2 0 V
2 0 V
2 0 V
2 1 V
2 0 V
2 0 V
2 0 V
2 0 V
2 0 V
2 1 V
2 0 V
2 0 V
2 0 V
2 0 V
2 0 V
2 0 V
2 0 V
2 1 V
2 0 V
2 0 V
2 0 V
2 0 V
2 0 V
2 0 V
2 0 V
2 0 V
2 0 V
2 0 V
2 0 V
2 0 V
2 0 V
2 0 V
2 -1 V
2 0 V
2 0 V
2 0 V
2 0 V
2 0 V
2 0 V
2 -1 V
2 0 V
2 0 V
2 0 V
2 0 V
2 0 V
2 -1 V
2 0 V
2 0 V
2 0 V
2 0 V
2 0 V
2 0 V
2 0 V
2 0 V
2 -1 V
2 0 V
2 0 V
2 0 V
2 0 V
2 0 V
2 0 V
2 0 V
2 0 V
2 0 V
2 0 V
2 0 V
2 0 V
2 0 V
2 0 V
2 0 V
2 0 V
2 0 V
2 0 V
2 0 V
2 0 V
2 0 V
2 0 V
2 0 V
2 0 V
2 0 V
2 0 V
2 0 V
2 1 V
2 0 V
2 0 V
2 0 V
2 0 V
2 0 V
2 0 V
2 0 V
2 1 V
2 0 V
2 0 V
2 0 V
2 0 V
2 0 V
2 0 V
2 1 V
2 0 V
2 0 V
2 0 V
2 0 V
2 0 V
2 0 V
2 0 V
2 0 V
2 0 V
2 0 V
2 0 V
2 0 V
2 -1 V
2 0 V
2 0 V
2 0 V
2 0 V
2 -1 V
2 0 V
2 0 V
2 -1 V
2 0 V
2 0 V
2 -1 V
2 -1 V
2 0 V
2 -1 V
2 -1 V
2 -1 V
2 -1 V
2 -1 V
2 -2 V
2 -1 V
2 -2 V
2 -1 V
2 -2 V
2 -2 V
2 -2 V
2 -2 V
2 -2 V
2 -2 V
2 -2 V
2 -3 V
2 -2 V
2 -3 V
2 -2 V
2 -3 V
2 -3 V
2 -2 V
2 -3 V
2 -3 V
2 -3 V
LT2
868 550 M
180 0 V
601 1257 M
2 0 V
2 0 V
2 0 V
2 1 V
2 0 V
2 0 V
2 1 V
2 0 V
2 0 V
2 0 V
2 1 V
2 0 V
2 0 V
2 0 V
2 1 V
2 0 V
2 0 V
2 0 V
2 1 V
2 0 V
2 0 V
2 0 V
2 1 V
2 0 V
2 0 V
2 0 V
2 1 V
2 0 V
2 0 V
2 0 V
2 0 V
2 0 V
2 1 V
2 0 V
2 0 V
2 0 V
2 0 V
2 0 V
2 0 V
2 0 V
2 0 V
2 0 V
2 -1 V
2 0 V
2 0 V
2 0 V
2 -1 V
2 0 V
2 -1 V
2 0 V
2 -1 V
2 0 V
2 -1 V
2 -1 V
2 0 V
2 -1 V
2 -1 V
2 0 V
2 -1 V
2 -1 V
2 -1 V
2 -1 V
2 -1 V
2 -1 V
2 -1 V
2 -1 V
2 -1 V
2 -1 V
2 -1 V
2 -1 V
2 -1 V
2 -1 V
2 -2 V
2 -1 V
2 -1 V
2 -2 V
2 -1 V
2 -1 V
2 -2 V
2 -1 V
2 -2 V
2 -1 V
2 -2 V
2 -1 V
2 -2 V
2 -2 V
2 -2 V
2 -1 V
2 -2 V
2 -2 V
2 -2 V
2 -2 V
2 -2 V
2 -2 V
2 -2 V
2 -2 V
2 -2 V
2 -2 V
2 -2 V
2 -2 V
2 -2 V
2 -2 V
2 -2 V
2 -3 V
2 -2 V
2 -2 V
2 -2 V
2 -2 V
2 -2 V
2 -2 V
2 -2 V
2 -2 V
2 -2 V
2 -2 V
2 -2 V
2 -2 V
2 -2 V
2 -2 V
2 -2 V
2 -2 V
2 -2 V
2 -1 V
2 -2 V
2 -2 V
2 -2 V
2 -1 V
2 -2 V
2 -2 V
2 -1 V
2 -2 V
2 -1 V
2 -2 V
2 -1 V
2 -1 V
2 -1 V
2 -2 V
2 -1 V
2 -1 V
2 -1 V
2 -1 V
2 -1 V
2 -1 V
2 -1 V
2 -1 V
2 -1 V
2 0 V
2 -1 V
2 -1 V
2 -1 V
2 -1 V
2 -1 V
2 -1 V
2 -1 V
2 0 V
2 -1 V
2 -1 V
2 -1 V
2 -1 V
2 -1 V
2 -1 V
2 -1 V
2 -1 V
2 -2 V
2 -1 V
2 -1 V
2 -1 V
2 -1 V
2 -1 V
2 -2 V
2 -1 V
2 -1 V
2 -1 V
2 -2 V
2 -1 V
2 -1 V
2 -1 V
2 -1 V
2 -2 V
2 -1 V
2 -1 V
2 -1 V
2 -2 V
2 -1 V
2 -1 V
2 -1 V
2 -1 V
2 -2 V
2 -1 V
2 -1 V
2 -1 V
2 -1 V
2 -2 V
2 -1 V
2 -1 V
2 -1 V
2 -2 V
2 -1 V
2 -1 V
2 -1 V
2 -2 V
2 -1 V
2 -1 V
2 -2 V
2 -1 V
2 -1 V
2 -2 V
2 -1 V
2 -1 V
2 -2 V
2 -1 V
2 -2 V
2 -1 V
2 -1 V
2 -2 V
2 -1 V
2 -2 V
2 -1 V
2 -2 V
2 -1 V
2 -1 V
2 -2 V
2 -1 V
2 -2 V
2 -1 V
2 -2 V
2 -1 V
2 -2 V
2 -1 V
2 -2 V
2 -1 V
2 -2 V
2 -2 V
2 -1 V
2 -2 V
3 -1 V
2 -2 V
2 -1 V
2 -2 V
2 -1 V
2 -2 V
2 -1 V
2 -2 V
2 -1 V
2 -2 V
2 -1 V
2 -2 V
2 -1 V
2 -2 V
2 -1 V
2 -1 V
2 -2 V
2 -1 V
2 -1 V
2 -2 V
2 -1 V
2 -2 V
2 -1 V
2 -1 V
2 -2 V
2 -1 V
2 -2 V
2 -1 V
2 -1 V
2 -2 V
2 -1 V
2 -2 V
2 -2 V
2 -1 V
2 -2 V
2 -1 V
2 -2 V
2 -2 V
2 -1 V
2 -2 V
2 -2 V
2 -1 V
2 -2 V
2 -1 V
2 -2 V
2 -2 V
2 -1 V
2 -2 V
2 -2 V
2 -1 V
2 -2 V
2 -1 V
2 -2 V
2 -2 V
2 -1 V
2 -2 V
2 -1 V
2 -2 V
2 -1 V
2 -2 V
2 -1 V
2 -2 V
2 -1 V
2 -2 V
2 -1 V
2 -2 V
2 -1 V
2 -2 V
2 -1 V
2 -2 V
2 -1 V
2 -2 V
2 -1 V
2 -1 V
2 -2 V
2 -1 V
2 -2 V
2 -1 V
2 -1 V
2 -2 V
2 -1 V
2 -1 V
2 -2 V
2 -1 V
2 -1 V
2 -2 V
2 -1 V
2 -2 V
2 -1 V
2 -1 V
2 -2 V
2 -1 V
2 -1 V
2 -2 V
2 -1 V
2 -1 V
2 -2 V
2 -1 V
2 -2 V
2 -1 V
2 -2 V
2 -1 V
2 -2 V
2 -1 V
2 -2 V
2 -1 V
2 -2 V
2 -1 V
2 -2 V
2 -1 V
2 -2 V
2 -1 V
2 -2 V
2 -1 V
2 -2 V
2 -2 V
2 -1 V
2 -2 V
2 -1 V
2 -2 V
2 -1 V
2 -2 V
2 -1 V
2 -2 V
2 -1 V
2 -2 V
2 -1 V
2 -1 V
2 -2 V
2 -1 V
2 -2 V
2 -1 V
2 -2 V
2 -1 V
2 -2 V
2 -1 V
2 -1 V
2 -2 V
2 -1 V
2 -2 V
2 -1 V
2 -2 V
2 -1 V
2 -2 V
2 -1 V
2 -1 V
2 -2 V
2 -1 V
2 -2 V
2 -1 V
2 -2 V
2 -1 V
2 -2 V
2 -1 V
2 -1 V
2 -2 V
2 -1 V
2 -1 V
2 -1 V
2 -2 V
2 -1 V
2 -1 V
2 -1 V
currentpoint stroke M
2 -1 V
2 -1 V
2 -1 V
2 -1 V
2 -1 V
2 -1 V
2 -1 V
2 -1 V
2 -1 V
2 -1 V
2 -1 V
2 -2 V
2 -1 V
2 -1 V
2 -1 V
2 -1 V
2 -1 V
2 -1 V
2 -1 V
2 -2 V
2 -1 V
2 -1 V
2 -1 V
2 -2 V
2 -1 V
2 -1 V
2 -1 V
2 -1 V
2 -2 V
2 -1 V
2 -1 V
2 -1 V
2 -2 V
2 -1 V
2 -1 V
2 -1 V
2 -1 V
2 -2 V
2 -1 V
2 -1 V
2 -1 V
2 -1 V
2 -2 V
2 -1 V
2 -1 V
2 -1 V
2 -1 V
2 -1 V
2 -1 V
2 -1 V
2 -1 V
2 -1 V
2 -1 V
2 -1 V
2 -1 V
2 -1 V
2 -1 V
2 -1 V
2 -1 V
2 -1 V
2 -1 V
2 -1 V
2 -1 V
2 -1 V
2 -1 V
2 -1 V
2 0 V
2 -1 V
2 -1 V
2 -1 V
2 -1 V
2 -1 V
2 -1 V
2 0 V
2 -1 V
2 -1 V
2 -1 V
2 -1 V
2 -1 V
2 -1 V
2 0 V
2 -1 V
2 -1 V
2 -1 V
2 -1 V
2 -1 V
2 -1 V
2 -1 V
2 -2 V
2 -1 V
2 -1 V
2 -1 V
2 -1 V
2 -1 V
2 -2 V
2 -1 V
2 -1 V
2 -1 V
2 -2 V
2 -1 V
2 -1 V
2 -1 V
2 -1 V
2 -1 V
2 -1 V
2 -1 V
2 -1 V
2 -1 V
2 -1 V
2 -1 V
2 -1 V
2 -1 V
2 -1 V
2 -1 V
2 -1 V
2 0 V
2 -1 V
2 -1 V
2 -1 V
2 -1 V
2 0 V
2 -1 V
2 -1 V
2 -1 V
2 -1 V
2 0 V
2 -1 V
2 -1 V
2 -1 V
2 0 V
2 -1 V
2 -1 V
2 -1 V
2 0 V
2 -1 V
2 -1 V
2 -1 V
2 -1 V
2 0 V
2 -1 V
2 -1 V
2 -1 V
2 -1 V
2 0 V
2 -1 V
2 -1 V
2 -1 V
2 -1 V
2 0 V
2 -1 V
2 -1 V
2 -1 V
2 -1 V
2 -1 V
2 0 V
2 -1 V
2 -1 V
2 -1 V
2 -1 V
2 0 V
2 -1 V
2 -1 V
2 -1 V
2 -1 V
2 -1 V
2 0 V
2 -1 V
2 -1 V
2 -1 V
2 -1 V
2 0 V
2 -1 V
2 -1 V
2 -1 V
2 -1 V
2 -1 V
2 -1 V
2 -1 V
2 0 V
2 -1 V
2 -1 V
2 -1 V
2 -1 V
2 -1 V
2 -1 V
2 -1 V
2 -1 V
2 -1 V
2 -1 V
2 -1 V
2 -1 V
2 -1 V
2 -1 V
2 -1 V
2 -1 V
2 -1 V
2 0 V
2 -1 V
2 -1 V
2 -1 V
2 -1 V
2 -1 V
2 -1 V
2 0 V
2 -1 V
2 -1 V
2 -1 V
2 0 V
2 -1 V
2 -1 V
2 -1 V
2 0 V
2 -1 V
2 -1 V
2 0 V
2 -1 V
2 -1 V
2 0 V
2 -1 V
2 -1 V
2 -1 V
2 -1 V
2 0 V
2 -1 V
2 -1 V
2 -1 V
2 -1 V
2 -1 V
2 0 V
2 -1 V
2 -1 V
2 -1 V
2 -1 V
2 -1 V
2 -1 V
2 -1 V
2 0 V
2 -1 V
2 -1 V
2 -1 V
2 -1 V
2 -1 V
2 -1 V
2 -1 V
2 -1 V
2 -1 V
2 0 V
2 -1 V
2 -1 V
2 -1 V
2 -1 V
2 -1 V
2 -1 V
2 -1 V
2 -1 V
2 -1 V
2 -1 V
2 -1 V
2 -1 V
2 -1 V
2 0 V
2 -1 V
2 -1 V
2 0 V
2 -1 V
2 -1 V
2 0 V
2 -1 V
2 -1 V
2 0 V
2 -1 V
2 0 V
2 -1 V
2 0 V
2 -1 V
2 0 V
2 -1 V
2 0 V
2 -1 V
2 -1 V
2 0 V
2 -1 V
2 0 V
2 -1 V
2 0 V
2 -1 V
2 -1 V
2 0 V
2 -1 V
2 -1 V
2 0 V
2 -1 V
2 0 V
2 -1 V
2 -1 V
2 0 V
2 -1 V
2 0 V
2 -1 V
2 -1 V
2 0 V
2 -1 V
2 0 V
2 -1 V
2 0 V
2 -1 V
2 0 V
2 -1 V
2 0 V
2 0 V
2 -1 V
2 0 V
2 -1 V
2 0 V
2 -1 V
2 0 V
2 0 V
2 -1 V
2 0 V
2 -1 V
2 0 V
2 -1 V
2 0 V
2 0 V
2 -1 V
3 0 V
2 -1 V
2 0 V
2 -1 V
2 -1 V
2 0 V
2 -1 V
2 0 V
2 -1 V
2 0 V
2 -1 V
2 0 V
2 -1 V
2 -1 V
2 0 V
2 -1 V
2 0 V
2 -1 V
2 0 V
2 -1 V
2 0 V
2 -1 V
2 0 V
2 -1 V
2 0 V
2 -1 V
2 0 V
2 -1 V
2 0 V
2 0 V
2 -1 V
2 0 V
2 -1 V
2 0 V
2 0 V
2 -1 V
2 0 V
2 -1 V
2 0 V
2 -1 V
2 0 V
2 0 V
2 -1 V
2 0 V
2 0 V
2 -1 V
2 0 V
2 0 V
2 -1 V
2 0 V
2 0 V
2 -1 V
2 0 V
2 0 V
2 -1 V
2 0 V
2 0 V
2 0 V
2 -1 V
2 0 V
2 0 V
2 0 V
2 -1 V
2 0 V
2 0 V
2 0 V
2 -1 V
2 0 V
2 -1 V
2 0 V
2 0 V
2 -1 V
2 0 V
2 -1 V
2 0 V
currentpoint stroke M
2 -1 V
2 0 V
2 -1 V
2 0 V
2 -1 V
2 -1 V
2 0 V
2 -1 V
2 0 V
2 -1 V
2 0 V
2 -1 V
2 -1 V
2 0 V
2 -1 V
2 0 V
2 -1 V
2 -1 V
2 0 V
2 -1 V
2 0 V
2 -1 V
2 -1 V
2 0 V
2 -1 V
2 0 V
2 -1 V
2 0 V
2 -1 V
2 0 V
2 -1 V
2 0 V
2 -1 V
2 0 V
2 -1 V
2 0 V
2 0 V
2 -1 V
2 0 V
2 0 V
2 0 V
2 0 V
2 -1 V
2 0 V
2 0 V
2 0 V
2 0 V
2 0 V
2 0 V
2 0 V
2 0 V
2 0 V
2 0 V
2 0 V
2 0 V
2 -1 V
2 0 V
2 0 V
2 0 V
2 0 V
2 0 V
2 0 V
2 -1 V
2 0 V
2 0 V
2 0 V
2 -1 V
2 0 V
2 0 V
2 -1 V
2 0 V
2 0 V
2 -1 V
2 0 V
2 0 V
2 -1 V
2 0 V
2 0 V
2 -1 V
2 0 V
2 0 V
2 -1 V
2 0 V
2 0 V
2 -1 V
2 0 V
2 0 V
2 0 V
2 -1 V
2 0 V
2 0 V
2 0 V
2 0 V
2 -1 V
2 0 V
2 0 V
2 0 V
2 0 V
2 0 V
2 0 V
2 0 V
2 0 V
2 0 V
2 0 V
2 0 V
2 0 V
2 0 V
2 0 V
2 0 V
2 0 V
2 0 V
2 0 V
2 0 V
2 0 V
2 1 V
2 0 V
2 0 V
2 0 V
2 0 V
2 -1 V
2 0 V
2 0 V
2 0 V
2 0 V
2 0 V
2 0 V
2 0 V
2 0 V
2 0 V
2 0 V
2 0 V
2 0 V
2 0 V
2 0 V
2 1 V
2 0 V
2 0 V
2 0 V
2 0 V
2 0 V
2 0 V
2 0 V
2 0 V
2 0 V
2 0 V
2 0 V
2 0 V
2 0 V
2 0 V
2 0 V
2 0 V
2 0 V
2 0 V
2 0 V
2 0 V
2 0 V
2 0 V
2 0 V
2 0 V
2 0 V
2 0 V
2 0 V
2 0 V
2 0 V
2 0 V
2 0 V
2 0 V
2 0 V
2 0 V
2 0 V
2 0 V
2 0 V
2 0 V
2 0 V
2 1 V
2 0 V
2 0 V
2 0 V
2 0 V
2 0 V
2 0 V
2 0 V
2 0 V
2 0 V
2 0 V
2 0 V
2 0 V
2 0 V
2 0 V
2 1 V
2 0 V
2 0 V
2 0 V
2 0 V
2 0 V
2 0 V
2 0 V
2 0 V
2 0 V
2 0 V
2 0 V
2 0 V
LT3
868 450 M
180 0 V
601 1161 M
2 1 V
2 0 V
2 1 V
2 1 V
2 1 V
2 1 V
2 0 V
2 1 V
2 1 V
2 1 V
2 1 V
2 1 V
2 1 V
2 1 V
2 0 V
2 1 V
2 1 V
2 1 V
2 1 V
2 1 V
2 1 V
2 0 V
2 1 V
2 1 V
2 0 V
2 1 V
2 0 V
2 1 V
2 0 V
2 1 V
2 0 V
2 0 V
2 0 V
2 0 V
2 0 V
2 0 V
2 0 V
2 0 V
2 0 V
2 -1 V
2 0 V
2 -1 V
2 -1 V
2 0 V
2 -1 V
2 -1 V
2 -1 V
2 -1 V
2 -1 V
2 -1 V
2 -1 V
2 -1 V
2 -2 V
2 -1 V
2 -1 V
2 -2 V
2 -1 V
2 -1 V
2 -2 V
2 -1 V
2 -2 V
2 -1 V
2 -1 V
2 -2 V
2 -1 V
2 -2 V
2 -1 V
2 -1 V
2 -1 V
2 -2 V
2 -1 V
2 -1 V
2 -1 V
2 -1 V
2 -1 V
2 -1 V
2 -1 V
2 -1 V
2 -1 V
2 -1 V
2 -1 V
2 -1 V
2 -1 V
2 -1 V
2 -1 V
2 -1 V
2 -1 V
2 -1 V
2 -1 V
2 -1 V
2 -1 V
2 -1 V
2 -1 V
2 -1 V
2 -1 V
2 -1 V
2 -1 V
2 -1 V
2 -1 V
2 -2 V
2 -1 V
2 -1 V
2 -2 V
2 -1 V
2 -2 V
2 -1 V
2 -2 V
2 -1 V
2 -2 V
2 -2 V
2 -2 V
2 -1 V
2 -2 V
2 -2 V
2 -2 V
2 -2 V
2 -2 V
2 -2 V
2 -2 V
2 -2 V
2 -1 V
2 -2 V
2 -2 V
2 -2 V
2 -2 V
2 -2 V
2 -1 V
2 -2 V
2 -2 V
2 -2 V
2 -1 V
2 -2 V
2 -1 V
2 -2 V
2 -1 V
2 -2 V
2 -1 V
2 -1 V
2 -2 V
2 -1 V
2 -1 V
2 -2 V
2 -1 V
2 -1 V
2 -2 V
2 -1 V
2 -1 V
2 -2 V
2 -1 V
2 -1 V
2 -2 V
2 -1 V
2 -2 V
2 -1 V
2 -2 V
2 -1 V
2 -2 V
2 -2 V
2 -1 V
2 -2 V
2 -2 V
2 -2 V
2 -2 V
2 -1 V
2 -2 V
2 -2 V
2 -2 V
2 -2 V
2 -2 V
2 -1 V
2 -2 V
2 -2 V
2 -2 V
2 -2 V
2 -1 V
2 -2 V
2 -2 V
2 -1 V
2 -2 V
2 -2 V
2 -1 V
2 -2 V
2 -2 V
2 -1 V
2 -2 V
2 -1 V
2 -2 V
2 -1 V
2 -2 V
2 -1 V
2 -2 V
2 -1 V
2 -2 V
2 -1 V
2 -1 V
2 -2 V
2 -1 V
2 -2 V
2 -1 V
2 -2 V
2 -1 V
2 -2 V
2 -1 V
2 -2 V
2 -1 V
2 -2 V
2 -1 V
2 -2 V
2 -1 V
2 -2 V
2 -1 V
2 -2 V
2 -1 V
2 -2 V
2 -1 V
2 -2 V
2 -1 V
2 -2 V
2 -1 V
2 -2 V
2 -1 V
2 -2 V
2 -1 V
2 -2 V
2 -1 V
2 -2 V
2 -1 V
2 -2 V
2 -1 V
2 -2 V
2 -1 V
2 -2 V
2 -2 V
2 -1 V
3 -2 V
2 -2 V
2 -1 V
2 -2 V
2 -2 V
2 -1 V
2 -2 V
2 -2 V
2 -1 V
2 -2 V
2 -2 V
2 -2 V
2 -1 V
2 -2 V
2 -2 V
2 -1 V
2 -2 V
2 -2 V
2 -1 V
2 -2 V
2 -2 V
2 -1 V
2 -2 V
2 -2 V
2 -1 V
2 -2 V
2 -1 V
2 -2 V
2 -2 V
2 -1 V
2 -2 V
2 -1 V
2 -2 V
2 -2 V
2 -1 V
2 -2 V
2 -1 V
2 -2 V
2 -1 V
2 -2 V
2 -1 V
2 -2 V
2 -1 V
2 -2 V
2 -2 V
2 -1 V
2 -2 V
2 -1 V
2 -2 V
2 -2 V
2 -1 V
2 -2 V
2 -1 V
2 -2 V
2 -2 V
2 -1 V
2 -2 V
2 -1 V
2 -2 V
2 -1 V
2 -2 V
2 -1 V
2 -2 V
2 -1 V
2 -2 V
2 -1 V
2 -2 V
2 -1 V
2 -1 V
2 -2 V
2 -1 V
2 -2 V
2 -1 V
2 -2 V
2 -1 V
2 -1 V
2 -2 V
2 -1 V
2 -2 V
2 -1 V
2 -2 V
2 -2 V
2 -1 V
2 -2 V
2 -1 V
2 -2 V
2 -2 V
2 -1 V
2 -2 V
2 -2 V
2 -1 V
2 -2 V
2 -2 V
2 -1 V
2 -2 V
2 -2 V
2 -1 V
2 -2 V
2 -1 V
2 -2 V
2 -2 V
2 -1 V
2 -2 V
2 -1 V
2 -2 V
2 -1 V
2 -2 V
2 -1 V
2 -2 V
2 -1 V
2 -1 V
2 -2 V
2 -1 V
2 -2 V
2 -1 V
2 -1 V
2 -2 V
2 -1 V
2 -1 V
2 -2 V
2 -1 V
2 -2 V
2 -1 V
2 -2 V
2 -1 V
2 -1 V
2 -2 V
2 -1 V
2 -2 V
2 -1 V
2 -2 V
2 -1 V
2 -2 V
2 -1 V
2 -2 V
2 -1 V
2 -2 V
2 -1 V
2 -2 V
2 -1 V
2 -1 V
2 -2 V
2 -1 V
2 -1 V
2 -1 V
2 -2 V
2 -1 V
2 -1 V
2 -1 V
2 -1 V
2 -2 V
2 -1 V
2 -1 V
2 -1 V
2 -1 V
2 -1 V
2 -1 V
2 -1 V
2 -1 V
2 -1 V
2 -1 V
2 -1 V
2 -1 V
currentpoint stroke M
2 -2 V
2 -1 V
2 -1 V
2 -1 V
2 -2 V
2 -1 V
2 -1 V
2 -1 V
2 -2 V
2 -1 V
2 -2 V
2 -1 V
2 -1 V
2 -2 V
2 -1 V
2 -2 V
2 -1 V
2 -2 V
2 -1 V
2 -1 V
2 -2 V
2 -1 V
2 -1 V
2 -2 V
2 -1 V
2 -1 V
2 -1 V
2 -2 V
2 -1 V
2 -1 V
2 -1 V
2 -1 V
2 -2 V
2 -1 V
2 -1 V
2 -1 V
2 -1 V
2 -1 V
2 -1 V
2 -2 V
2 -1 V
2 -1 V
2 -1 V
2 -1 V
2 -1 V
2 -1 V
2 -1 V
2 -1 V
2 -1 V
2 -1 V
2 -1 V
2 -1 V
2 -1 V
2 -1 V
2 -1 V
2 -1 V
2 -1 V
2 -1 V
2 -1 V
2 -1 V
2 -1 V
2 0 V
2 -1 V
2 -1 V
2 -1 V
2 -1 V
2 0 V
2 -1 V
2 -1 V
2 -1 V
2 -1 V
2 0 V
2 -1 V
2 -1 V
2 -1 V
2 -1 V
2 -1 V
2 -1 V
2 -1 V
2 -1 V
2 -1 V
2 -1 V
2 -1 V
2 -1 V
2 -1 V
2 -1 V
2 -1 V
2 -1 V
2 -2 V
2 -1 V
2 -1 V
2 -1 V
2 -1 V
2 -2 V
2 -1 V
2 -1 V
2 -1 V
2 -1 V
2 -1 V
2 -1 V
2 -2 V
2 -1 V
2 -1 V
2 -1 V
2 -1 V
2 -1 V
2 0 V
2 -1 V
2 -1 V
2 -1 V
2 -1 V
2 -1 V
2 -1 V
2 -1 V
2 -1 V
2 -1 V
2 -1 V
2 -1 V
2 -1 V
2 0 V
2 -1 V
2 -1 V
2 -1 V
2 -1 V
2 -1 V
2 -1 V
2 -1 V
2 -1 V
2 -1 V
2 -1 V
2 -1 V
2 -1 V
2 0 V
2 -1 V
2 -1 V
2 -1 V
2 -1 V
2 -1 V
2 -1 V
2 0 V
2 -1 V
2 -1 V
2 -1 V
2 -1 V
2 -1 V
2 -1 V
2 -1 V
2 0 V
2 -1 V
2 -1 V
2 -1 V
2 -1 V
2 -1 V
2 -1 V
2 -1 V
2 0 V
2 -1 V
2 -1 V
2 -1 V
2 -1 V
2 0 V
2 -1 V
2 -1 V
2 0 V
2 -1 V
2 -1 V
2 0 V
2 -1 V
2 -1 V
2 0 V
2 -1 V
2 -1 V
2 0 V
2 -1 V
2 0 V
2 -1 V
2 -1 V
2 0 V
2 -1 V
2 0 V
2 -1 V
2 -1 V
2 0 V
2 -1 V
2 -1 V
2 0 V
2 -1 V
2 -1 V
2 -1 V
2 0 V
2 -1 V
2 -1 V
2 -1 V
2 0 V
2 -1 V
2 -1 V
2 -1 V
2 -1 V
2 0 V
2 -1 V
2 -1 V
2 -1 V
2 -1 V
2 -1 V
2 -1 V
2 -1 V
2 -1 V
2 0 V
2 -1 V
2 -1 V
2 -1 V
2 -1 V
2 -1 V
2 -1 V
2 -1 V
2 -1 V
2 -1 V
2 -1 V
2 -1 V
2 -1 V
2 -1 V
2 -1 V
2 -1 V
2 -1 V
2 -1 V
2 -1 V
2 -1 V
2 -1 V
2 -1 V
2 -1 V
2 -1 V
2 -1 V
2 -1 V
2 -1 V
2 0 V
2 -1 V
2 -1 V
2 -1 V
2 -1 V
2 -1 V
2 -1 V
2 0 V
2 -1 V
2 -1 V
2 -1 V
2 -1 V
2 -1 V
2 -1 V
2 0 V
2 -1 V
2 -1 V
2 -1 V
2 -1 V
2 -1 V
2 0 V
2 -1 V
2 -1 V
2 -1 V
2 -1 V
2 -1 V
2 -1 V
2 -1 V
2 0 V
2 -1 V
2 -1 V
2 -1 V
2 -1 V
2 -1 V
2 -1 V
2 0 V
2 -1 V
2 -1 V
2 -1 V
2 0 V
2 -1 V
2 -1 V
2 0 V
2 -1 V
2 -1 V
2 0 V
2 -1 V
2 -1 V
2 0 V
2 -1 V
2 0 V
2 -1 V
2 0 V
2 -1 V
2 0 V
2 -1 V
2 0 V
2 0 V
2 -1 V
2 0 V
2 -1 V
2 0 V
2 -1 V
2 0 V
2 0 V
2 -1 V
2 0 V
2 -1 V
2 0 V
2 0 V
2 -1 V
2 0 V
2 -1 V
2 0 V
2 0 V
2 -1 V
2 0 V
2 -1 V
2 0 V
2 -1 V
2 0 V
2 -1 V
2 0 V
2 -1 V
2 0 V
2 -1 V
2 -1 V
2 0 V
2 -1 V
2 0 V
2 -1 V
3 -1 V
2 0 V
2 -1 V
2 -1 V
2 0 V
2 -1 V
2 0 V
2 -1 V
2 -1 V
2 0 V
2 -1 V
2 0 V
2 -1 V
2 0 V
2 -1 V
2 0 V
2 -1 V
2 0 V
2 0 V
2 -1 V
2 0 V
2 -1 V
2 0 V
2 -1 V
2 0 V
2 0 V
2 -1 V
2 0 V
2 -1 V
2 0 V
2 -1 V
2 0 V
2 -1 V
2 0 V
2 -1 V
2 0 V
2 -1 V
2 0 V
2 -1 V
2 0 V
2 -1 V
2 0 V
2 0 V
2 -1 V
2 0 V
2 -1 V
2 0 V
2 -1 V
2 0 V
2 0 V
2 -1 V
2 0 V
2 0 V
2 -1 V
2 0 V
2 0 V
2 -1 V
2 0 V
2 0 V
2 0 V
2 -1 V
2 0 V
2 0 V
2 -1 V
2 0 V
2 0 V
2 0 V
2 -1 V
2 0 V
2 0 V
2 -1 V
2 0 V
2 0 V
2 -1 V
2 0 V
currentpoint stroke M
2 0 V
2 -1 V
2 0 V
2 -1 V
2 0 V
2 -1 V
2 0 V
2 -1 V
2 0 V
2 -1 V
2 0 V
2 -1 V
2 -1 V
2 0 V
2 -1 V
2 0 V
2 -1 V
2 -1 V
2 0 V
2 -1 V
2 0 V
2 -1 V
2 0 V
2 -1 V
2 0 V
2 -1 V
2 0 V
2 -1 V
2 0 V
2 -1 V
2 0 V
2 -1 V
2 0 V
2 -1 V
2 0 V
2 0 V
2 -1 V
2 0 V
2 -1 V
2 0 V
2 -1 V
2 0 V
2 -1 V
2 0 V
2 0 V
2 -1 V
2 0 V
2 -1 V
2 0 V
2 -1 V
2 0 V
2 0 V
2 -1 V
2 0 V
2 0 V
2 -1 V
2 0 V
2 0 V
2 -1 V
2 0 V
2 0 V
2 0 V
2 0 V
2 -1 V
2 0 V
2 0 V
2 0 V
2 0 V
2 0 V
2 0 V
2 0 V
2 -1 V
2 0 V
2 0 V
2 0 V
2 0 V
2 0 V
2 0 V
2 0 V
2 0 V
2 0 V
2 0 V
2 0 V
2 0 V
2 1 V
2 0 V
2 0 V
2 0 V
2 0 V
2 0 V
2 0 V
2 0 V
2 0 V
2 -1 V
2 0 V
2 0 V
2 0 V
2 0 V
2 0 V
2 0 V
2 0 V
2 0 V
2 0 V
2 0 V
2 -1 V
2 0 V
2 0 V
2 0 V
2 0 V
2 0 V
2 0 V
2 -1 V
2 0 V
2 0 V
2 0 V
2 0 V
2 0 V
2 0 V
2 -1 V
2 0 V
2 0 V
2 0 V
2 0 V
2 0 V
2 0 V
2 0 V
2 -1 V
2 0 V
2 0 V
2 0 V
2 0 V
2 0 V
2 0 V
2 0 V
2 0 V
2 0 V
2 0 V
2 0 V
2 0 V
2 -1 V
2 0 V
2 0 V
2 0 V
2 0 V
2 0 V
2 0 V
2 0 V
2 0 V
2 0 V
2 0 V
2 0 V
2 0 V
2 0 V
2 0 V
2 0 V
2 0 V
2 0 V
2 0 V
2 0 V
2 0 V
2 0 V
2 0 V
2 0 V
2 0 V
2 1 V
2 0 V
2 0 V
2 0 V
2 0 V
2 0 V
2 0 V
2 0 V
2 0 V
2 1 V
2 0 V
2 0 V
2 0 V
2 0 V
2 0 V
2 1 V
2 0 V
2 0 V
2 0 V
2 0 V
2 1 V
2 0 V
2 0 V
2 0 V
2 0 V
2 1 V
2 0 V
2 0 V
2 0 V
2 0 V
2 0 V
2 1 V
2 0 V
2 0 V
2 0 V
2 0 V
2 0 V
2 0 V
stroke
grestore
end
showpage
}
\put(808,450){\makebox(0,0)[r]{4}}
\put(808,550){\makebox(0,0)[r]{3}}
\put(808,650){\makebox(0,0)[r]{2}}
\put(808,750){\makebox(0,0)[r]{1}}
\put(1603,0){\makebox(0,0){$r \;\;\; [{\mathcal L}]$}}
\put(240,1018){%
\special{ps: gsave currentpoint currentpoint translate
270 rotate neg exch neg exch translate}%
\makebox(0,0)[b]{\shortstack{$\Delta F \;\;\; [{\mathcal C}^2/{\mathcal L}^2]$}}%
\special{ps: currentpoint grestore moveto}%
}
\put(2453,151){\makebox(0,0){8}}
\put(2221,151){\makebox(0,0){7}}
\put(1990,151){\makebox(0,0){6}}
\put(1758,151){\makebox(0,0){5}}
\put(1527,151){\makebox(0,0){4}}
\put(1295,151){\makebox(0,0){3}}
\put(1063,151){\makebox(0,0){2}}
\put(832,151){\makebox(0,0){1}}
\put(600,151){\makebox(0,0){0}}
\put(540,1602){\makebox(0,0)[r]{$10^{-5}$}}
\put(540,1219){\makebox(0,0)[r]{$10^{-6}$}}
\put(540,835){\makebox(0,0)[r]{$10^{-7}$}}
\put(540,452){\makebox(0,0)[r]{$10^{-8}$}}
\end{picture}